\documentclass[a4paper,fleqn]{cas-sc}

\usepackage{amsmath,amssymb}
\usepackage{float}
\usepackage{caption}
\usepackage{subcaption}
\usepackage{mathtools,amsfonts}
\usepackage{colortbl}
\definecolor{Black}{rgb}{0,0,0}

\usepackage{framed}
\usepackage{changepage}

\usepackage[utf8x]{inputenc}
\usepackage{tgpagella}
\usepackage[right]{lineno}

\usepackage{array}

\usepackage{graphicx}

\usepackage[square,sort,comma,numbers]{natbib}

\begin{document}
	\let\WriteBookmarks\relax
	\def\floatpagepagefraction{1}
	\def\textpagefraction{.001}
	\shorttitle{Modeling Bottom-Up and Top-Down Attention with a Neurodynamic Model of V1}
	\shortauthors{David Berga \& Xavier Otazu}

	\title [mode = title]{Modeling Bottom-Up and Top-Down Attention with a Neurodynamic Model of V1}                      
	%\tnotemark[1,2]
	
	%\tnotetext[1]{This document is the results of the research project funded by the National Science Foundation.}
	
	%\tnotetext[2]{The second title footnote which is a longer text matter to fill through the whole text width and overflow into another line in the footnotes area of the first page.}

	\author[1]{David Berga}[%type=editor,
	%auid=000,
	%bioid=1,
	%prefix=Sir,
	%role=Researcher,
	orcid=0000-0001-7543-2770
	]
	\cormark[1]
	\cortext[cor1]{Corresponding author}
	\ead{dberga@cvc.uab.es}
	\author[1,2]{Xavier Otazu}[%type=editor,
	%auid=000,
	%bioid=1,
	%prefix=Sir,
	%role=Researcher,
	orcid=0000-0002-4982-791X
	]
	%\fnmark[1]
	
	%\credit{Conceptualization of this study, Methodology, Software}
	
	\address[1]{Computer Vision Center, Edifici O, Campus UAB, 08193 Bellaterra (Barcelona), Spain}
	
	%\fnmark[2]
	\address[2]{Universitat Aut\`onoma de Barcelona, Edifici Q, Campus UAB, 08193 Bellaterra (Barcelona), Spain}

	\begin{abstract}
		Previous studies suggested that lateral interactions of V1 cells are responsible, among other visual effects, of bottom-up visual attention (alternatively named visual salience or saliency). Our objective is to mimic these connections with a neurodynamic network of firing-rate neurons in order to predict visual attention. Early visual subcortical processes (i.e. retinal and thalamic) are functionally simulated. An implementation of the cortical magnification function is included to define the retinotopical projections towards V1, processing neuronal activity for each distinct view during scene observation. Novel computational definitions of top-down inhibition (in terms of inhibition of return and selection mechanisms), are also proposed to predict attention in Free-Viewing and Visual Search tasks. Results show that our model outpeforms other biologically-inpired models of saliency prediction while predicting visual saccade sequences with the same model. We also show how temporal and spatial characteristics of inhibition of return can improve prediction of saccades, as well as how distinct search strategies (in terms of feature-selective or category-specific inhibition) can predict attention at distinct image contexts.
\vspace{-1.5em}
	\end{abstract}
	

	\maketitle

%	\linenumbers
	
	%\input{header_elsevier.tex}

\vspace{-1.5em}
% Use "Eq" instead of "Equation" for equation citations.
\section{Introduction}

%brief visual attention literature
The human visual system (HVS) structure has evolved in a way to efficiently discriminate redundant information \cite{Shannon1948,Barlow2001RedundancyRR,Zhaoping2016}. In order to filter or select the information to be processed in higher areas of visual processing in the brain, the HVS guides eye movements towards regions that appear to be visually conspicuous or distinct in the scene. This phenomenon was observed during visual search tasks \cite{Treisman1980,Wolfe1989}, where detecting early visual features (such as orientation, color or size) was done in parallel (pre-attentively) or required either a serial "binding" step depending on scene context. Koch \& Ullman \cite{Koch1987} came up with the hypothesis that neuronal mechanisms involved in selective visual attention generate a unique "master" map from visual scenes, coined with the term "saliency map". From that, Itti, Koch \& Niebur \cite{Itti1998} presented a computational implementation of the aforementioned framework (IKN), inspired by the early mechanisms of the HVS. It was done by extracting properties of the image as feature maps (using a pyramid of difference-of-gaussian filters at distinct orientations, color and intensity), obtaining feature-wise conspicuity by computing  center-surround differences as receptive field responses and integrating them on a unique map using winner-take-all mechanisms. Such framework served as a starting point for saliency modeling \cite{Borji2013c,Zhang2013}, which derived in a myriad of computational models, that differed in their computations but conserved a similar pipeline. From a biological perspective, further hypotheses suggested that primates' visual system structure was mainly connected to the efficient coding principle. Later studies considered that maximizing information of scenes was the key factor on forming visual feature representations. To test that, Bruce \& Tsotsos \cite{Bruce2005} implemented a saliency model (AIM) by extracting sparse representations of image statistics (using independent component analysis). These representations were found to be remarkably similar to cells in V1, which follow similar spatial properties to Gabor filters \cite{Olshausen1996}. 

While the current concept of saliency maps is to predict probabilities of specific spatial locations as candidates of eye movements, it is also crucial to understand how to predict individual fixations or saccade sequences (also named "scanpaths"). Scanpath predictions were formerly done through probabilistic measures of saccade amplitude statistics. These followed a similar heavy-tailed distribution to a Cauchy-Levy (in reference to random walks or "Levy flights", minimizing global uncertainty) \cite{Brockmann1999}, with highest probability of fixations at a low saccade amplitude. This procedure was implemented in Boccignone \& Ferraro's model \cite{Boccignone2004}, taking saliency from IKN. Later, LeMeur \& Liu \cite{LeMeur2015} proposed a more biologically-plausible approach, accounting for oculomotor biases and inhibition of return effects. It used a graph-based saliency model (GBVS, also inspired by IKN) \cite{Harel2006}, with a higher probability to catch grouped fixations (which tend to be in stimulus center).

%Bell1997
In order to evaluate model predictions with eye movement data, certain patterns underlying human eye movement behavior need to be accounted for a more detailed description and analysis of visual attention. These effects are found to be dependent on context, discriminability, temporality, task and memory during scene viewing and visual search \cite{Bruce2015, Berga2018a}. Attention and spatial selection, therefore, is also dependent on the neuronal activations from both bottom-up and top-down mechanisms. These processes are known to compete \cite{Desimone1995} to form a unique representation, termed priority map \cite{Egeth1997}. These hypotheses suggest that attention is separated in distinct stages (pre-attentive as bottom-up and attentive as top-down) and that contributions towards guiding eye movements are simultaneously affected by distinct mechanisms in the HVS \cite{WhiteMunoz2011}. This competition for visual priority is biased by a term called relevance (as opposed to saliency), where top-down attention is driven by task demands, working and semantic memory as well as episodic memory, emotion and motivation (3 of which seem to be unique for each individual and momentum)\cite{rolls2008memory}. At that end, it is stated \cite{Tsotsos1995,Huang2007} that visual selection relies on activations from higher-level layers towards lower-level receptive fields. Therefore, modelization of attention should consider as well the influences of task and many other top-down effects.

\subsection{Objectives}

Initial hypotheses by Li \cite{Li2002,zhaoping2014understanding} suggested that visual saliency is processed by the lateral interactions of V1 cells. In their work, pyramidal cells and interneurons in the primary visual cortex (V1, Brodmann Area 17 or striate cortex) and their horizontal intracortical connections are seen to modulate activity in V1. Li's neurodynamic model \cite{Li1998} of excitatory and inhibitory firing-rate neurons was able to determine how contextual influences of visual scenes contribute to the formation of saliency. In this model, interactions between neurons tuned to specific orientation sensitivities served as predictors of pop-out effects and search asymmetries \cite{Li1999}. Li's neurodynamic model was later extended by Penacchio et al. \cite{Penacchio2013} proposing the aforementioned lateral interactions to also be responsible for brightness induction mechanisms. By considering neuron orientation selectivity at distinct spatial scales, this model can act as a contrast enhancement mechanism of a particular visual area depending of induced activity from surrounding regions. Latest work from Berga \& Otazu \cite{Berga2018b} has shown that the same model (without changing its parametrization) is able to predict saliency using real and synthetic color images. We propose to extend the model providing saliency computations with foveation, concerning distinct viewpoints during scene observation (mapping retinal projections towards V1 retinotopy) as a main hypothesis for predicting visual scanpaths. Furthermore, we also test how the model is able to provide predictions considering recurrent feedback mechanisms of already visited regions, as well as from visual feature and exemplar search tasks with top-down inhibition mechanisms.

\subsection{A unified model of V1 predicts several perceptual processes}

Here we present a novel neurodynamic model of visual attention and we remark its biological plausability as being able to simultaneously reproduce other effects such as Brightness Induction \cite{Penacchio2013}, Chromatic Induction \cite{Cerda2016} and Visual Discomfort \cite{Penacchio2016} effects in previous work. Brightness and Chromatic induction stand for the variation of perceived luminance and color of a visual target depending on its luminance and/or chromatic properties as well as for its surrounding area respectively. Thus, a visual target can be perceived as being different (contrast) or similar (assimilation) to its physical properties by varying its surrounding context. With the simulations of our model, the output of V1's neuronal activity (coded as firing-rates during several cycles of excitatory-inhibitory V1 interneuron interactions), is used as predictor of induction and saliency respectively. These responses will act as a contrast enhancement mechanism, which for the case of saliency, are integrated towards projections in the superior colliculus (SC) for eye movement control. Therewith, our model has also been able to reproduce visual discomfort, as relative contrast energy of particular region on a scene is found to produce hyperexcitability in V1 \cite{Penacchio2015,Le2017}, one of possible causes of producing certain conditions such as malaise, nausea or even migraine. Previous neurodynamic \cite{Deco2004,Gu2007,Chevallier2010,CoenCagli2012,Chang2014,Mari2018} and saliency models \cite{Borji2013c,Zhang2013,Bylinskii2015} have been able to predict eye movements. However, most of these models have been built specifically for visual saliency in a free-viewing task, a characteristic that denies their biological plausibility for modeling distinct visual processing mechanisms or other visual processes simultaneously. On behalf of model biological plasusibility on V1 function and its computations, we present a unified model of lateral connections in V1, able to predict attention (both in free-viewing and visual search) from real and synthetic color images while mimicking physiological properties of the neural circuitry stated previously. %Chang2014
%Visual scenes are projected to the retinal photoreceptors (RP), processed by retinal ganglion cells (RGC), and later projected from lateral geniculate nucleus (LGN) pathways towards V1 receptive fields (RF).

%Similarly to previous BiWaM, CiWaM and SIM . 

\section{Model}

\subsection{Retinal and LGN responses}

\begin{figure}[h]
	\centering
	\begin{subfigure}{0.30\linewidth}
		\includegraphics[width=\linewidth]{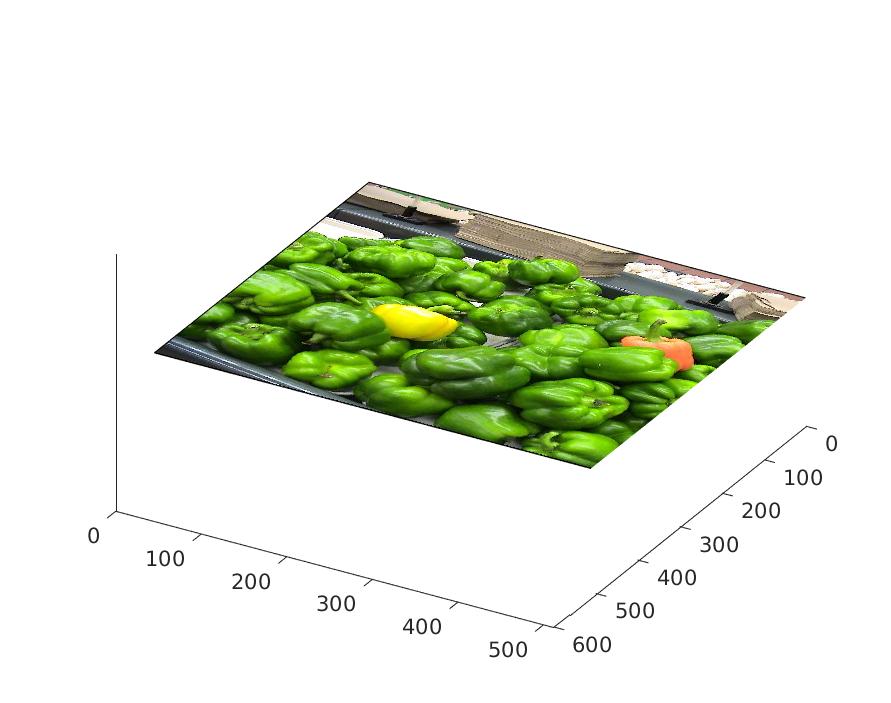}
		\caption*{\centering Image }
	\end{subfigure}
	\begin{subfigure}{0.30\linewidth}
		\includegraphics[width=\linewidth]{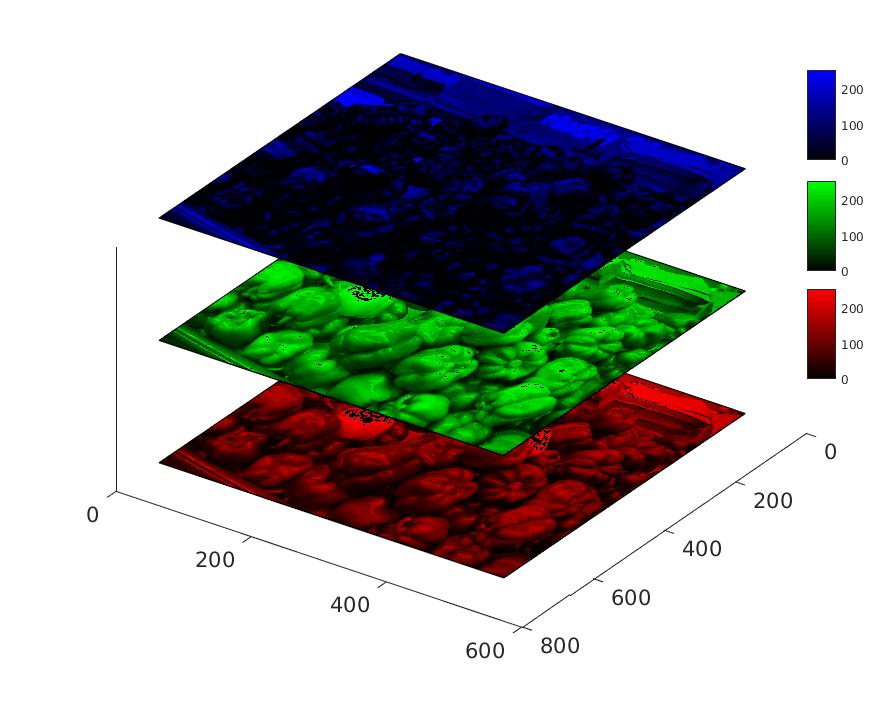}
		\caption*{\centering RGB components}
	\end{subfigure}
	\\
	\begin{subfigure}{0.30\linewidth}
		\includegraphics[width=\linewidth]{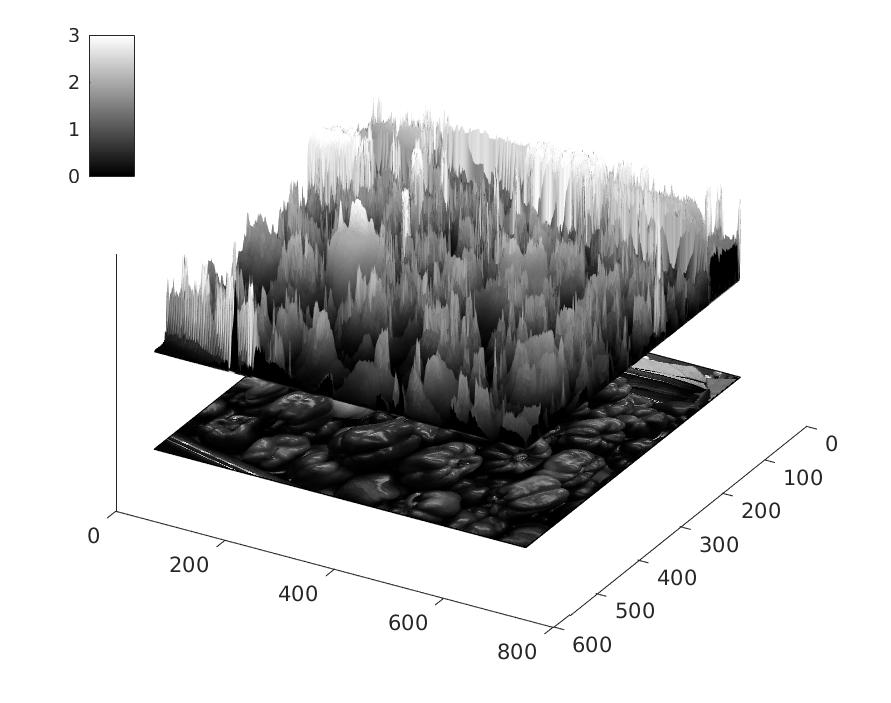}
		\caption*{\centering L* (M-)}
	\end{subfigure}
	\begin{subfigure}{0.30\linewidth}
		\includegraphics[width=\linewidth]{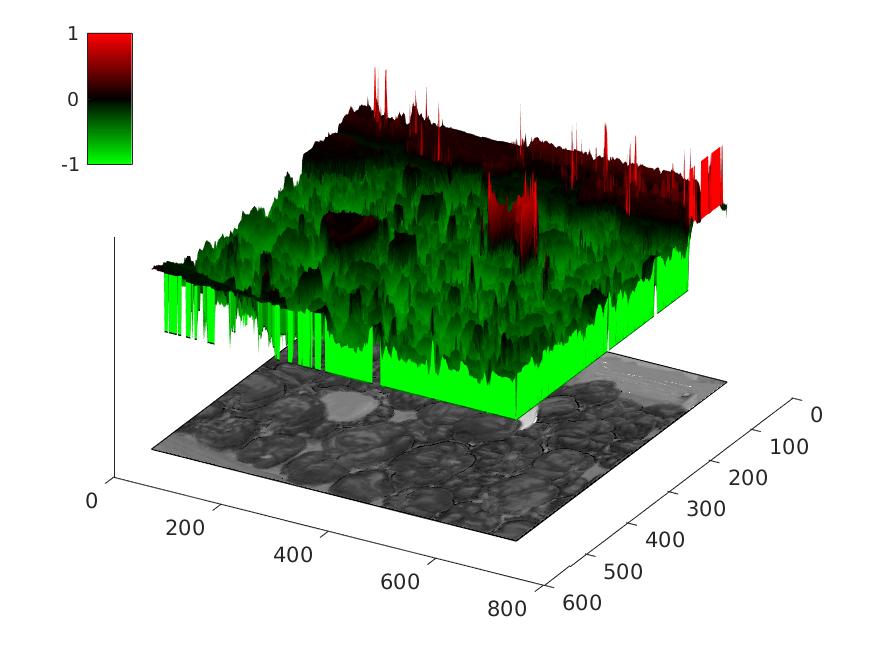}
		\caption*{\centering a* (P-)}
	\end{subfigure}
	\begin{subfigure}{0.30\linewidth}
		\includegraphics[width=\linewidth]{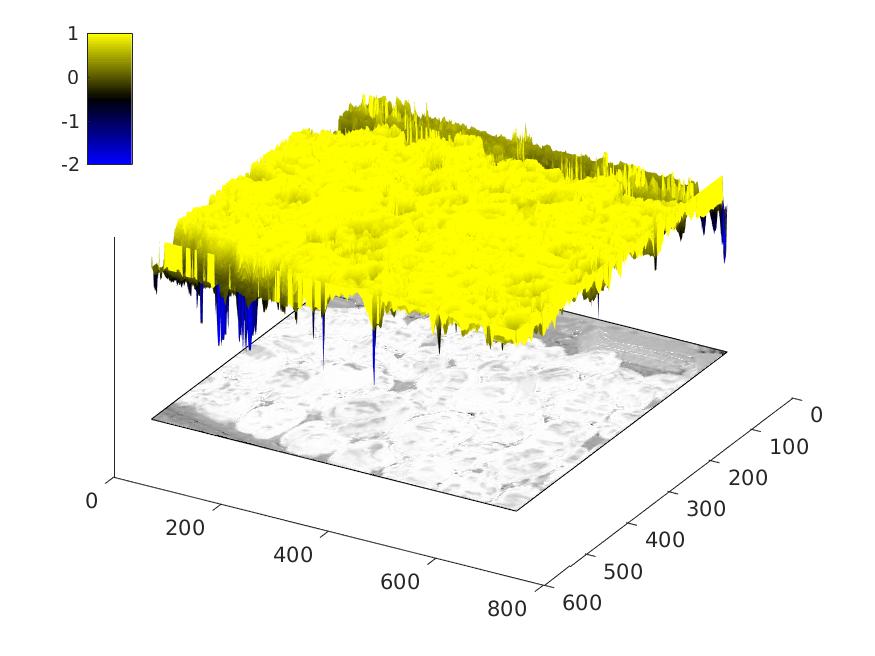}
		\caption*{\centering b* (K-)}
	\end{subfigure}
	\caption{Example of CIELAB components of color opponencies given a sample image, corresponding to $L^*$ (Intensity), $a^*$ (Red-Green) and $b^*$ (Blue-Yellow).}
	\label{fig:opp}
\end{figure}

The HVS perceives the light at distinct wavelengths of the visual spectrum and separates them to distinct channels for further processing in the cortex. First, retinal photoreceptors (or RP, corresponding to rod and cone cells) are photosensitive to luminance (rhodopsin-pigmented) and color (photopsin-pigmented) \cite{Solomon2007,Imamoto2014}. Mammal cone cells are photosensitive to distinct wavelengths between a range of $\sim400-700nm$, corresponding to three cell types, measured to be maximally responsive to Long (L, $\lambda_{max}\simeq560$nm), Medium (M, $\lambda_{max}\simeq530$nm) and Short (S, $\lambda_{max}\simeq430$nm) wavelengths respectively \cite{Stockman1993}. RP signals are received by retinal ganglion cells (or RGC) forming an opponent process \cite{Sincich2005}. This opponent process allows to model midget, bistratified and parasol cells as "Red vs Green", "Blue vs Yellow", and "Light vs Dark" channels. In order to simulate these chromatic and light intensity opponencies using digital images, we transformed the RGB color space to the CIELAB ($Lab$ or $L^*a^*b^*$) space (including a gamma correction of $\gamma_{RGB}$=1/2.2), as exemplified in \hyperref[fig:opp]{Fig. \ref*{fig:opp}}. %Billmeyer1983

\vspace{-2em}
\noindent

\vspace{-2em}
\footnotesize
\begin{align}
\begin{split} 
L^*=R+G+B,\\ 
a^*=\frac{R-G}{L^*},\\ 
b^*=\frac{R+G-2B}{L^*}.
\end{split}
\end{align}
\normalsize

The $L^*$, $a^*$ and $b^*$ channels form a cubic color space \cite{Lennie1990} with RGB opponencies ($+L$=lighter, $-L$=darker, $+a$=reddish, $-a$=greenish, $+b$=yellowish and $-b$=blueish).

Later, receptive fields in RGC \cite{Sincich2005} are activated in a center-surround fashion, receiving ON-OFF responses, being connected to horizontal (H-cell) and bipolar cell (B-cell) upstream circuitry. B-cells are hyperpolarized (OFF) or depolarized (ON) according to RP activity. In conjunction, H-cells send excitatory (center) and inhibitory feedback (surround) to RP. Midget (R-G), bistratified (B-Y) and parasol (L-D) RGC signals are sent through the optic nerve towards Parvo-, Konio- and Magno-cellular pathways in LGN respectively. %Nassi2009

\subsection{V1 Hypercolumnar organization}

RGC center-surround responses are sent to LGN and projected to V1 cells. V1's cortical hypercolumns encode similar features of orientation-selective cells at different spatial frequencies. Simple cells found in V1 receptive fields (RFs) are sensitive to center-surround responses at distinct orientations, whereas complex cells overlap ON and OFF regions (and can be modeled as a combination of simple cell responses). Parvo- (P- or $\beta$), Konio- (K- or $\gamma$) and Magno-cellular (M- or $\alpha$) pathways send signals separately towards distinct layers of the striate cortex (correspondingly projecting to $4C\beta$ \& 6 from "P-", $2/3$ \& 4A from "K-" and $4C\alpha$ \& 6 from "M-" cell pathways) for parallel and recurrent processing in V1.  %, obtaining distinct responses for position and width summation. 

\begin{figure}[h!] %xo: cambiar contrast dels colormaps DWT
	\centering
	\begin{subfigure}{0.65\linewidth}
		\includegraphics[width=\textwidth]{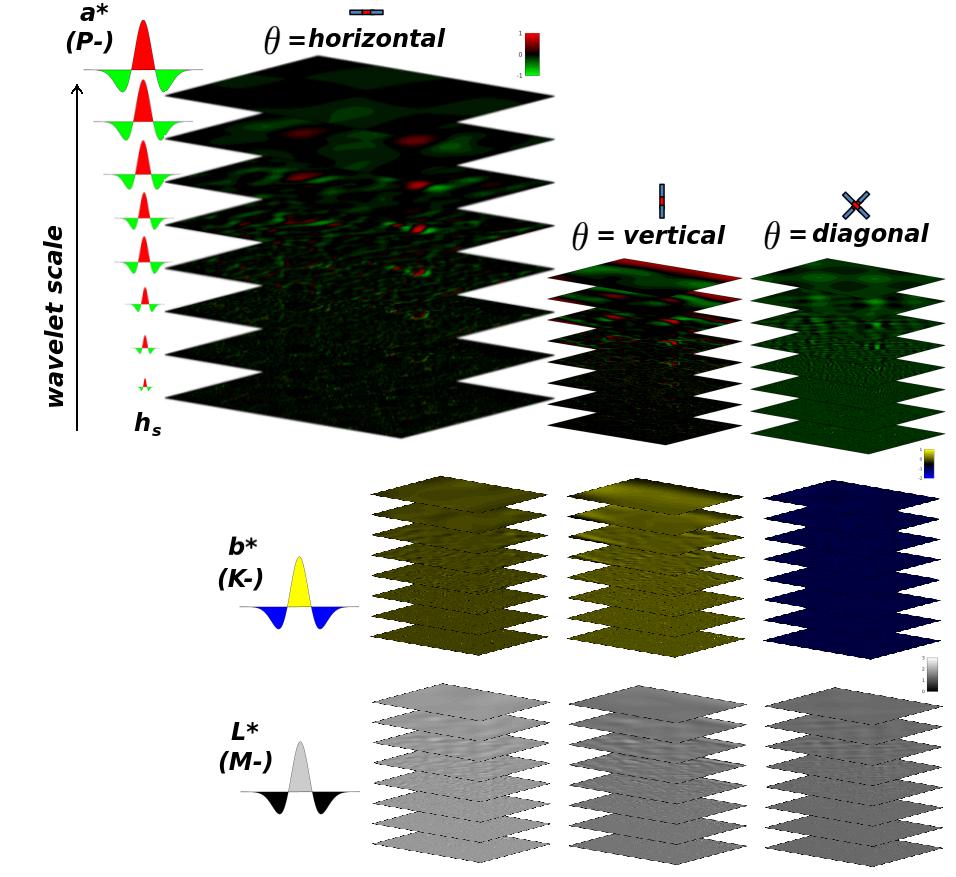}
	\end{subfigure}
	\caption{Representation of wavelet coefficients ($\omega_{iso\theta}$), in conjunction with the output of "a-trous" wavelet transform applied to components ($o=L^*,a^*,b^*$) shown in \hyperref[fig:opp]{Fig. \ref*{fig:opp}}.}
	\label{fig:dwt}
\end{figure}

We modeled the input to V1's simple cell responses with a 2D "a-trous" wavelet transform \cite{GonzlezAudcana2005}. Discrete wavelet transforms allow to process signals by extracting information of orientation and scale-dependent features in the visual space (feature maps), which we used for filtering each of the aforementioed opponencies separately, shown in \hyperref[fig:dwt]{Fig. \ref*{fig:dwt}}. Although these computations cannot be considered exact to each separate process of RGC and LGN, the transform seemingly resembles bottom-up activity projected to V1. The "a-trous" transform is undecimated and invertible, and allows to perform a transform where its basis functions remain similar to Gabor filters.

The "a trous" wavelet transform can be defined as:
%alinear formulas
\footnotesize
%\[h_1=\frac{1}{16}\begin{bmatrix}
%1&4&6&4&1
%\end{bmatrix}\]
\begin{align}
\begin{split}
\omega_{s,h}=c_{s-1}-c_{s,h}, \label{eq:wavelets2} \\
\omega_{s,v}=c_{s-1}-c_{s,v}, \\
\omega_{s,d}=c_{s-1}-(c_{s,h}\otimes h_s' + \omega_{s,h}+\omega_{s,v}), \\
c_s = c_{s-1} - (\omega_{s,h}+\omega_{s,v}+\omega_{s,d}). 
\end{split}
\end{align}

\vspace{-1em}
\noindent
\normalsize
where

\vspace{-1em}
\footnotesize
\begin{align}
\begin{split} 
%c_{s}= c_{s-1} \otimes h_s \label{eq:wavelets1}\\
%c_{s}'= c_{s-1} \otimes h_s'
c_{s,h}= c_{s-1} \otimes h_s, \label{eq:wavelets1}\\
c_{s,v}= c_{s-1} \otimes h_s'.
\end{split}
\end{align}
\normalsize

By transposing the wavelet filter ($h_s$, expressed in \hyperref[fig:dwt]{Fig. \ref*{fig:dwt}}) and dilating it at distinct spatial scales ($s=1...S$), we can obtain a set of wavelet approximation planes ($c_{s,\theta}$), that are combined for calculating wavelet coefficients ($\omega_{s,\theta}$) at distinct orientation selectivities ($\theta=h,v,d$). From these equations, three orientation selectivities can be extracted, corresponding to horizontal ($\theta_h\simeq\ $   $\{0\pm30||180\pm30\}$º), vertical ($\theta_v\simeq$ $\{90\pm30||270\pm30\}$º) and diagonal ($\theta_d\simeq\{45\pm15||135\pm15||225\pm15||315\pm15\}$º) angles. For the case of scale features, sensititivies to size (in degree of visual angle) correspond to $2^{s_0(s-1)}/\{pxva\}$, where "$pxva$" is the number of pixels for each degree of visual angle according to experimentation, and $s_0$=8, is the minimum size of the wavelet filter ($h_0$) defining the first the scale frequency sensitivity. Initial $c_0=I_o$ is obtained from the CIE L*a*b* components and $c_n$ corresponds to the residual plane of the last wavelet component (e.g. $s=n$). The image inverse ($I'_o$) can be obtained by integrating the wavelet $\omega_{s,\theta}$ and residual planes $c_n$: %($h_s$)

\vspace{-1em}
\footnotesize
\begin{align}
\begin{split}
I_o'=\sum_{s=1, \theta=h,v,d}^{n} \omega_{s,\theta} + c_n. \label{eq:wavelets3}
\end{split}
\end{align}
\normalsize

%The image inverse ($I'_o$) corresponds to applying a feature fusion (reversing the feature extraction procedure) by considering both wavelet ($\omega_s$) and residual ($c_s$) planes, which summated form the original signal ($I_o$) using the inverse wavelet transform.

%\paragraph{Orientation Features}
% \begin{itemize}
%  \item Horizontal orientations correspond to features with a phase angle between $\{...\}$ and $\{...13/12\pi\}$ %-15..15,165-195
%\item Vertical orientations correspond to feature angles of $\{5\pi/12...7\pi/12\}$ and $\{17/12\pi...19/12\pi\}$
%\item Diagonal orientations correspond to feature angles of $\{5\pi/12...7\pi/12\}$ and $\{17/12\pi...19/12\pi\}$
%\end{itemize}
%\paragraph{Scale Features}
%\paragraph{Luminance and Chromatic Features}

\subsection{Cortical mapping} \label{sec:cortical}

\begin{figure}[H] 
	\centering
	\begin{subfigure}{0.50\linewidth}
		\centering
		\includegraphics[width=0.43\linewidth,height=2cm]{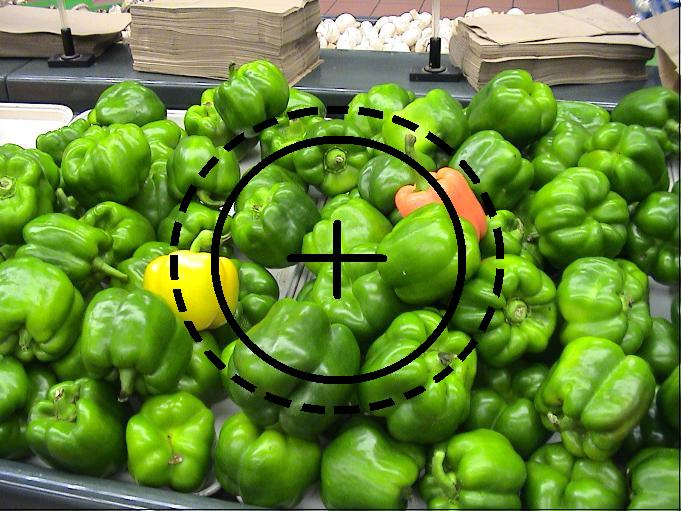} $\Rightarrow$
		\includegraphics[width=0.43\linewidth,height=2cm]{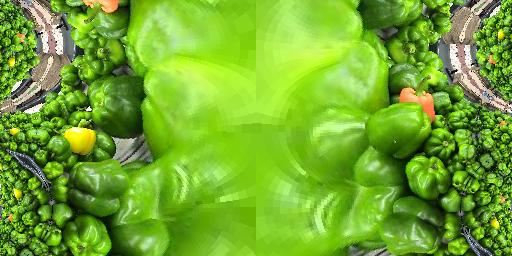} \\
		\includegraphics[width=0.43\linewidth,height=2cm]{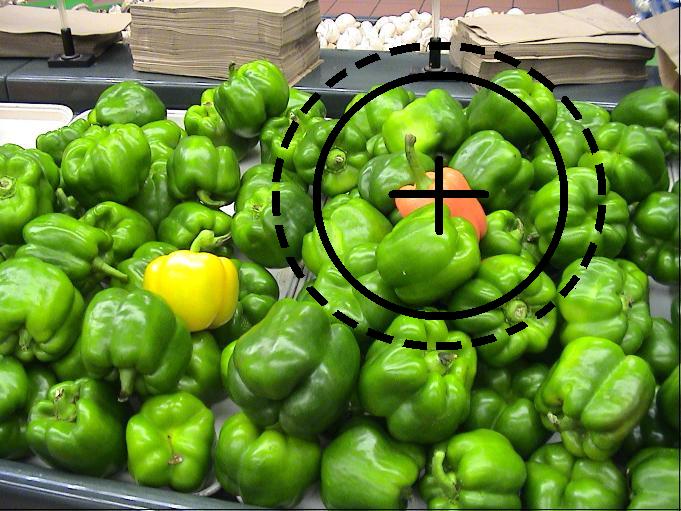} $\Rightarrow$
		\includegraphics[width=0.43\linewidth,height=2cm]{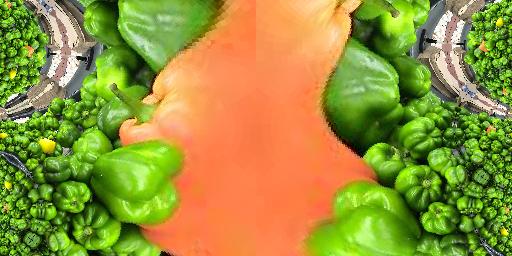} \\
		\includegraphics[width=0.43\linewidth,height=2cm]{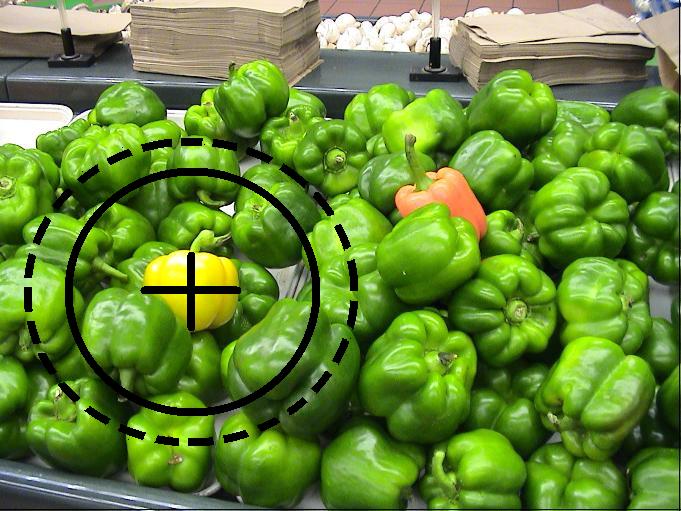} $\Rightarrow$
		\includegraphics[width=0.43\linewidth,height=2cm]{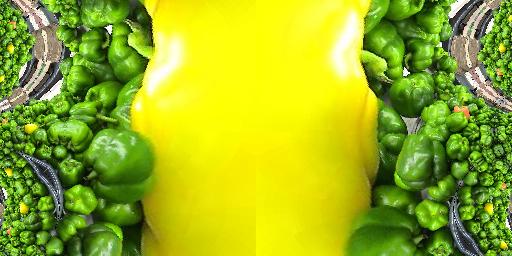}
		%\caption*{\centering \textbf{B}}
	\end{subfigure}
	\begin{subfigure}{0.40\linewidth}
		\centering 
		\includegraphics[width=0.47\linewidth,height=2.3cm]{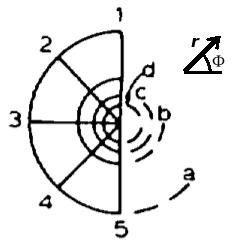} \\ 
		\footnotesize\begin{equation} W(r,\Phi)=\lambda \hspace{1mm} log(r e^{i\Phi}+e_0)\label{eq:cortical1},\end{equation}
		\vspace{-2em}
		\footnotesize\begin{equation} Z(X, Yi)=e^{(W/\lambda)}-e_0\label{eq:cortical2}.\end{equation} 
		\includegraphics[width=0.47\linewidth,height=2.3cm]{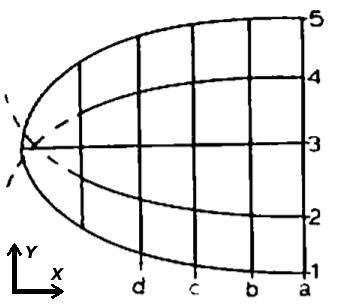} 
		%\caption*{\centering \textbf{A}}
	\end{subfigure}
	\vspace{-1em}
	\caption{\textbf{Left:} Examples of applying the cortical magnification function (transforming the visual space to the cortical space) at distinct views of the image presented in \hyperref[fig:opp]{Fig. \ref*{fig:opp}}. \textbf{Right:} Illustration of how polar coordinates (Z-plane) of azimuth $\Phi=(1,2,3,4,5)$ in the left visual field at distinct eccentricities $r=(d,c,b,a)$ are transformed to the cortical space (W-plane) in mm (X and Yi axis values). \hyperref[eq:cortical1]{Equations \ref*{eq:cortical1} \& \ref*{eq:cortical2}} express the monopole direct and inverse cortical mapping transformations (parameters set as $\lambda=12$mm and $e_0=1$deg \cite[Section~2.3.1]{zhaoping2014understanding}). Illustration sketch was adapted from E.L. Schwartz \cite{Schwartz1977}, \textit{Biol.Cybernetics 25}, p.184. \footnotesize Copyright (1977) by Springer-Verlag. %\footnotesize Illustration sketch was adapted from ``Spatial mapping in the primate sensory projection: Analytic structure and relevance to perception", 1977, by E.L. Schwartz, \textit{Biol.Cybernetics 25}, p.184. Copyright (1977) by Springer-Verlag \cite{Schwartz1977}.
	}
\vspace{-1.5em}
	\label{fig:magnification}
\end{figure}

The human eye is composed by RP but these are not homogeneously or equally distributed along the retina, contrarily to digital cameras. RP are distributed as a function of eccentricity with respect to the fovea (or central vision)\cite{Strasburger2011}. Fovea's diameter is known to comprise $\sim$5deg of diameter in the visual field, extended by the parafovea ($\sim$5-9deg), the perifovea ($\sim$9-17deg) and the macula ($\sim$17deg). Central vision is known to provide maximal resolution at $\sim$1deg of the fovea, whereas in periphery ($\sim$60-180- deg) there is lower resolution for the retinotopic positions that are further away from the fovea. These effects are known to affect color, shape, grouping and motion perception of visual objects (even at few degrees of eccentricity), making performance on attentional mechanisms eccentricity-dependent \cite{Carrasco2006b}. Axons from the nasal retina project to the contralateral LGN, whereas the ones from the temporal retina are connected with the ipsilateral LGN. These projections \cite{Wandell2007} make the left visual field send inputs of the LGN towards the right V1 hemifield (\hyperref[fig:magnification]{Fig. \ref*{fig:magnification}-Right}), similarly for the case of the right visual field to the left hemifield of V1. 

\vspace{-1em}

We have modeled these projections with a cortical magnification function \cite{Schwartz1977}\cite[Section~2.3.1]{zhaoping2014understanding} using 128 mm of simulated cortical surface (see an example in \hyperref[fig:magnification]{Fig. \ref*{fig:magnification}-Left}). The visual space is transformed to a cortically-magnified space (with its correspondence of millimeter for each degree of visual angle) with a logarithmic mapping function. The pixel-wise cartesian visual space is transformed to polar coordinates in terms of eccentricity and azimuth for a specific foveation instance, then transformed to coordinates in mm of cortical space. %wandell1995foundations
%Cones are mostly distributed in the center of the fovea whereas rods are distributed in most part away from the fovea  
Acknowledging that the visual space for digital images is represented with either a squared or rectangular shape, we computed the continuation of cortical coordinates by symmetrically mirroring existing coordinates of the image with their correspondence of visual space outside boundaries in the cortical space. In that manner, we exclude possible effects of zero-padding over recurrent processing while preserving 2D shapes for our feature representations. For this case, these effects were minimized by the inverse and repeating the same process at specific interaction cycles. Schwartz's mapping has been applied over the wavelet coefficients represented in \hyperref[fig:dwt]{Fig. \ref*{fig:dwt}}, as basis functions are convolved in the visual space, later magnified to the cortical space for representing V1 signals. These signals will serve as input to excitatory pyramidal cells, projected to their respective iso-orientation domains at distinct RF sizes.

\subsection{V1 Neuronal Dynamics}

Li's hypotheses suggest that V1 computations are responsible of generating a bottom-up saliency map \cite{Li2002,zhaoping2014understanding}. These hypotheses state that intracortical interactions between orientation-selective neurons in V1 are able to explain contextually-dependent perceptual effects present in pre-attentive vision \cite{Li1998,Li1999,Li2000,Zhaoping2003,Zhaoping2007,Zhaoping2015}, relative to contour integration, visual segmentation, visual search asymmetries, figure-ground and border effects, among others. Pop-out effects that form the saliency map are believed to be the result of horizontal connections in V1, that interact with each other locally and reciprocally. These connections are formed by excitatory cells and inhibitory interneurons \cite{Gilbert1992,Weliky1995}, processing information from pyramidal cell signals in layers of V1. Spatial organization of these cells accounts for selectivity in their orientation columns, their RF size and axonal field localization. The aforementioned interactions between orientation-selective cells was defined by Li's model \cite{Li1998} of excitatory-inhibitory firing-rate neural dynamics, later extended by Penacchio et al. \cite{Penacchio2013}. Here, contrast enhancement or suppression in neural responses emerge from lateral connections as an induction mechanism. Latest implementation done by Berga \& Otazu \cite{Berga2018b} for saliency prediction used colour images, where chromatic (P-,K-) and luminance (M-) opponent channels were individually processed in order to compute firing-rate dynamics of each pathway separately. With cortical magnification, each gaze can significantly vary contextual information and therefore the output of the model. 

\begin{figure}[H]
	\centering
	\begin{subfigure}{0.40\linewidth} %posar imatge en mes resolucio
		\includegraphics[width=\linewidth, height=5cm]{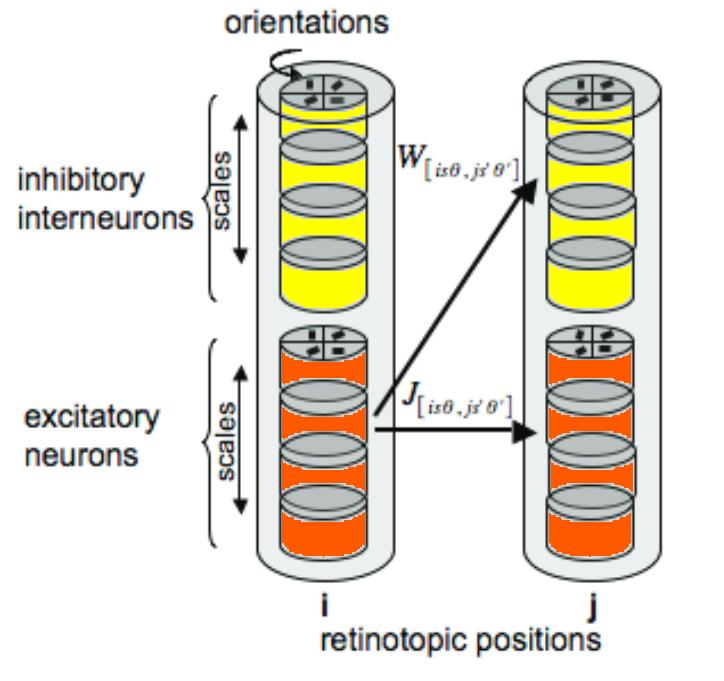}
		%\caption*{\centering \textbf{A}}
	\end{subfigure}
	\begin{subfigure}{0.44\linewidth}
		\includegraphics[width=\linewidth, height=5cm]{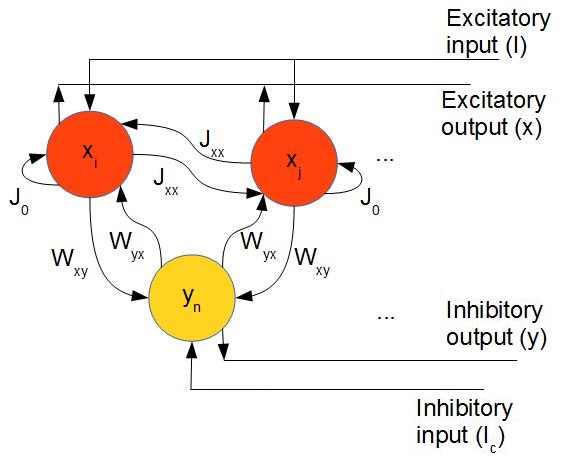} 
		%\caption*{\centering \textbf{B}}
	\end{subfigure}
	\caption{\textbf{Left:} Representation of cortical hypercolumns with scale and orientation selectivity interactions. \textbf{Right:} Model's intracortical excitatory-inhibitory interactions, membrane potentials (\textbf{\textcolor{orange}{orange}} "$\dot{x}$" for excitatory and \textbf{\textcolor{yellow}{yellow}} "$\dot{y}$" for inhibitory) and connectivities ("$J$" for monosynaptic excitation and "$W$" for dysynaptic inhibition).}
	\label{fig:model}
	\vspace{-1em}
\end{figure}

\begin{figure}[h]
	\begin{adjustwidth}{-0.4in}{-0.75in}
		\centering
		\includegraphics[clip, trim=0cm 20.75cm 0cm 0.90cm, width=1\linewidth, height=8cm]{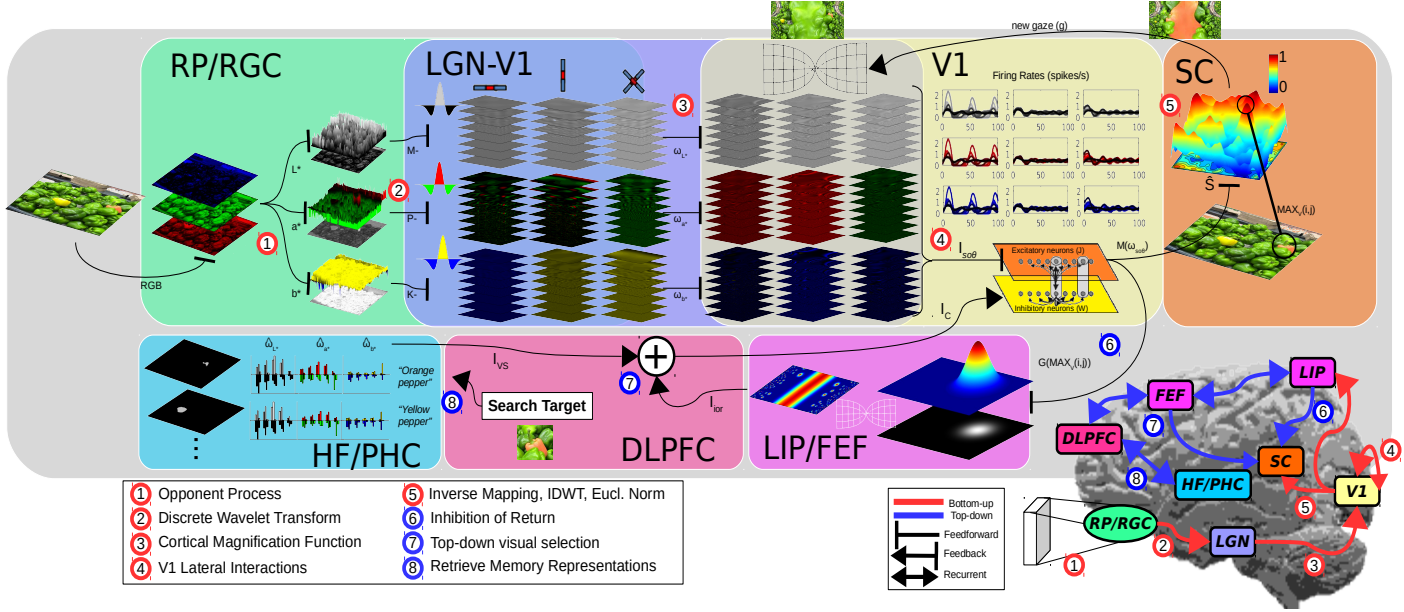}
	\end{adjustwidth}
\vspace{-1em}
	\caption{Diagram illustrating how visual information is processed by NSWAM-CM, including a brain drawing of each bottom-up and top-down mechanisms and their localization in the cortex (\textbf{Bottom-Right}).}
	\label{fig:nswamcm}
	\vspace{-1em}
\end{figure}

Our excitatory-inhibitory model\footnote{Model implementation in MATLAB: \url{https://github.com/dberga/NSWAM}} is described in \hyperref[tab:model]{Table \ref*{tab:model}}. Horizontal connections (lateral and reciprocal) are schematized in \hyperref[fig:model]{Fig. \ref*{fig:model}} and \hyperref[tab:model]{Table \ref*{tab:model}C}, where excitatory cells have self-directed ($J_0$) and monosynaptic connections ($J$) between each other, whereas dysynaptically connected through ($W$) inhibitory interneurons. Axonal field projections (window) follow a concentric toroid of radius $\Delta_s=15\times 2^{s-1}$ and radial distance $\Delta_\theta$ (accounting for RF size $d_s$ and radial distance $\beta$). Membrane potentials of excitatory ($\dot{x}_{is\theta}$) and inhibitory ($\dot{y}_{is\theta}$) cells are obtained with partial derivative equations defined in \hyperref[tab:model]{Table \ref*{tab:model}D}, composed by a chain of functions that consider firing-rates (obtained by piece-wise linear functions $g_x$ and $g_y$) and membrane potentials from previous membrane cycles (modulated by $\alpha_x$, $\alpha_y$ constants), current lateral connection potentials ($J$ and $W$) and spread of inhibitory activity within hypercolumns ($\psi$). Background inputs ($I_{noise}$ and $I_{norm}$) correspond to simulating random noise and divisive normalization signals (i.e. accounting for local nonorientation-specific cortical normalization and nonlinearities). Top-down inhibitory control mechanisms ($I_c$) are further explained in \hyperref[tab:model]{Table \ref*{tab:model}E} and in \hyperref[sec:topdown]{Section \ref*{sec:topdown}}. See the whole model pipeline in \hyperref[fig:nswamcm]{Fig. \ref*{fig:nswamcm}}. %comentar surround suppression?
%; $\tau\equiv10$ cycles

%Further details of model equations and parameters are specified in \hyperref[tab:model]{Table \ref*{tab:model}}.
% and \cite[\href{https://doi.org/10.1371/journal.pone.0064086.s001}{Supporting Information S1}]{Penacchio2013}\textbf{}
% \cite{Lund1975,Rockland1983}
%posar referencia del Heeger 1992 de la divisive normalization?
%The continuation of this model includes retinotopic cortical mapping projections (\hyperref[fig:magnification]{Fig. \ref*{fig:magnification}}), allowing to compute V1 dynamics upon distinct views of the visual scene.
%(we started the first view at the center of the image)

 %incorporar exemple de dinamiques amb la imatge indicada

Input signals ($I^{t}_{i;so\theta}$) have been defined as the wavelet coefficients ($\omega^{t}_{iso\theta}$), splitted between ON and OFF components (representing ON and OFF-center cell signals from RGC and LGN) depending on the value polarity ($+$ for positive and $-$ for negative coefficient values) from the RF. These signals are processed separately during $10\tau$ ($\tau=1$ membrane time $=10 ms$), including a rest interval (using an empty input) of $3\tau$ to simulate intervals between each saccade shift. The model output has been computed as the firing-rate average $g_x$ of the ON and OFF components ($M(\omega^{t+}_{iso\theta})$ and $M(\omega^{t-}_{iso\theta})$) during the whole viewing time, corresponding to a total of 10 membrane time (being the mean of $g_x$ for a specific range of $t$).

\begin{figure}[H]
	%\begin{adjustwidth}{-2in}{0in}
	\begin{subfigure}{.20\linewidth}
		\includegraphics[clip, trim=1cm 1cm 15cm 25cm, width=\linewidth, height=3.6cm]{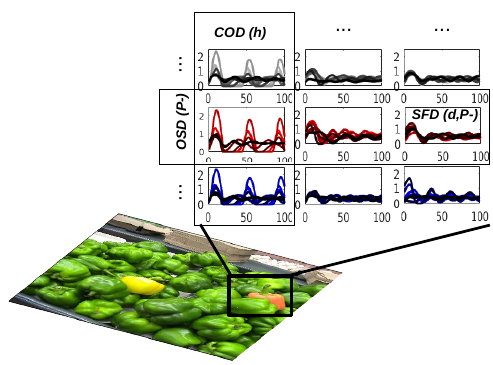}
		%\caption*{\centering Spatial Frequency Dynamics}
	\end{subfigure}
	\begin{subfigure}{.25\linewidth}
		\includegraphics[width=\linewidth,height=3.6cm]{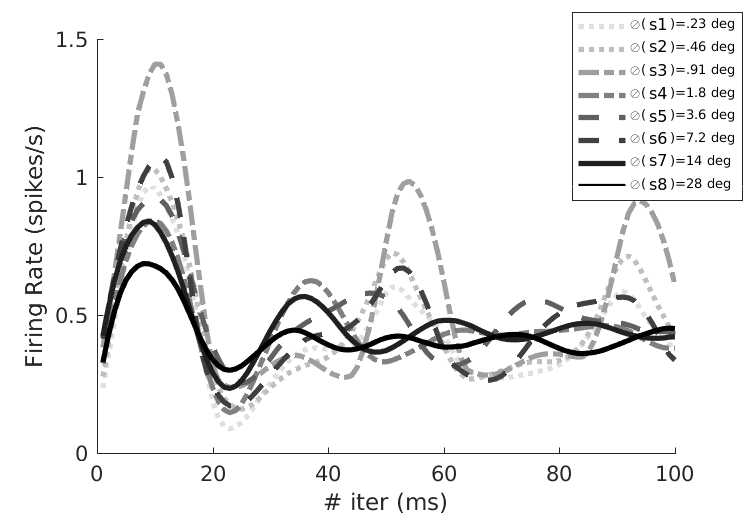}
		%\caption*{\centering Spatial Frequency Dynamics}
	\end{subfigure}
	\begin{subfigure}{.25\linewidth}
		\includegraphics[width=\linewidth,height=3.6cm]{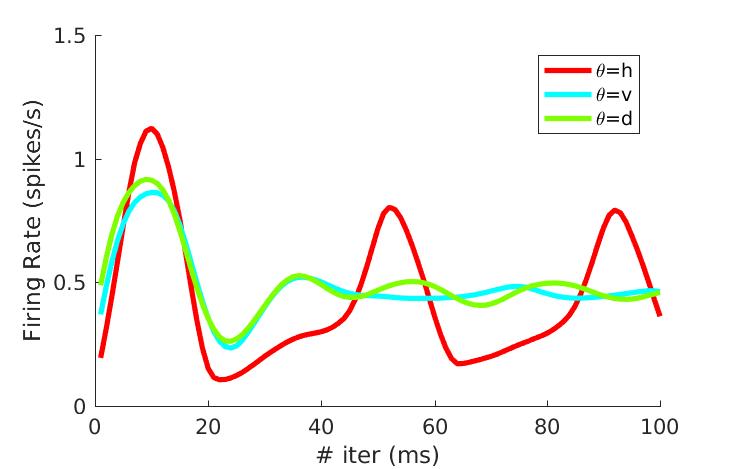}
		%\caption*{\centering Orientation Selectivity Dynamics}
	\end{subfigure}
	\begin{subfigure}{.25\linewidth}
		\includegraphics[width=\linewidth,height=3.6cm]{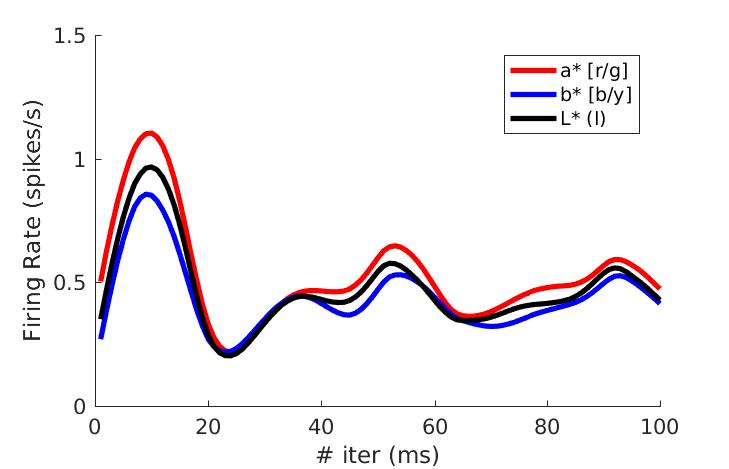}
		%\caption*{\centering Chromatic Opponency Dynamics}
	\end{subfigure}
	\caption{Firing rates plotted for 10 membrane time (100 iterations) accounting for neurons (ON+OFF values) inside a specific region (\textbf{1st col.}). Mean firing rates for all scales (Spatial Frequency Dynamics, \textbf{2nd col.}), orientations (Orientation Selectivity Dynamics, \textbf{3rd col.}), and color channels (Chromatic Opponency Dynamics, \textbf{4th col.}).}
	%\end{adjustwidth}
	\label{fig:dynamics}
	\vspace{-1em}
\end{figure}

\noindent
Combining the output of all components by

\footnotesize
\begin{equation} \label{eq:model6}
\hat{S}^{t}_{i;o}=\sum_{s=1..S;\theta=h,v,d}^{n_s}M(\omega^{t+}_{iso\theta})+\sum_{s=1..S;\theta=h,v,d}^{n_s}M(\omega^{t-}_{iso\theta})+c_i\quad,
\end{equation}
\normalsize

\noindent
we can describe the changes of the model (resulting from the simulated lateral interactions of V1) with respect the original wavelet coefficients $\omega^{t}_{iso\theta}$. Our result ($S^t_{i;o}$) will define the saliency map as an average conspicuity map or feature-wise distinctiveness (RF firing rates across scales and orientations for each pathway). These changes in firing-rate alternatively define the contrast enhancement seen on the brightness and chromatic induction cases \cite{Penacchio2013,Cerda2016,Penacchio2016}, where the model output is combined with the wavelet coefficients \{$M(\omega^{t}_{iso})\omega^{t}_{iso}$\} instead. The network is in total, composed of $1.18\times10^6$ neurons (accounting for 3 opponent channels, both ON/OFF polarities and RF sizes of $128\times64\times3\times$8).

\begin{table}[H]%[!ht]
%\begin{adjustwidth}{-1in}{0in}
%\centering
\caption{Overview of the model, following Nordlie et. al.'s format \cite{Nordlie2009}. Further explanation for model variables and parameters is in \cite[\href{https://doi.org/10.1371/journal.pone.0064086.s001}{Supporting Information S1}]{Penacchio2013}.} 
\label{tab:model}
\vspace{2mm}
\begin{tabular}{| p{2.5cm} | p{12.5cm} |}
\hline
\rowcolor{Black} \textbf{\textcolor{white}{A}} & \multicolumn{1}{|c|}{ \textbf{\textcolor{white}{Model Summary}}} \\ \hline
Populations & Excitatory ($x$), Inhibitory ($y$)\\ \hline
Topology & -- \\ \hline
Connectivity & Feedforward: one-to-all, Feedback: one-to-all, \newline Lateral: all-to-all (including self-connections) \\ \hline
Neuron model & Dynamic rate model\\ \hline
Channel model & -- \\ \hline
Synapse model & Piece-wise linear synapse\\ \hline
Plasticity & --\\ \hline
Input & External current in lower ($I$) or higher ($I_c$) cortical areas and random noise ($I_0$) \\ \hline
Measurements & Firing-rate ($g_x$ and $g_y$)\\
\hline
\end{tabular}
\vspace{2mm}
\begin{tabular}{| p{2.5cm} | p{4.5cm} | p{7.5cm} |}
\hline
\rowcolor{Black} \textbf{\textcolor{white}{B}} & \multicolumn{2}{|c|}{\textbf{\textcolor{white}{Populations}}} \\ \hline
\textbf{Name} & \textbf{Elements} & \textbf{Size} \\ \hline
$x$ & Sigmoidal-like neuron & $K_x = M \times N \times \Theta \times S$ = $64 \times 128 \times 3 \times 8$
%\newline \scriptsize where $M \times N$ corresponds to the cortical surface (64x128), $\Theta=3$ and $S=8$ 
\\ \hline
$y$ & Sigmoidal-like neuron & $K_y = K_x$ \\
\hline
\end{tabular}
\vspace{2mm}
\begin{tabular}{| p{2.5cm} | p{1.5cm} | p{1.5cm} | p{8.5cm} |}
\hline
\rowcolor{Black} \textbf{\textcolor{white}{C}} & \multicolumn{3}{|c|}{\textbf{\textcolor{white}{Connectivity}}} \\ \hline
\textbf{Name} & \textbf{Source} & \textbf{Target} & \textbf{Pattern} \\ \hline
$J_{xx}$ & $x$ & $x$ & Excitatory, toric, all to all, non-plastic\\ \hline
$J_0$ & $x$ & $x$ &  Excitatory, constant $J_0=0.8$ \\ \hline
$W_{xy}$ & $x$ & $y$ &  Inhibitory, toric, all to all, non-plastic \\ \hline
$W_{yx}$ & $y$ & $x$ &  Inhibitory, toric, all to all, non-plastic  \\
\hline
\end{tabular}
\vspace{2mm}
\begin{tabular}{| p{2.5cm} | p{12.5cm} |}
\hline
\rowcolor{Black} \textbf{\textcolor{white}{D}} & \multicolumn{1}{|c|}{ \textbf{\textcolor{white}{Neuron and Synapse Model}}}  \\ \hline
\textbf{Name} & V1 neuron\\
\hline
\textbf{Type} & Dynamic rate model \\
\hline
\textbf{Synaptic dynamics}
& \label{eq:connections} \vspace{-2em} %alternativa equation es $$
\footnotesize\begin{equation}
J_{[is\theta,js'\theta']}= 
    \lambda(\Delta_s)  0.126  e^{(-\beta / d_s )^2 - 2(\beta / d_s)^7 - d_s^2/90}
    \end{equation}\normalsize\\
& \vspace{-2em}
\footnotesize\begin{equation}
W_{[is\theta,js'\theta']}= 
    \lambda(\Delta_s)  0.14  (1-e^{-0.4(\beta / d_s)^{1.5}})e^{-(\Delta_\theta/(\pi/4))^{1.5}}
\end{equation}\normalsize 
%$\psi(\Delta_s, \Delta_\theta)=\lambda(\Delta_s) cos(\Delta_\theta)^3$; %$\Delta_\theta=(\theta-\theta')$; $\Delta_s=(s-s')=10·2.2^{s-1}$ \\ 
\\ \hline
\textbf{Membrane \newline potential} \label{eq:mempotentials}
& \vspace{-2em}
\footnotesize\begin{equation}
\begin{multlined}
\dot{x}_{is\theta} = -\alpha_x x_{is\theta}-g_y(y_{is\theta})-\sum_{\Delta_s,\Delta_\theta \neq 0}\psi(\Delta_s,\Delta_\theta)g_y(y_{is}+\Delta_{s\theta}+\Delta_\theta) \\ +J_0 g(x_{is\theta})+ \sum_{j\neq i,s',\theta'} J_{[is\theta,js'\theta']}g_x(x_{js'\theta'})+I_{is\theta}+I_0,
\end{multlined}
\end{equation}\normalsize\\
 &  \vspace{-2em}
\footnotesize\begin{equation}
\begin{multlined} 
\dot{y}_{is\theta} = -\alpha_y y_{is\theta}-g_x(x_{is\theta}) +\sum_{j\neq i,s',\theta'} W_{[is\theta,js'\theta']}g_x(x_{js'\theta'})+I_c
\end{multlined}
\end{equation}\normalsize \\
\hline
\end{tabular}
\vspace{2mm}
\begin{tabular}{| p{2.5cm} | p{12.5cm} |}
\hline
\rowcolor{Black} \textbf{\textcolor{white}{E}} & \multicolumn{1}{|c|}{ \textbf{\textcolor{white}{Input}}} \\ \hline
\textbf{Type} & \textbf{Description} \\ \hline
Sensory  \newline \scriptsize (bottom-up)& Input to excitatory neurons, $I^{t}_{i;o}=\omega^{t}_{iso\theta}$\\ \hline
Control \newline \scriptsize (top-down) & Input to inhibitory interneurons, $I_c=1.0 + I_{noise}+I_{vs}+I_{ior}$\\
%Recurrent \scriptsize (noise) & $\begin{aligned} I_0=0.85 + I_{norm} + I_{noise}, \\ \scriptsize I_{noise}=\textit{N}(0; 0.1, 0.1),\\ I_{norm}=-2.0 (\frac{\sum_{j\in A_i}+\sum_{\theta'} g_x(x_{j\theta'})}{\sum_{j\in A_i} 1})^2 \end{aligned}$ \\
\hline
\end{tabular}
%\raggedright
\begin{tabular}{| p{2.5cm} | p{12.5cm} |}
\hline
\rowcolor{Black} \textbf{\textcolor{white}{F}} & \multicolumn{1}{|c|}{ \textbf{\textcolor{white}{Measurements}}} \\ \hline
\multicolumn{2}{|l|}{Mean Firing-rate of excitatory neurons for $\tau$=10 membrane time ($M(\omega^{p=[+,-]}_{iso\theta})$).}  \\
\hline
\end{tabular}
%\end{adjustwidth}
\end{table}

%$J_{[is\theta,js'\theta']}= \begin{cases}
%    \lambda(\Delta_s)  0.126  e^{(-\beta / d_s )^2 - 2(\beta / d_s)^7 - d_s^2/90},& \text{, if } ( 0 < d_s <= \Delta_s \quad and \quad  \beta < \pi/2.69) \\ & \quad or \quad  [(0 < d_s <= \Delta_s \quad and \quad  \beta < \pi/1.1) \\ & \quad and\quad  |\theta_1| < \pi/5.9 \quad and\quad  |\theta_1|% < \pi/5.9] \\
%    0              & \text{, otherwise}
%\end{cases}$

%$W_{[is\theta,js'\theta']}= \begin{cases} 
%    0 & \text{, if } d_s <= \Delta_s \quad or \quad \beta > \pi/1.1 \\ & \quad %or \quad |\Delta\theta| <= \pi/3 \\ & \quad or \quad |\theta_1| > \pi/11.999 \\
%    \lambda(\Delta_s)  0.14  (1-e^{-0.4(\beta / %d_s)^{1.5}})e^(\Delta_\theta/(\pi/4))^{1.5}              & \text{, otherwise}
%\end{cases}$

 %cambiar formulas J i W per les abreviades
%altes exemples:
%https://hal.inria.fr/hal-01052817/document
%https://hal.inria.fr/hal-01109483/document

\subsection{Projections to the SC}

Latest hypotheses about neural correlates of saliency \cite{Veale2017,White2017b} state that the superior colliculus is responsible for encoding visual saliency and to guide eye movements \cite{WhiteMunoz2011,Schiller2001}. Acknowledging that the superficial layers of the SC (sSC) receive inputs from the early stages of visual processing (V1, retina), the SC selects these as the root of bottom-up activity to be selected in the intermediate and deep layers (iSC, dSC). In accordance to the previous stated hypotheses \cite{Li2002}, saccadic eye movements modulated by saliency therefore are computed by V1 activity, whereas recurrent and top-down attention is suggested to be processed by neural correlates in the parieto-frontal cortex and basal ganglia. All these projections are selected as a winner-take-all mechanism in SC \cite{Li1999,Li2002,zhaoping2014understanding} to a unique map, where retinotopic positions with the highest activity will be considered as candidates to the corresponding saccade locations. These activations in the SC are transmitted to guide vertical and horizontal saccade visuomotor nerves \cite{Horn2012}. We have defined the higher active neurons (\hyperref[eq:saliency2a]{Equation \ref*{eq:saliency2a}}) as the locations for saccades in the visual space (i,j) by decoding the inverse of the cortical magnification (\hyperref[eq:cortical2]{Equation \ref*{eq:cortical2}}) of their retinotopic position ("$i$" neuron at X,Yi).  %\cite{Schiller1974}Krauzlis2013,\cite{Tehovnik2003}\cite{Wang2015}\cite{Awh2006}\cite{Steinmetz2012}\cite{White2017a}

\vspace{-0.75em}
\footnotesize
\begin{equation}
MAX_W(X,Yi)= arg max(\hat{S}) \rightarrow MAX_Z(r,\Phi) \rightarrow MAX_V(i,j),
\label{eq:saliency2a}
\end{equation}
\normalsize

%iso-feature suppression
%phisiological quantity, behavioral/perceptual quantity

The behavioral quantity of the unique 2D saliency map has been defined by computing the inverse of the previous processes using the model output for each pathway separately. Retinotopic positions have been transformed to coordinates in the visual space using the inverse of the cortical magnification function (\hyperref[eq:cortical2]{Equation \ref*{eq:cortical2}}). Output signals (V1 sensitivities to orientation and spatial frequencies) are integrated by computing the inverse discrete wavelet transform to obtain unique maps for each channel opponency (\hyperref[eq:wavelets3]{Equation \ref*{eq:wavelets3}}). A unique representation (\hyperref[eq:saliency2]{Equation \ref*{eq:saliency2}}) of final neuronal responses for each pathway (P-, K- and M- as $a^*$, $b^*$ and $L^*$) is generated with the euclidean norm (adding responses of all channels as in Murray et al.\cite{Murray2011} model). The resulting map is later normalized by the variance (\hyperref[eq:saliency3]{Equation \ref*{eq:saliency3}}) of the firing rate \cite[Chapter~5]{zhaoping2014understanding}. This map represents the final saliency map, that describes the probability distribution of fixation points in certain areas of the image. In addition to this estimation, the saliency map has been convolved with a gaussian filter simulating a smoothing caused by the deviations of $\sigma=1$ deg given from eye tracking experimentation, recommended by LeMeur \& Baccino \cite{LeMeur2012}. 

\vspace{-0.5em}
\footnotesize
\begin{equation}
%S_i=MAX(I'^{t}_{i;o1}, I'^{t}_{i;o2}, I'^{t}_{i;o3}),
\hat{S}_i=\sqrt[]{\hat{S}_{i;a^*} + \hat{S}_{i;b^*} + \hat{S}_{i;L^*}},
\label{eq:saliency2}
\end{equation}
\vspace{-1em}

\begin{equation} 
z_{i}(\hat{S})=\frac{\hat{S}_i-\mu_{\hat{S}}}{\sigma_{\hat{S}}},
\label{eq:saliency3}
\end{equation}
\normalsize

%LINE PLOTS 3D
%https://es.mathworks.com/matlabcentral/fileexchange/35262-matlab-plot-gallery-line-plot-3d
%https://es.mathworks.com/matlabcentral/answers/107590-3d-plot-of-detach-cube-matrix-sequential-2d-plot-in-one-3d-ribbon-alternative

%FIGURE INSIDE FIGURE
%https://es.mathworks.com/matlabcentral/answers/60376-how-to-make-an-inset-of-matlab-figure-inside-the-figure

%SLICE PLOTS IMAGES
%http://matlabnewbie.blogspot.com/2013/07/recently-i-read-post-from-dr.html

\subsection{Attention as top-down inhibition} \label{sec:topdown}

An additional purpose of our work is the modeling of attentional mechanisms beyond pre-attentive visual selection. Instead of analyzing the scene serially, the visual brain uses a set of attentional biases to recognize objects, their relationships and their importance with respect to the task, all given in a set of visual representations. Similarly to the saliency map, the priority map can be interpreted as a unique 2D representation for eye movement guidance formed in the SC, here including top-down (not guided by the stimulus itself) and recurrent information as visual relevance. This phenomenon suggests that executive, long-term and short-term/working memory correlates also direct eye movement control \cite{PierrotDeseilligny2004,WhiteMunoz2011}. Previous hypotheses model these properties by forming the priority map through selective tuning \cite{Tsotsos1995,tsotsos2016}. Selective tuning explains attention mechanisms as a hierarchy of winner-take-all processes. This hypothesis suggests that top-down attention can be simulated by spatially inhibiting specific layers of processing. Latest hypotheses \cite{Ahmadlou2018} confirm that striate cortical activity gain can be modulated by SC responses, with additional modulations arising from pulvinar to extrastriate visual areas. In addition, it has also been stated \cite{Yan2018} that V1 influences both saliency and top-down learning during visual detection tasks. By functionally simulating the aforementioned top-down mechanisms as inhibitory gates of top-down feedback control in our model \cite{Li1998}, we are able to perform task-specific visual selection (VS) and inhibition of return (IoR) mechanisms.  %\cite{Ullman1984}\cite{Tsotsos1988}Tsotsos2014

\paragraph{Top-down selection:} Goal-directed or memory-guided saccades imply executive control mechanisms that account for task requirements during stimulus perception. The dorsolateral prefrontal cortex (DLPFC) is known to be responsible for short-term spatial memory, to retrieve long-term memory signals of object representations (through projections towards the para- and hippocampal formations) as well as to perform reflective saccade inhibition, among other functions. These inhibitory signals, later projected to the frontal eye field (FEF), are able to direct gaze during search and smooth pursuit tasks \cite{PierrotDeseilligny2003,PierrotDeseilligny2004,schall2009} (also suggested to be crucial for planning intentional or endogenously-guided saccades), where its signals are sent to the SC. By feeding our model with inhibitory signals ($I_{c}$ shown in \hyperref[fig:model]{Fig. \ref*{fig:model}} and \hyperref[tab:model]{Table \ref*{tab:model}E}) we can simulate top-down feedback control mechanisms in V1 (initially proposed by Li \cite[Sec. 3.7]{Li1998}). In this case, a new term $I_{\{vs\}}$ is added to the top-down inhibition of our V1 cortical signals that will be projected to the SC during each gaze. 
%https://www.researchgate.net/publication/239924247_Top-down_facilitation_of_visual_object_recognition_Object-based_and_context-based_contributions
%https://www2.le.ac.uk/centres/csn/publications-1/Publications/Neuropsychologia2016Final.pdf

\begin{equation}
I_{\{vs\}}=\alpha_{\{vs\}}\cdot
\begin{cases}
arg max_{p,s,o,\theta}(\omega)
%\Omega_{p\subset\{+,-\},s\subset\{1..S\},o\subset\{L^*,b^*,a^*\},\theta\subset\{h,v,d\}}
& , \text{feature-selective }(VS_M)
\\
%\frac{\sum\limits_{i=1}^N{\omega}_{pso\theta}}{N}
(\sum\limits_{i=1}^N{\omega}_{pso\theta})/N
& , \text{category-specific }(VS_C)
\end{cases}
\label{eq:topdown}
\end{equation}

In this implementation, we can perform distinct search tasks such as feature search (by manually selecting the features, or selecting features with maximal responses, similarly to a boolean selection \cite{Huang2007}), exemplar and categorical object search (by processing the mean of responses $\hat{\omega}$ from wavelet coefficients of a single or several image samples "N"). These low-level computations would serve as cortical activations to be stored as weights in our low-level memory representations, that will be used as inhibitory modulation for the task execution. %lo del DLPFC de cara al nostre model esta agafat amb pinces, ja que les projeccions top-down a V1 no estan clares (també s'haurien de comentar les del pulvinar). fer referencia a una altra literatura de top-down i V1?  fer referencia al iconic bottleneck? %
% This term is modulated with a constant factor $\alpha_{\{vs\}}$.

%top-down inhibitory: sallin & bullier 1995

% \begin{align} %cambiar notacio de la G?
%\begin{split}
%I_{vs,manual}=\alpha_{vs}( \Omega_{p=(+,-),s=(1..S),o=(a*,b*,L*),\theta=(h,v,d)}), \\ %I_{vs,learned}=\alpha_{vs}( \hat{\omega}_{pso\theta})
%\label{eq:topdown}
%\end{split}
%\end{align}

% \begin{figure}[H] %utilitzar la mateixa figura
%    \centering
%    \includegraphics[width=1\linewidth,height=2cm]{example-image-c}
%    \caption{Feature and Categorical search}
%\end{figure} 
%\cite{Unema2005}\cite{Follet2011}\cite{Pannasch2008}
\paragraph{Inhibition of Return:} During scene viewing, saccadic eye movements show distinct patterns of fixations \cite{Eisenberg2016}, directed by exploratory purposes or either towards putting the attentional focus on specific objects in the scene. For the former case, the HVS needs to ignore already visited regions (triggering anti-saccades away from these memorized regions, as a consequence of inhibition) during a period of time before gazing again towards them. This phenomena is named inhibition of return \cite{Godijn2002}, and similarly involves extracting sensory information and short-term memory during scene perception. As mentioned before, DLPFC is responsible of memory-guided saccades, and this function might be done in conjunction with the parietal cortex and the FEF. The parietal areas (LIP and PEF)\cite{PierrotDeseilligny2003,PierrotDeseilligny2004,Bisley2006} are known to be responsible of visuospatial integration and preparation of saccade sequences. These areas conjunctively interact with the FEF and DLPFC for planning these reflexive visually-guided saccades. Acknowledging that LIP receives inputs from FEF and DLPFC, the role of each cannot be disentangled as a unique functional correlate for the IoR. Following the above, we have modeled return mechanisms as top-down cortical inhibition feedback control accounting for previously-viewed saccade locations. Thus, we added an inhibition input $I_{\{IoR\}}$ at the start of each saccade, which will determine our IoR mechanism: %\cite{Posner1984}

\begin{align} %cambiar notacio de la G? %posar formula de decay de 100 iter?
\begin{split}
I_{\{IoR\}}^{g,t=0}=MAX(\hat{S}) \cdot G(MAX_V(x,y)) +I_{\{IoR\}}^{g-1}, \\
I_{\{IoR\}}^{g,t>0}= \alpha_{\{IoR\}} (I_{\{IoR\}}^{t-1}) \prod\limits_{i=1}^{10\tau} e^{log(\beta_{\{IoR\}})/\tau}.
\label{eq:ior}
\end{split}
\end{align}

This term is modulated with a constant power factor $\alpha_{\{IoR\}}$ and a decay factor $\beta_{\{IoR\}}$, which in every cycle will progressively reduce inhibition. The spatial region of the IoR has been defined as a gaussian function centered to the previous gaze (g), with a spatial standard deviation $\sigma_{\{IoR\}}$ dependent on a specific spatial scale and a peak with an amplitude of the maximal RF firing rate of our model's output ($\hat{S}$). Inhibitory activity is accumulated to the same map and can be shown how is progressively reduced during viewing time (\hyperref[fig:resultsior1]{Fig. \ref{fig:resultsior1}}). Alternatively illustrated in Itti et al.'s work \cite{Itti1998}, the IoR can be applied to static saliency models by substracting the accumulated inhibitory map to the saliency map during each gaze ($\hat{S} - I^g_{\{IoR\}}$).

\section{Materials and Methods}

\subsection{Procedure}

Experimental data has been extracted from eye tracking experimentation. Four datasets were analyzed, corresponding to 120 real indoor and outdoor images (Toronto \cite{Bruce2005}), 40 nature scene images (KTH \cite{Kootstra2011}), 100 synthetic image patterns ($CAT2000_{P}$ \cite{CAT2000}) and 230 psychophysical images (SID4VAM \cite{Berga2018a,Berga_2019_ICCV}). Generically, experimentation for these type of datasets \cite{Winkler2013} capture fixations from about 5 to 55 subjects, looking at a monitor inside a luminance controlled room while being restrained with a chin rest, located at a relative distance of 30-40 pixels per degree of visual angle ($pxva$). The tasks performed mostly consist of freely looking at each image during 5000 ms, looking at the "most salient objects" or searching for specific objects of interest. We have selected these datasets to evaluate prediction performance at distinct scene contexts. Indicators of psychophysical consistency of the models has been presented, evaluating prediction performance upon fixation number and feature contrast. Visual search performance has been evaluated by computing predictions of locating specific objects of interest. For the case of stimuli from real image contexts (\hyperref[fig:results_search2]{Fig. \ref*{fig:results_search2}}) we have used salient object segmented regions from Toronto's dataset \cite{Bruce2005}, extracted from Li et al. \cite{Li2014}. Finally, for the case of evaluating fixations performed with synthetic image patterns, we used fixations from SID4VAM's psychophysical stimuli. 
%A benchmark for the aforementioned predictions of fixation regions and saccadic eye movement locations will be provided in the next subsections. 

\subsection{Model evaluation}

Current eye tracking experimentation represent indicators of saliency as the probability of fixations on certain regions of an image\footnote{Code for computing metrics: \url{https://github.com/dberga/saliency}}. Metrics used in saliency benchmarks \cite{Bylinskii2015} consider all fixations during viewing time with same importance, making saliency hypotheses unclear of which computational procedures perform best using real image datasets. Previous psychophysical studies \cite{Bruce2015,Berga2018a} revealed that fixations guided by bottom-up attention are influenced by the type of features that appear in the scene and their relative feature contrast. From these properties, the order of fixations and the type of task can drive specific eye movement patterns and center biases, relevant in this case. 

The AUC metric (Area Under ROC/Receiver Operant Characteristic) represents a score of a curve comprised of true positive values (TP) against false positive (FP) values. The TP are set as human fixations inside a region of the saliency map, whereas FP are those predicted saliency regions that did not fall on human fixation instances. For our prediction evaluation we computed the sAUC (shuffled AUC), where FP are expressed as TP from fixations of other image instances. This metric prioritizes model consistency and penalizes for prediction biases that appear over eye movement datasets, such as oculomotor and center biases (not driven by pre-attentional factors). We also calculated the Information Gain (InfoGain) metric for model evaluation, which compares FP in the probability density distribution of human fixations with the model prediction, while substracting a baseline distribution of the center bias (all fixations grouped together in a single map). Saliency metrics, largely explained by Bylinskii et al. \cite{Bylinskii2018}, usually compare model predictions with human fixations during the whole viewing time, regardless of fixation order. In our study is also represented the evolution of prediction scores for each gaze. For the case of scanpaths, we evaluated saccade sequences by analyzing saccade amplitude (SA) and saccade landing (SL) statistics. These are calculated using euclidean distance between fixation coordinates (distance between saccade length for SA and distance between locations of saccades for SL). 

Initial investigations on visual attention \cite{Treisman1980,Wolfe1989} during visual search tasks formulated that reaction times of finding a target (defined in a region of interest/ROI) among a set of distractors are dependent on set size as well as target-distractor feature contrast. In order to evaluate performance on visual search, we utilised two metrics that account for the ground truth mask of specific regions for search and the saliency map (in this context, it could be considered as a "relevance" map) or predicted saccade coordinates (from locations with highest neuronal activity). The Saliency Index (SI) \cite{Spratling2012,Berga2018a,Berga_2019_ICCV} calculates the amount of energy of a saliency map inside a ROI ($S_t$) with respect to the one outside ($S_b$), calculated as: $SI=(S_t-S_b)/S_b$. For the case of saccades in visual search, we considered to calculate the probability of fixations inside the ROI (PFI). 

\section{Results}

\subsection{Results on predicting Saliency} \label{sec:results_saliency}

\begin{figure}[H]%[!ht]
  \begin{adjustwidth}{-0.4in}{-2in}
  %\begin{adjustwidth}{-2in}{-2in}
  \begin{subfigure}{0.5\linewidth}
  \begin{tabular}{cccc}
  Model & sAUC & InfoGain & \parbox[c]{1em}{\includegraphics[width=0.40in,height=0.30in]{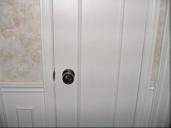}\includegraphics[width=0.40in,height=0.30in]{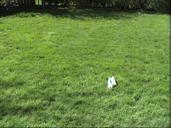}\includegraphics[width=0.40in,height=0.30in]{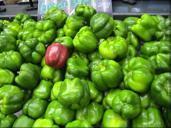}\includegraphics[width=0.40in,height=0.30in]{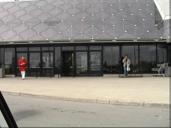}\includegraphics[width=0.40in,height=0.30in]{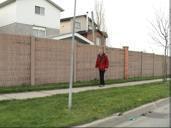}}\\
  \hline
  Human Fix. & .904 & 2.42 & \parbox[c]{1em}{\includegraphics[width=0.40in,height=0.30in]{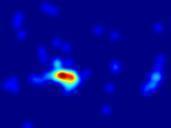}\includegraphics[width=0.40in,height=0.30in]{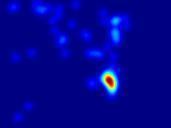}\includegraphics[width=0.40in,height=0.30in]{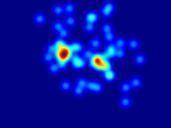}\includegraphics[width=0.40in,height=0.30in]{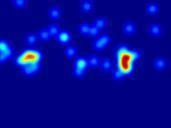}\includegraphics[width=0.40in,height=0.30in]{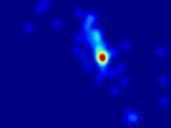}}\\
  %Gaussian & 1 & 1 & \parbox[c]{1em}{\includegraphics[width=0.40in,height=0.30in]{example-image-a}\includegraphics[width=0.40in,height=0.30in]{example-image-a}\includegraphics[width=0.40in,height=0.30in]{example-image-a}\includegraphics[width=0.40in,height=0.30in]{example-image-a}\includegraphics[width=0.40in,height=0.30in]{example-image-a}}\\
  IKN \cite{Itti1998} & .649 & -.024* & \parbox[c]{1em}{\includegraphics[width=0.40in,height=0.30in]{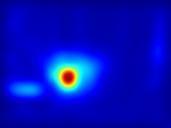}\includegraphics[width=0.40in,height=0.30in]{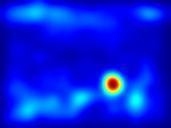}\includegraphics[width=0.40in,height=0.30in]{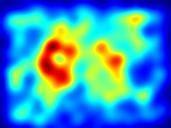}\includegraphics[width=0.40in,height=0.30in]{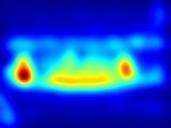}\includegraphics[width=0.40in,height=0.30in]{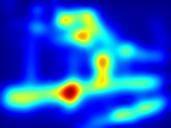}}\\
  AIM \cite{Bruce2005} & .663* & -.579 & \parbox[c]{1em}{\includegraphics[width=0.40in,height=0.30in]{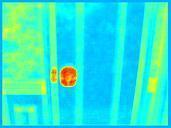}\includegraphics[width=0.40in,height=0.30in]{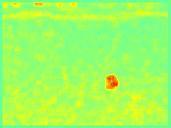}\includegraphics[width=0.40in,height=0.30in]{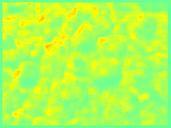}\includegraphics[width=0.40in,height=0.30in]{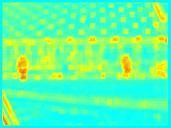}\includegraphics[width=0.40in,height=0.30in]{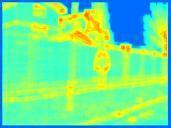}}\\
  %SWAM \cite{Berga2018b} & .716 & .654 & \parbox[c]{1em}{\includegraphics[width=0.40in,height=0.30in]{figures/qualitative/tsotsos/SWAM/30.jpg}\includegraphics[width=0.40in,height=0.30in]{figures/qualitative/tsotsos/SWAM/27.jpg}\includegraphics[width=0.40in,height=0.30in]{figures/qualitative/tsotsos/SWAM/72.jpg}\includegraphics[width=0.40in,height=0.30in]{figures/qualitative/tsotsos/SWAM/77.jpg}\includegraphics[width=0.40in,height=0.30in]{figures/qualitative/tsotsos/SWAM/94.jpg}}\\
  %SIM \cite{Murray2011} & .744 & .705 & \parbox[c]{1em}{\includegraphics[width=0.40in,height=0.30in]{figures/qualitative/tsotsos/SIM/30.jpg}\includegraphics[width=0.40in,height=0.30in]{figures/qualitative/tsotsos/SIM/27.jpg}\includegraphics[width=0.40in,height=0.30in]{figures/qualitative/tsotsos/SIM/72.jpg}\includegraphics[width=0.40in,height=0.30in]{figures/qualitative/tsotsos/SIM/77.jpg}\includegraphics[width=0.40in,height=0.30in]{figures/qualitative/tsotsos/SIM/94.jpg}}\\
  NSWAM & .631 & -.552 & \parbox[c]{1em}{\includegraphics[width=0.40in,height=0.30in]{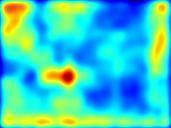}\includegraphics[width=0.40in,height=0.30in]{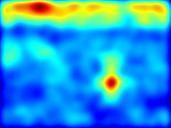}\includegraphics[width=0.40in,height=0.30in]{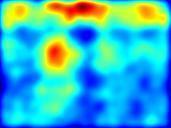}\includegraphics[width=0.40in,height=0.30in]{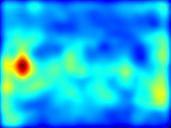}\includegraphics[width=0.40in,height=0.30in]{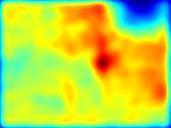}}\\
  NSWAM-CM1 & .636 & -.818 & \parbox[c]{1em}{\includegraphics[width=0.40in,height=0.30in]{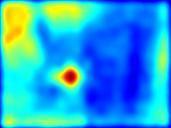}\includegraphics[width=0.40in,height=0.30in]{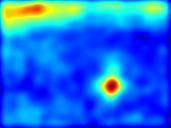}\includegraphics[width=0.40in,height=0.30in]{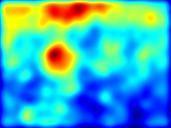}\includegraphics[width=0.40in,height=0.30in]{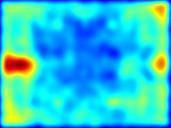}\includegraphics[width=0.40in,height=0.30in]{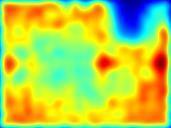}}\\
  NSWAM-CM2 & .644 & -.738 & \parbox[c]{1em}{\includegraphics[width=0.40in,height=0.30in]{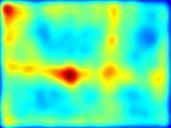}\includegraphics[width=0.40in,height=0.30in]{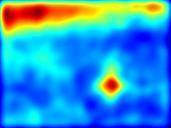}\includegraphics[width=0.40in,height=0.30in]{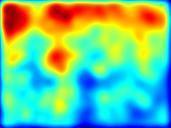}\includegraphics[width=0.40in,height=0.30in]{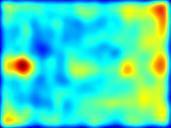}\includegraphics[width=0.40in,height=0.30in]{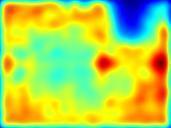}}\\
  NSWAM-CM5 & .650 & -.701 & \parbox[c]{1em}{\includegraphics[width=0.40in,height=0.30in]{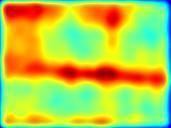}\includegraphics[width=0.40in,height=0.30in]{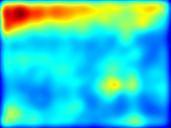}\includegraphics[width=0.40in,height=0.30in]{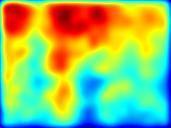}\includegraphics[width=0.40in,height=0.30in]{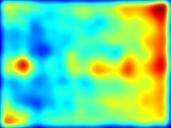}\includegraphics[width=0.40in,height=0.30in]{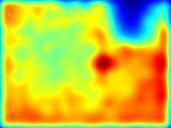}}\\
  NSWAM-CM10 & .655 & -.692 & \parbox[c]{1em}{\includegraphics[width=0.40in,height=0.30in]{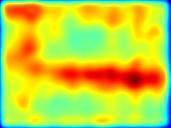}\includegraphics[width=0.40in,height=0.30in]{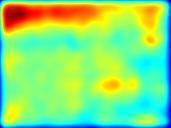}\includegraphics[width=0.40in,height=0.30in]{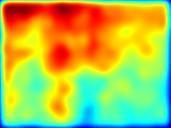}\includegraphics[width=0.40in,height=0.30in]{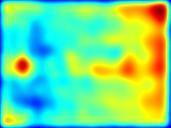}\includegraphics[width=0.40in,height=0.30in]{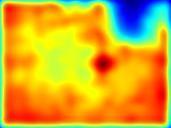}}\\
  \end{tabular} %\hspace{5mm} 
  \end{subfigure}
  \begin{subfigure}{0.5\linewidth}
    \includegraphics[width=2.8in,height=1.5in]{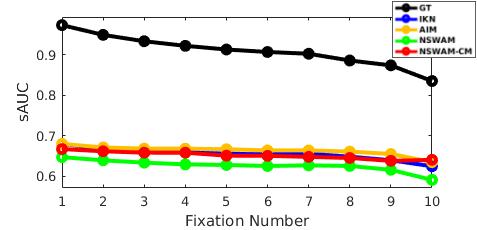} \\
    \includegraphics[width=2.8in,height=1.5in]{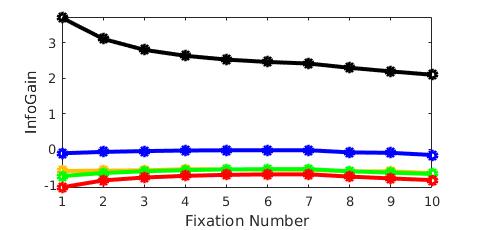}
    %\caption*{grafica de AUC i sAUC vs fixation num.}
    \end{subfigure}
  \end{adjustwidth}
  \caption{Results on saliency for Toronto (Bruce \& Tsotsos \cite{Bruce2005}) Eye Tracking Dataset. \textbf{Left}: Saliency metric scores. \textbf{Middle}: Examples of saliency maps. \textbf{Right}: Shuffled scores per fixation number.}
  \label{fig:saliency1}
\end{figure}

\begin{figure}[H]%[!ht]
  \begin{adjustwidth}{-0.4in}{-2in}
  \begin{subfigure}{0.5\linewidth}
  \begin{tabular}{cccc}
  Model & sAUC & InfoGain & \parbox[c]{1em}{\includegraphics[width=0.40in,height=0.30in]{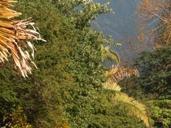}\includegraphics[width=0.40in,height=0.30in]{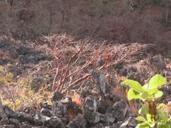}\includegraphics[width=0.40in,height=0.30in]{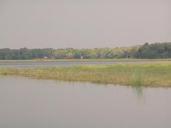}\includegraphics[width=0.40in,height=0.30in]{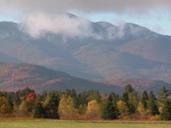}\includegraphics[width=0.40in,height=0.30in]{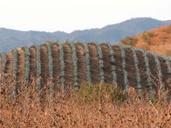}}\\
  \hline
  Human Fix. & .822 & 1.41 & \parbox[c]{1em}{\includegraphics[width=0.40in,height=0.30in]{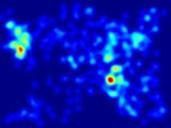}\includegraphics[width=0.40in,height=0.30in]{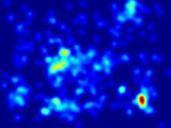}\includegraphics[width=0.40in,height=0.30in]{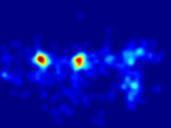}\includegraphics[width=0.40in,height=0.30in]{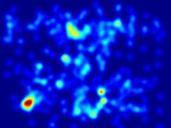}\includegraphics[width=0.40in,height=0.30in]{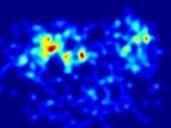}}\\
  %Gaussian & 1 & 1 & \parbox[c]{1em}{\includegraphics[width=0.40in,height=0.30in]{example-image-a}\includegraphics[width=0.40in,height=0.30in]{example-image-a}\includegraphics[width=0.40in,height=0.30in]{example-image-a}\includegraphics[width=0.40in,height=0.30in]{example-image-a}\includegraphics[width=0.40in,height=0.30in]{example-image-a}}\\
  IKN \cite{Itti1998} & .551 & -.172 & \parbox[c]{1em}{\includegraphics[width=0.40in,height=0.30in]{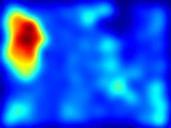}\includegraphics[width=0.40in,height=0.30in]{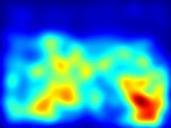}\includegraphics[width=0.40in,height=0.30in]{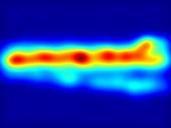}\includegraphics[width=0.40in,height=0.30in]{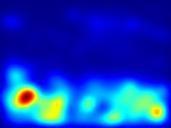}\includegraphics[width=0.40in,height=0.30in]{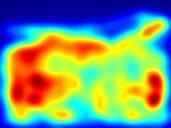}}\\
  AIM \cite{Bruce2005} & .552 & -.509 & \parbox[c]{1em}{\includegraphics[width=0.40in,height=0.30in]{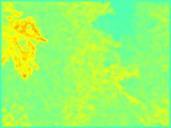}\includegraphics[width=0.40in,height=0.30in]{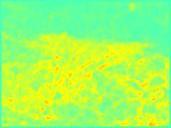}\includegraphics[width=0.40in,height=0.30in]{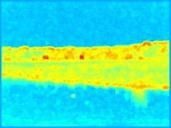}\includegraphics[width=0.40in,height=0.30in]{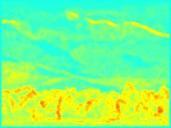}\includegraphics[width=0.40in,height=0.30in]{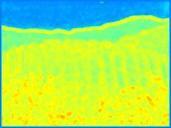}}\\
  %SWAM & .596 & .574 & \parbox[c]{1em}{\includegraphics[width=0.40in,height=0.30in]{figures/qualitative/kootstra/SWAM/nature_03.jpg}\includegraphics[width=0.40in,height=0.30in]{figures/qualitative/kootstra/SWAM/nature_16.jpg}\includegraphics[width=0.40in,height=0.30in]{figures/qualitative/kootstra/SWAM/nature_37.jpg}\includegraphics[width=0.40in,height=0.30in]{figures/qualitative/kootstra/SWAM/nature_42.jpg}\includegraphics[width=0.40in,height=0.30in]{figures/qualitative/kootstra/SWAM/nature_18.jpg}}\\
  %SIM \cite{Murray2011} & .584 & .573 & \parbox[c]{1em}{\includegraphics[width=0.40in,height=0.30in]{figures/qualitative/kootstra/SIM/nature_03.jpg}\includegraphics[width=0.40in,height=0.30in]{figures/qualitative/kootstra/SIM/nature_16.jpg}\includegraphics[width=0.40in,height=0.30in]{figures/qualitative/kootstra/SIM/nature_37.jpg}\includegraphics[width=0.40in,height=0.30in]{figures/qualitative/kootstra/SIM/nature_42.jpg}\includegraphics[width=0.40in,height=0.30in]{figures/qualitative/kootstra/SIM/nature_18.jpg}}\\
  NSWAM & .565 & -.168* & \parbox[c]{1em}{\includegraphics[width=0.40in,height=0.30in]{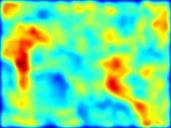}\includegraphics[width=0.40in,height=0.30in]{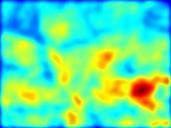}\includegraphics[width=0.40in,height=0.30in]{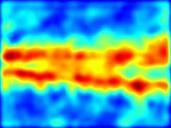}\includegraphics[width=0.40in,height=0.30in]{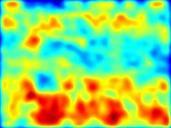}\includegraphics[width=0.40in,height=0.30in]{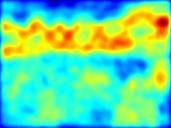}}\\
  NSWAM-CM1 & .564 & -.227 & \parbox[c]{1em}{\includegraphics[width=0.40in,height=0.30in]{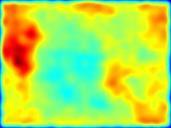}\includegraphics[width=0.40in,height=0.30in]{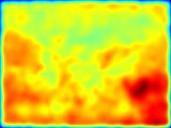}\includegraphics[width=0.40in,height=0.30in]{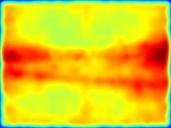}\includegraphics[width=0.40in,height=0.30in]{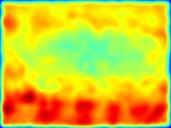}\includegraphics[width=0.40in,height=0.30in]{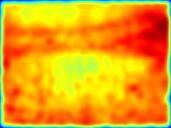}}\\
  NSWAM-CM2 & .566 & -.213 & \parbox[c]{1em}{\includegraphics[width=0.40in,height=0.30in]{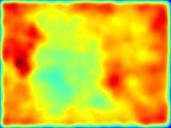}\includegraphics[width=0.40in,height=0.30in]{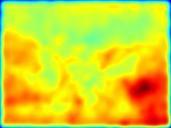}\includegraphics[width=0.40in,height=0.30in]{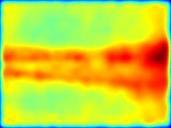}\includegraphics[width=0.40in,height=0.30in]{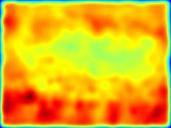}\includegraphics[width=0.40in,height=0.30in]{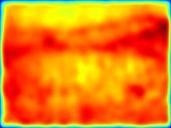}}\\
  NSWAM-CM5 & .566 & -.211 & \parbox[c]{1em}{\includegraphics[width=0.40in,height=0.30in]{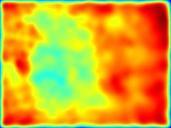}\includegraphics[width=0.40in,height=0.30in]{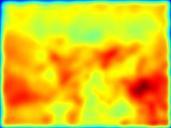}\includegraphics[width=0.40in,height=0.30in]{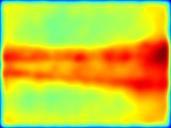}\includegraphics[width=0.40in,height=0.30in]{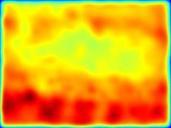}\includegraphics[width=0.40in,height=0.30in]{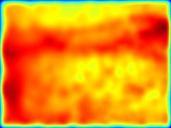}}\\
  NSWAM-CM10 & .567* & -.209 & \parbox[c]{1em}{\includegraphics[width=0.40in,height=0.30in]{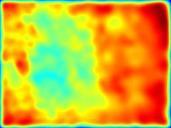}\includegraphics[width=0.40in,height=0.30in]{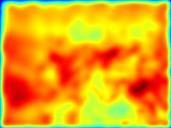}\includegraphics[width=0.40in,height=0.30in]{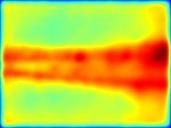}\includegraphics[width=0.40in,height=0.30in]{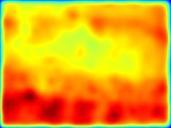}\includegraphics[width=0.40in,height=0.30in]{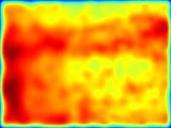}}\\
  \end{tabular} %\hspace{5mm} 
  \end{subfigure}
  \begin{subfigure}{0.5\linewidth}
    \includegraphics[width=2.8in,height=1.5in]{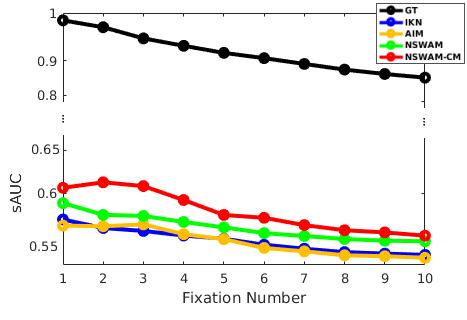}\\
    \includegraphics[width=2.8in,height=1.5in]{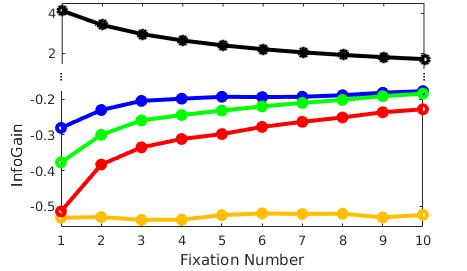}
    %\caption*{grafica de AUC i sAUC vs fixation num.}
    \end{subfigure}
  \end{adjustwidth}
  \caption{Results on saliency for KTH (Kootra et al'.s \cite{Kootstra2011}) Eye Tracking Dataset. \textbf{Left}: Saliency metric scores. \textbf{Middle}: Examples of saliency maps. \textbf{Right}: Shuffled scores per fixation number.}
\label{fig:saliency2}
\end{figure}

\begin{figure}[H]%[!ht]
  \begin{adjustwidth}{-0.4in}{-2in}
  \begin{subfigure}{0.5\linewidth}
  \begin{tabular}{cccc}
  Model & sAUC & InfoGain & \parbox[c]{1em}{\includegraphics[width=0.40in,height=0.30in]{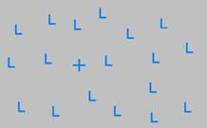}\includegraphics[width=0.40in,height=0.30in]{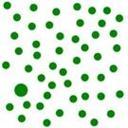}\includegraphics[width=0.40in,height=0.30in]{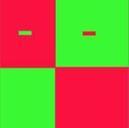}\includegraphics[width=0.40in,height=0.30in]{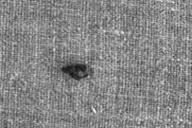}\includegraphics[width=0.40in,height=0.30in]{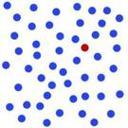}}\\
  \hline
  Human Fix. & .623 & .777 & \parbox[c]{1em}{\includegraphics[width=0.40in,height=0.30in]{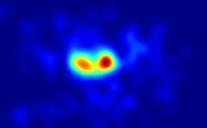}\includegraphics[width=0.40in,height=0.30in]{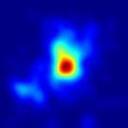}\includegraphics[width=0.40in,height=0.30in]{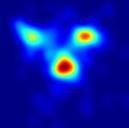}\includegraphics[width=0.40in,height=0.30in]{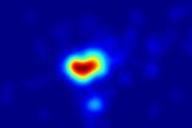}\includegraphics[width=0.40in,height=0.30in]{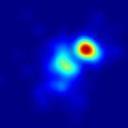}}\\
  %Gaussian & 1 & 1 & \parbox[c]{1em}{\includegraphics[width=0.40in,height=0.30in]{example-image-a}\includegraphics[width=0.40in,height=0.30in]{example-image-a}\includegraphics[width=0.40in,height=0.30in]{example-image-a}\includegraphics[width=0.40in,height=0.30in]{example-image-a}\includegraphics[width=0.40in,height=0.30in]{example-image-a}}\\
  IKN \cite{Itti1998} & .562 & -.724* & \parbox[c]{1em}{\includegraphics[width=0.40in,height=0.30in]{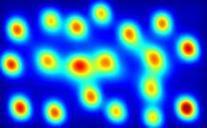}\includegraphics[width=0.40in,height=0.30in]{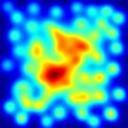}\includegraphics[width=0.40in,height=0.30in]{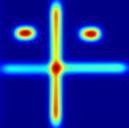}\includegraphics[width=0.40in,height=0.30in]{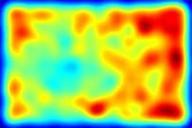}\includegraphics[width=0.40in,height=0.30in]{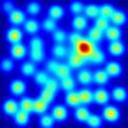}}\\
  AIM \cite{Bruce2005} & .544 & -6.49 & \parbox[c]{1em}{\includegraphics[width=0.40in,height=0.30in]{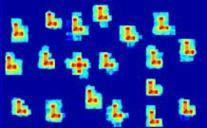}\includegraphics[width=0.40in,height=0.30in]{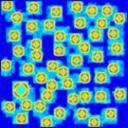}\includegraphics[width=0.40in,height=0.30in]{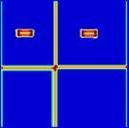}\includegraphics[width=0.40in,height=0.30in]{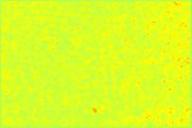}\includegraphics[width=0.40in,height=0.30in]{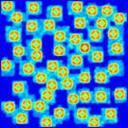}}\\
  %SWAM & .602 & .571 & \parbox[c]{1em}{\includegraphics[width=0.40in,height=0.30in]{figures/qualitative/cat2000_nopad/SWAM/051.jpg}\includegraphics[width=0.40in,height=0.30in]{figures/qualitative/cat2000_nopad/SWAM/009.jpg}\includegraphics[width=0.40in,height=0.30in]{figures/qualitative/cat2000_nopad/SWAM/043.jpg}\includegraphics[width=0.40in,height=0.30in]{figures/qualitative/cat2000_nopad/SWAM/157.jpg}\includegraphics[width=0.40in,height=0.30in]{figures/qualitative/cat2000_nopad/SWAM/003.jpg}}\\
  %SIM \cite{Murray2011} & .578 & .566 & \parbox[c]{1em}{\includegraphics[width=0.40in,height=0.30in]{figures/qualitative/cat2000_nopad/SIM/051.jpg}\includegraphics[width=0.40in,height=0.30in]{figures/qualitative/cat2000_nopad/SIM/009.jpg}\includegraphics[width=0.40in,height=0.30in]{figures/qualitative/cat2000_nopad/SIM/043.jpg}\includegraphics[width=0.40in,height=0.30in]{figures/qualitative/cat2000_nopad/SIM/157.jpg}\includegraphics[width=0.40in,height=0.30in]{figures/qualitative/cat2000_nopad/SIM/003.jpg}}\\
  NSWAM & .567* & -1.01 & \parbox[c]{1em}{\includegraphics[width=0.40in,height=0.30in]{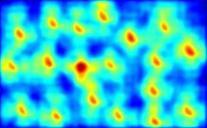}\includegraphics[width=0.40in,height=0.30in]{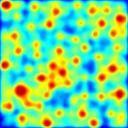}\includegraphics[width=0.40in,height=0.30in]{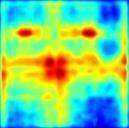}\includegraphics[width=0.40in,height=0.30in]{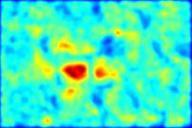}\includegraphics[width=0.40in,height=0.30in]{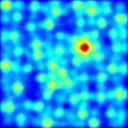}}\\
  NSWAM-CM1 & .561 & -1.24 & \parbox[c]{1em}{\includegraphics[width=0.40in,height=0.30in]{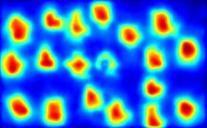}\includegraphics[width=0.40in,height=0.30in]{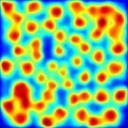}\includegraphics[width=0.40in,height=0.30in]{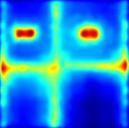}\includegraphics[width=0.40in,height=0.30in]{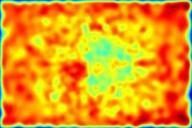}\includegraphics[width=0.40in,height=0.30in]{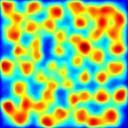}}\\
  NSWAM-CM2 & .563 & -1.14 & \parbox[c]{1em}{\includegraphics[width=0.40in,height=0.30in]{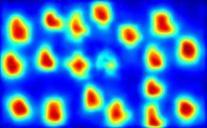}\includegraphics[width=0.40in,height=0.30in]{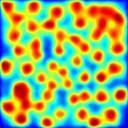}\includegraphics[width=0.40in,height=0.30in]{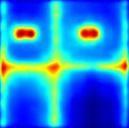}\includegraphics[width=0.40in,height=0.30in]{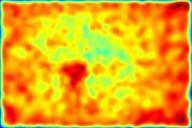}\includegraphics[width=0.40in,height=0.30in]{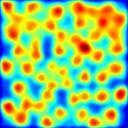}}\\
  NSWAM-CM5 & .565 & -1.09 & \parbox[c]{1em}{\includegraphics[width=0.40in,height=0.30in]{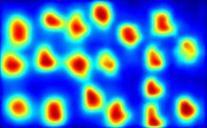}\includegraphics[width=0.40in,height=0.30in]{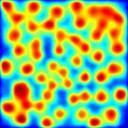}\includegraphics[width=0.40in,height=0.30in]{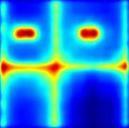}\includegraphics[width=0.40in,height=0.30in]{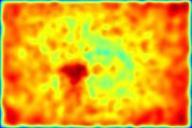}\includegraphics[width=0.40in,height=0.30in]{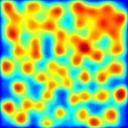}}\\
  NSWAM-CM10 & .567* & -1.07 & \parbox[c]{1em}{\includegraphics[width=0.40in,height=0.30in]{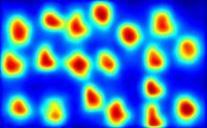}\includegraphics[width=0.40in,height=0.30in]{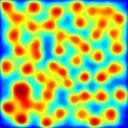}\includegraphics[width=0.40in,height=0.30in]{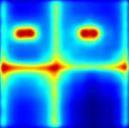}\includegraphics[width=0.40in,height=0.30in]{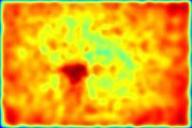}\includegraphics[width=0.40in,height=0.30in]{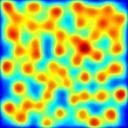}}\\
  \end{tabular} %\hspace{5mm} 
  \end{subfigure}
  \begin{subfigure}{0.5\linewidth}
    \includegraphics[width=2.8in,height=1.5in]{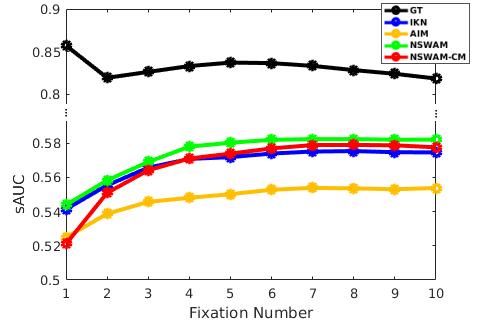}\\
    \includegraphics[width=2.8in,height=1.5in]{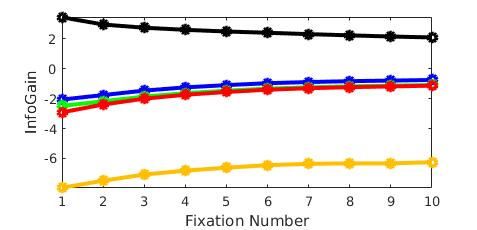}
    %\caption*{grafica de AUC i sAUC vs fixation num.}
    \end{subfigure}
  \end{adjustwidth}
  \caption{Results on saliency for $CAT2000_{Pattern}$ (Borji \& Itti \cite{CAT2000})  Dataset. \textbf{Left}: Saliency metric scores. \textbf{Middle}: Examples of saliency maps. \textbf{Right}: Shuffled scores per fixation number. }
\label{fig:saliency3}
\end{figure}

\begin{figure}[H]%[!ht]
  \begin{adjustwidth}{-0.4in}{-2in}
  \begin{subfigure}{0.5\linewidth}
  \begin{tabular}{cccc}
  Model & sAUC & InfoGain & \parbox[c]{1em}{\includegraphics[width=0.40in,height=0.30in]{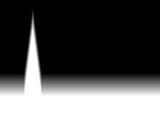}\includegraphics[width=0.40in,height=0.30in]{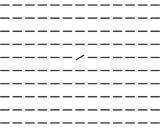}\includegraphics[width=0.40in,height=0.30in]{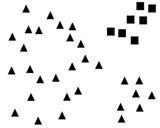}\includegraphics[width=0.40in,height=0.30in]{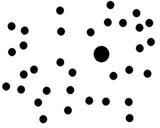}\includegraphics[width=0.40in,height=0.30in]{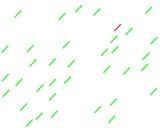}}\\
  \hline
  Human Fix. & .860 & 2.80 & \parbox[c]{1em}{\includegraphics[width=0.40in,height=0.30in]{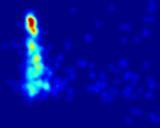}\includegraphics[width=0.40in,height=0.30in]{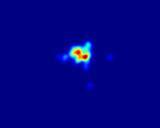}\includegraphics[width=0.40in,height=0.30in]{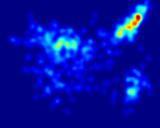}\includegraphics[width=0.40in,height=0.30in]{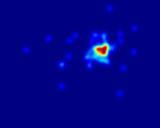}\includegraphics[width=0.40in,height=0.30in]{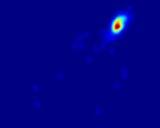}}\\
  %Gaussian & 1 & 1 & \parbox[c]{1em}{\includegraphics[width=0.40in,height=0.30in]{example-image-a}\includegraphics[width=0.40in,height=0.30in]{example-image-a}\includegraphics[width=0.40in,height=0.30in]{example-image-a}\includegraphics[width=0.40in,height=0.30in]{example-image-a}\includegraphics[width=0.40in,height=0.30in]{example-image-a}}\\
  IKN \cite{Itti1998} & .608 & -.233 & \parbox[c]{1em}{\includegraphics[width=0.40in,height=0.30in]{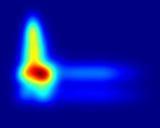}\includegraphics[width=0.40in,height=0.30in]{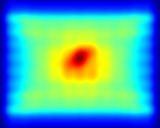}\includegraphics[width=0.40in,height=0.30in]{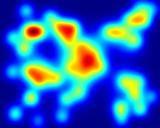}\includegraphics[width=0.40in,height=0.30in]{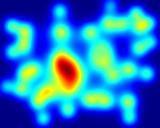}\includegraphics[width=0.40in,height=0.30in]{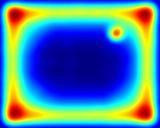}}\\
  AIM \cite{Bruce2005} & .557 & -18.2 & \parbox[c]{1em}{\includegraphics[width=0.40in,height=0.30in]{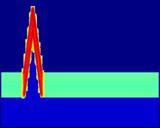}\includegraphics[width=0.40in,height=0.30in]{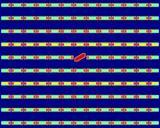}\includegraphics[width=0.40in,height=0.30in]{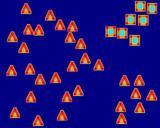}\includegraphics[width=0.40in,height=0.30in]{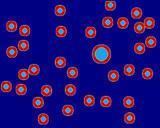}\includegraphics[width=0.40in,height=0.30in]{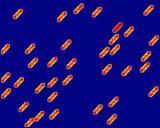}}\\
  %SWAM & .618 & .601 & \parbox[c]{1em}{\includegraphics[width=0.40in,height=0.30in]{figures/qualitative/sid4vam/SWAM/d1Bfv1BdefaultB15.jpg}\includegraphics[width=0.40in,height=0.30in]{figures/qualitative/sid4vam/SWAM/d2Bvs9BlinearB30.jpg}\includegraphics[width=0.40in,height=0.30in]{figures/qualitative/sid4vam/SWAM/d1Bfv5BproximityGdissimilarB5p8333.jpg}\includegraphics[width=0.40in,height=0.30in]{figures/qualitative/sid4vam/SWAM/d1Bvs6BdefaultB5.jpg}\includegraphics[width=0.40in,height=0.30in]{figures/qualitative/sid4vam/SWAM/d2Bvs1BfeatureB34.jpg}}\\
  %SIM \cite{Murray2011} & .641 & .619 & \parbox[c]{1em}{\includegraphics[width=0.40in,height=0.30in]{figures/qualitative/sid4vam/SIM/d1Bfv1BdefaultB15.jpg}\includegraphics[width=0.40in,height=0.30in]{figures/qualitative/sid4vam/SIM/d2Bvs9BlinearB30.jpg}\includegraphics[width=0.40in,height=0.30in]{figures/qualitative/sid4vam/SIM/d1Bfv5BproximityGdissimilarB5p8333.jpg}\includegraphics[width=0.40in,height=0.30in]{figures/qualitative/sid4vam/SIM/d1Bvs6BdefaultB5.jpg}\includegraphics[width=0.40in,height=0.30in]{figures/qualitative/sid4vam/SIM/d2Bvs1BfeatureB34.jpg}}\\
  NSWAM & .622* & -.149 & \parbox[c]{1em}{\includegraphics[width=0.40in,height=0.30in]{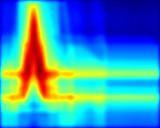}\includegraphics[width=0.40in,height=0.30in]{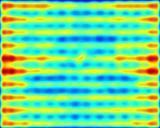}\includegraphics[width=0.40in,height=0.30in]{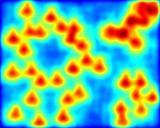}\includegraphics[width=0.40in,height=0.30in]{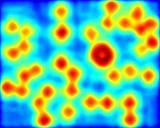}\includegraphics[width=0.40in,height=0.30in]{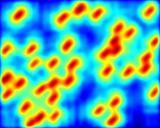}}\\
  NSWAM-CM1 & .617 & -.204 & \parbox[c]{1em}{\includegraphics[width=0.40in,height=0.30in]{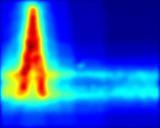}\includegraphics[width=0.40in,height=0.30in]{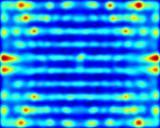}\includegraphics[width=0.40in,height=0.30in]{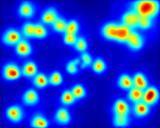}\includegraphics[width=0.40in,height=0.30in]{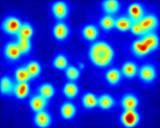}\includegraphics[width=0.40in,height=0.30in]{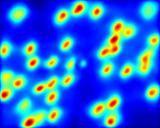}}\\
  NSWAM-CM2 & .622* & -.164 & \parbox[c]{1em}{\includegraphics[width=0.40in,height=0.30in]{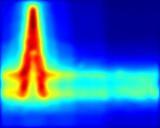}\includegraphics[width=0.40in,height=0.30in]{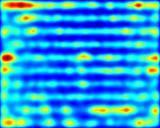}\includegraphics[width=0.40in,height=0.30in]{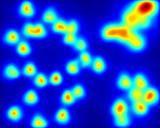}\includegraphics[width=0.40in,height=0.30in]{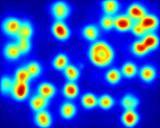}\includegraphics[width=0.40in,height=0.30in]{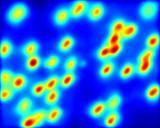}}\\
  NSWAM-CM5 & .620 & -.139 & \parbox[c]{1em}{\includegraphics[width=0.40in,height=0.30in]{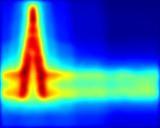}\includegraphics[width=0.40in,height=0.30in]{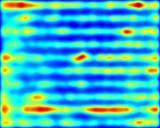}\includegraphics[width=0.40in,height=0.30in]{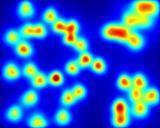}\includegraphics[width=0.40in,height=0.30in]{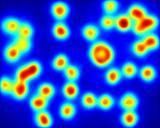}\includegraphics[width=0.40in,height=0.30in]{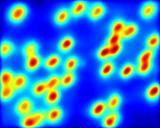}}\\
  NSWAM-CM10 & .618 & -.131* & \parbox[c]{1em}{\includegraphics[width=0.40in,height=0.30in]{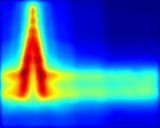}\includegraphics[width=0.40in,height=0.30in]{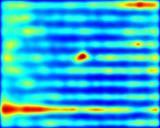}\includegraphics[width=0.40in,height=0.30in]{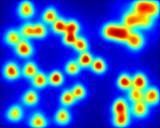}\includegraphics[width=0.40in,height=0.30in]{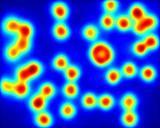}\includegraphics[width=0.40in,height=0.30in]{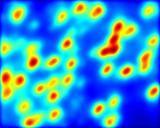}}\\
  \end{tabular} %\hspace{5mm} 
  \end{subfigure}
  \begin{subfigure}{0.5\linewidth}
    \includegraphics[width=2.8in,height=1.5in]{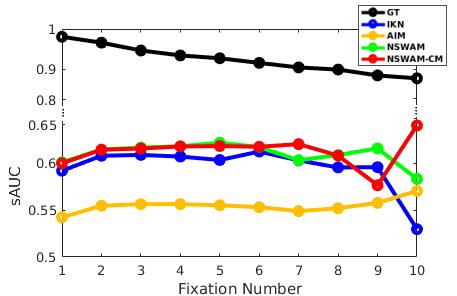}\\
    \includegraphics[width=2.8in,height=1.5in]{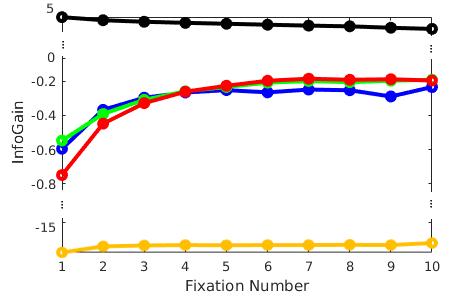}
    %\caption*{grafica de AUC i sAUC vs fixation num.}
    \end{subfigure}
  \end{adjustwidth}
  \caption{Results on saliency for SID4VAM (Berga et al. \cite{Berga2018a}) Eye Tracking Dataset. \textbf{Left}: Saliency metric scores. \textbf{Middle}: Examples of saliency maps. \textbf{Right}: Shuffled scores per fixation number. }
\label{fig:saliency4}
\end{figure}

In this section, probability density maps (GT) have been generated using fixation data of all participants from Toronto, KTH, CAT2000 and SID4VAM eye tracking datasets (model scores and examples in \hyperref[fig:saliency1]{Figs \ref*{fig:saliency1}-\ref*{fig:saliency4}}). Several saliency predictions have been computed from different biologically-inspired models. Our Neurodynamic Saliency Wavelet Model has been computed without (NSWAM) and with foveation (NSWAM-CM), as a mean of cortically-mapped saliency computations through a loop of 1, 2, 5 and 10 saccades. The loop consists on obtaining a saliency map for each view of the scene, and obtaining an unique map for each saccade instance by computing the mean of all saliency maps.
%(\hyperref[sec:cortical]{Cortical Mapping \ref*{sec:cortical}})
%(whether participants are free to view the scene, search or detect a particular object or characteristic in the scene)
%(with a gaussian function of $\sigma=1$ deg \cite{LeMeur2012}) 

Based on the shuffled metric scores, traditional saliency models such as AIM overall score higher on real scene images (\hyperref[fig:saliency1]{Fig. \ref*{fig:saliency1}}), scoring $sAUC_{AIM}$=.663, and $InfoGain_{IKN}$=.024. For the case of nature images (\hyperref[fig:saliency2]{Fig. \ref*{fig:saliency2}}), our non-foveated and foveated versions of the model (NSWAM and NSWAM-CM) scored highest on both metrics ($InfoGain_{NSWAM}$=.168 and $sAUC_{NSWAM-CM10}$=.567). As mentioned before, fixation center biases are present when the task and/or stimulus do not induce regions that are enough salient to produce bottom-up saccades. In addition, in real image datasets (Toronto and KTH), not all images contain particularly salient regions. This is seemingly presented in our models' saliency maps from 1st to 10th fixations (\hyperref[fig:saliency1]{Figs. \ref*{fig:saliency1}-\ref*{fig:saliency2}, rows 5-8}), where salient regions are presented to be less evident across fixation order.

In synthetic image patterns ($CAT2000_P$), both of our model versions outperforms other models \\$sAUC_{NSWAM,NSWAM-CM}$=.567. Center biases are present in such dataset (\hyperref[fig:saliency3]{see Fig. \ref*{fig:saliency3}, "Human Fix." heatmaps}), seemingly reproduced by IKN in the illustration ($InfoGain_{IKN}$=-.724). For the case of SID4VAM dataset (\hyperref[fig:saliency4]{Fig. \ref*{fig:saliency4}}), salient regions are labeled with specific feature type and contrast, and fixation patterns present lower center biases (due to mainly being based a singleton search type of task with a unique salient target with random location). Our model presents highest scores on both metrics ($sAUC_{NSWAM,NSWAM-CM2}$=.622 and $InfoGain_{NSWAM-CM10}$=-.131).

In \hyperref[fig:saliency1]{Figs. \ref*{fig:saliency1}-\ref*{fig:saliency4}} are compared the average score per gaze of human fixations and saliency model predictions. It can be observed that prediction scores for all models decrease as a function of gaze number. Scores of probability density distributions of human fixations (in comparison to fixation locations) decrease around 10\% the sAUC after 10 saccades. This decrease of performance is not reproduced by any of the presented models, instead, most of them show a flat or slightly increasing slopes for the case of sAUC scores and logarithmically increasing scores for InfoGain. NSWAM and NSWAM-CM present similar results upon fixation number.

\defcitealias{Itti1998}{IKN}
\defcitealias{Bruce2005}{AIM}

 %posar barra de colormap

\defcitealias{Itti1998}{IKN}
\defcitealias{Bruce2005}{AIM}

 %posar barra de colormap

\defcitealias{Itti1998}{IKN}
\defcitealias{Bruce2005}{AIM}

 %posar barra de colormap

\defcitealias{Itti1998}{IKN}
\defcitealias{Bruce2005}{AIM}

 %posar barra de colormap

%Temporality of saccades has a big impact of whether fixations are guided through bottom-up and/or top-down activity. 

\begin{figure}[H]
	\centering
	\begin{adjustwidth}{-0.4in}{-2in}
		\begin{subfigure}{0.4\linewidth}
			\includegraphics[width=\linewidth]{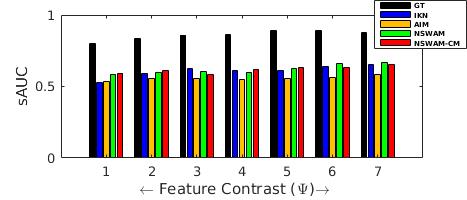} 
		\end{subfigure}
		\begin{subfigure}{0.4\linewidth}
			\includegraphics[width=\linewidth]{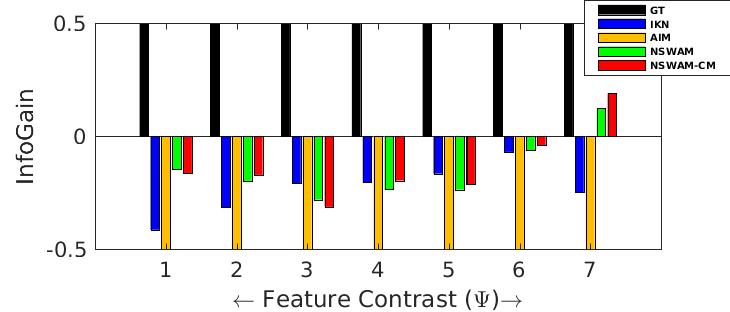}
		\end{subfigure}
	\end{adjustwidth}
	\caption{sAUC and InfoGain scores for each relative target-distractor feature contrast}
	\label{fig:saliency_psi}
\end{figure}

In SID4VAM, stimuli are categorized with specific difficulty (according to the relative target-distractor feature contrast). With these, we computed the score for each relative contrast instance ($\Psi$)  in \hyperref[fig:saliency_psi]{Fig. \ref*{fig:saliency_psi}}. After computing every low-level stimulus instance with the presented models and evaluating results with the same metrics, our saliency model (NSWAM and NSWAM-CM) presents better performance than AIM and IKN and also increases score at higher feature contrasts.

%fer una grafica i exemples comparant el center bias? (exemples clars a cat2000)

\subsubsection{Discussion}

Quantitatively, systematic tendencies in free-viewing (center biases, inter-participant differences, etc.\cite{Tatler2005}) should not be likely to be considered as indicators of saliency. Although shuffled metrics try to penalize for these effects, benchmarks do not compensate for these tendencies from model evaluations (these are particular for each dataset task and stimulus properties). Acknowledging that first saccades determine bottom-up eye movement guidance \cite{Antes1974, Zhaoping2012}, it is a phenomenon also present in our experimental data (in terms of the decrease of performance with respect fixation region probability compared to fixation locations). In that aspect, evaluating first fixations with more importance could define new benchmarks for saliency modeling, similarly with stimuli where feature contrast in salient objects is quantified. Ideal conditions (following the Weber law) determine that if there is less difficulty for finding the salient region (higher target-distractor contrast), saliency will be focused on that region. Conversely, fixations would be distributed on the whole scene if otherwise. Our model presents better performance than other biologically-inspired ones accounting for these basis.

\subsection{Results on predicting scanpaths} \label{sec:results_scanpath}

\begin{figure}[H]
	\centering
	\begin{adjustwidth}{-0.4in}{-2in}
		\makebox[0.20\linewidth]{Toronto}
		\makebox[0.20\linewidth]{KTH}
		\makebox[0.20\linewidth]{CAT2000$_P$}
		\makebox[0.20\linewidth]{SID4VAM}\\
		\includegraphics[width=.20\linewidth,height=2.75cm]{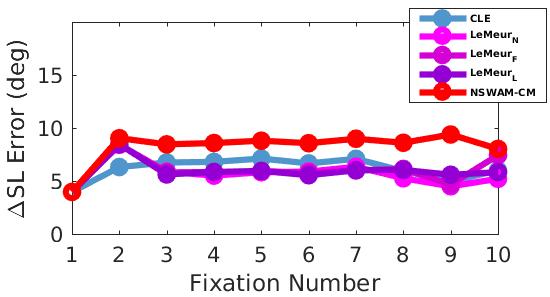}
		\includegraphics[width=.20\linewidth,height=2.75cm]{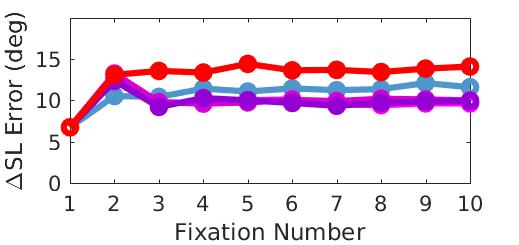}
		\includegraphics[width=.20\linewidth,height=2.75cm]{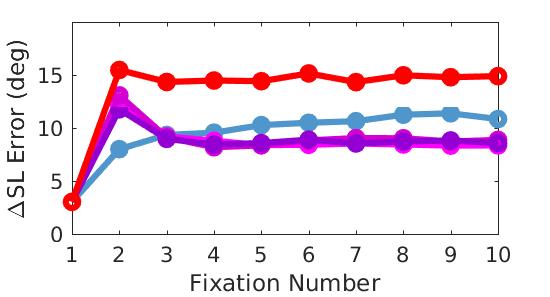}
		\includegraphics[width=.20\linewidth,height=2.75cm]{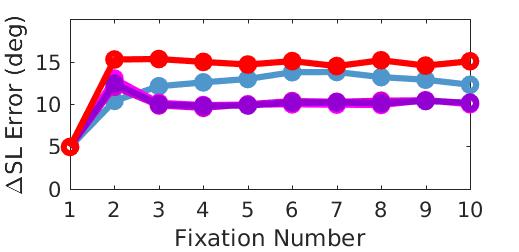}\\
		\includegraphics[width=.20\linewidth,height=2.75cm]{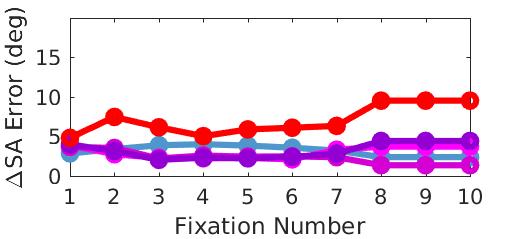}
		\includegraphics[width=.20\linewidth,height=2.75cm]{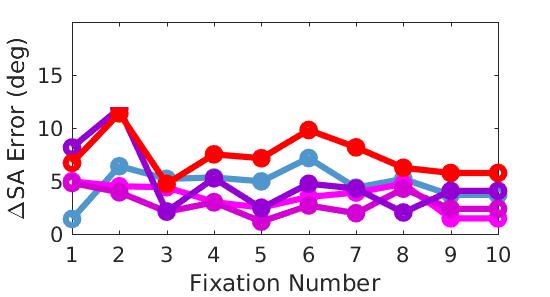}
		\includegraphics[width=.20\linewidth,height=2.75cm]{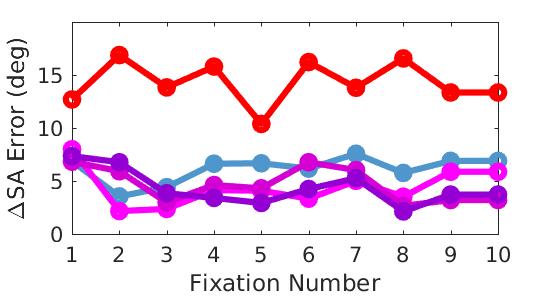}
		\includegraphics[width=.20\linewidth,height=2.75cm]{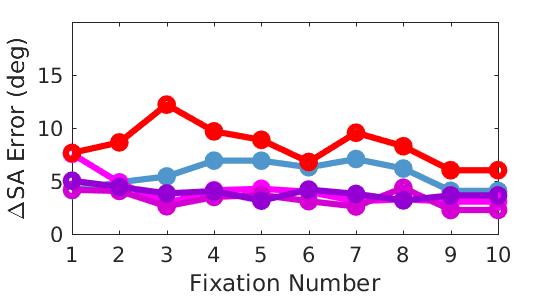}\\
		\includegraphics[width=.20\linewidth,height=2.75cm]{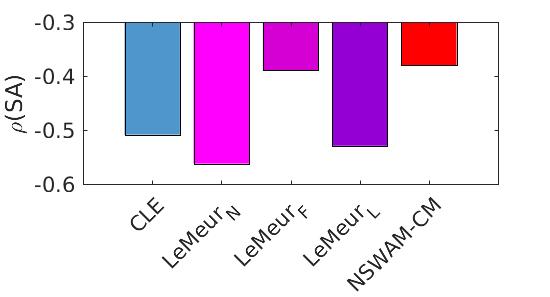}
		\includegraphics[width=.20\linewidth,height=2.75cm]{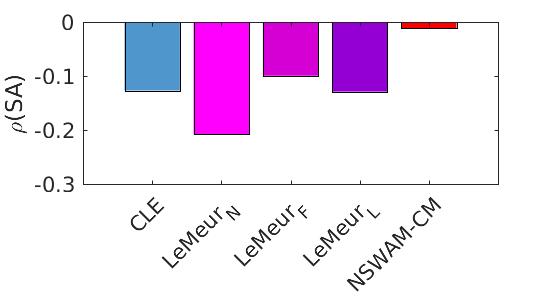}
		\includegraphics[width=.20\linewidth,height=2.75cm]{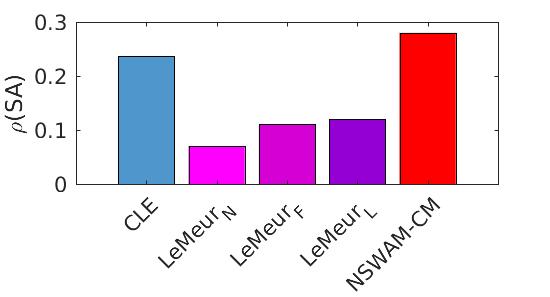}
		\includegraphics[width=.20\linewidth,height=2.75cm]{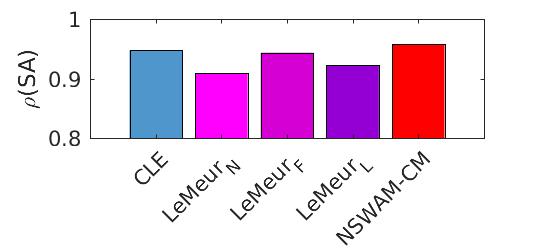}\\
	\end{adjustwidth}
	\caption{\textbf{1st row:} Prediction errors in Saccade Landing ($\Delta$SL) for real indoor/outdoor (Toronto), nature (KTH) and synthetic (CAT2000$_P$ and SID4VAM) image datasets. \textbf{2nd row:} Prediction errors in Saccade Amplitude ($\Delta$SA) on same datasets.  \textbf{3rd row:} Correlations of Saccade Amplitude ($\rho$SA) with respect human fixations.}
	\label{fig:results_scanpath}
\end{figure}

Illustration of scanpaths from datasets presented in \hyperref[sec:results_saliency]{Section \ref*{sec:results_saliency}} were computed with scanpath models in \hyperref[fig:examples_scanpaths]{Fig.  \ref*{fig:examples_scanpaths}}. Scanpaths are predicted by NSWAM-CM during the first 10 saccades, by selecting maximum activity of our model for every saccade. We have plotted our model's performance in addition to Boccignone\&Ferraro's and LeMeur\&Liu's predictions (\hyperref[fig:results_scanpath]{Fig. \ref*{fig:results_scanpath}}). Saccade statistics show an initial increment of saccade amplitude, decreasing as a function of fixation number. Errors of SA and SL ($\Delta$SA and $\Delta$SL) are calculated as absolute differences between model predictions and human fixations. Values of $\Delta$SL appear to be lower and similar for all models during initial fixations. 

Prediction errors are shown to be sustained or increasing for CLE and NSWAM-CM (maybe due to their lack of processing higher level features, experimental center biases, etc.). Errors on $\Delta$SA predictions are lower for LeMeur\&Liu's model, retaining similar saccades (except for synthetic images of SID4VAM). Although these errors are representative in terms of saccade sequence, we also computed correlations of models' SA with GT ($\rho$SA). In this last case, NSWAM-CM presents most higher correlation values for all datasets ($\rho$SA$_{Toronto}$=-.38, $p$=.09; $\rho$SA$_{KTH}$=.012, $p$=.96; $\rho$SA$_{CAT2000_P}$=.28, $p$=.16; $\rho$SA$_{SID4VAM}$=.96, $p$=1.26$\times 10^{-71}$) than other models. Most of them seem to accurately predict SA for SID4VAM (which contains mostly visual search psychophysical image patterns), with $\rho$SA between .7 and .8. Our scanpath model tend to predict eye movements with large mean saccade amplitudes
\{$M(SA)_{Toronto}$ = 7.8$\pm$3.5; $M(SA)_{KTH}$ = 13$\pm$6.1; $M(SA)_{CAT2000_P}$ = 15.7$\pm$6.7; $M(SA)_{SID4VAM}$ = 15.7$\pm$6.9 deg\}, whereas human fixations combine both short and large saccades \{$M(SA)_{Toronto}$ = 4.6$\pm$1; $M(SA)_{KTH}$ = 6.7$\pm$.5; $M(SA)_{CAT2000_P}$ = 5.1$\pm$.9; $M(SA)_{SID4VAM}$ = 5.8$\pm$1.5 deg\}. In that aspect, our prediction errors might arise from not correctly predicting focal fixations.

%(saccade-to-saccade differences for SA; between human fixation coordinates and predictions for SL)
%results are found in figures/quantitative/scanpaths (model order=GT, CLE, LeMeurNFL, NSWAM-CM)
%Although it gave promising results on predicting scanpaths at distinct real image contexts (faces, webpages, landscapes...), we claim the importance of foveation mechanisms.

\definecolor{fuchsia}{RGB}{148,0,211}
\definecolor{darkorchid}{RGB}{211,0,211}

%SCANPATHS
\begin{figure}[H]
	\centering
	
	\begin{adjustwidth}{-0.4in}{-2in}
		\makebox[5em]{"Real"}\includegraphics[width=.14\linewidth]{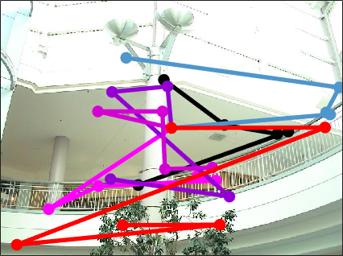}
		\includegraphics[width=.14\linewidth]{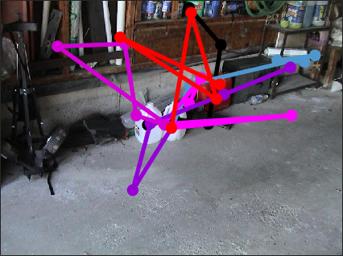}
		\includegraphics[width=.14\linewidth]{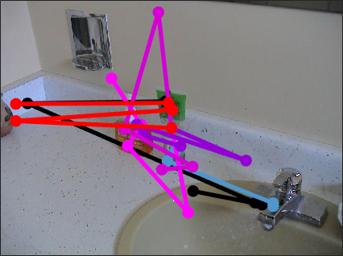}
		\includegraphics[width=.14\linewidth]{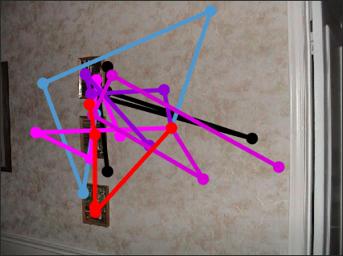}
		\includegraphics[width=.14\linewidth]{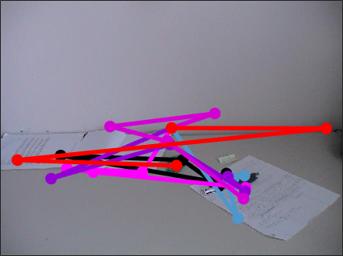}\\
		\makebox[5em]{"Nature"}\includegraphics[width=.14\linewidth]{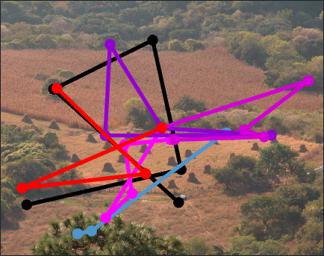}
		\includegraphics[width=.14\linewidth]{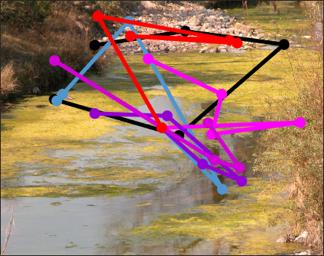}
		\includegraphics[width=.14\linewidth]{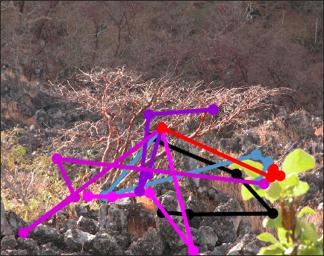}
		\includegraphics[width=.14\linewidth]{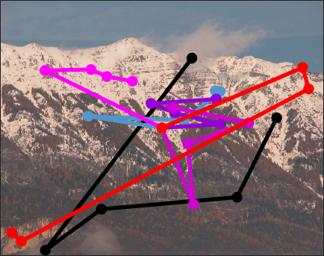}
		\includegraphics[width=.14\linewidth]{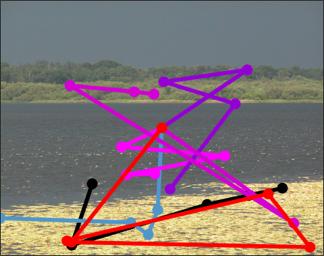}\\
		\makebox[5em]{"Synthetic"}\includegraphics[width=.14\linewidth]{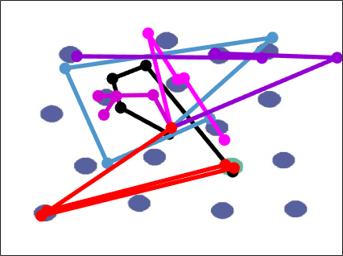}
		\includegraphics[width=.14\linewidth]{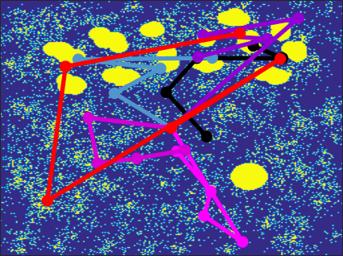}
		\includegraphics[width=.14\linewidth]{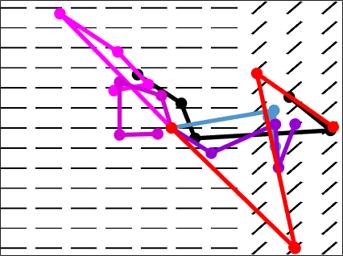}
		\includegraphics[width=.14\linewidth]{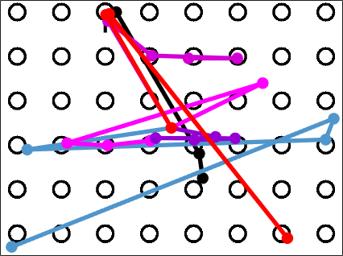}
		\includegraphics[width=.14\linewidth]{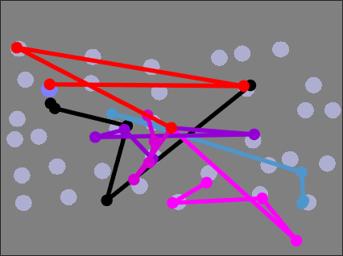}
	\end{adjustwidth}
	\caption{Examples of visual scanpaths for a set of real (\textbf{1st row}, \cite{Bruce2005}), nature (\textbf{2nd row},\cite{Kootstra2011}) and synthetic (\textbf{3rd row},\cite{CAT2000,Berga2018a,Berga_2019_ICCV}) images. Model scanpaths correspond to \textbf{Human Fixations (single sample)}, \textbf{\textcolor{cyan}{CLE} \cite{Boccignone2004}}, \textbf{\textcolor{magenta}{LeMeur$_{Natural}$}}, \textbf{\textcolor{darkorchid}{LeMeur$_{Faces}$}}, \textbf{\textcolor{fuchsia}{LeMeur$_{Landscapes}$}} \textbf{\cite{LeMeur2015}} and \textbf{\textcolor{red}{NSWAM-CM} (ours)}.}
	\label{fig:examples_scanpaths}
	
\end{figure}

\begin{figure}[H]
	\centering
	\begin{adjustwidth}{-0.4in}{-2in}
		\begin{subfigure}{0.20\linewidth}
			\includegraphics[width=\linewidth,height=2.75cm]{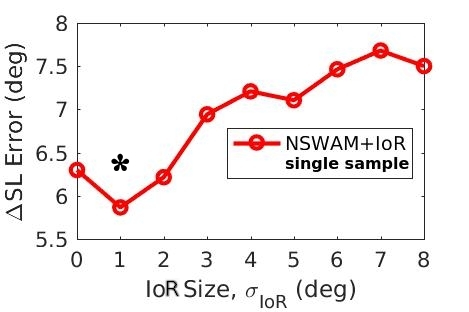}
		\end{subfigure}
		\begin{subfigure}{0.20\linewidth}
			\includegraphics[width=\linewidth,height=2.75cm]{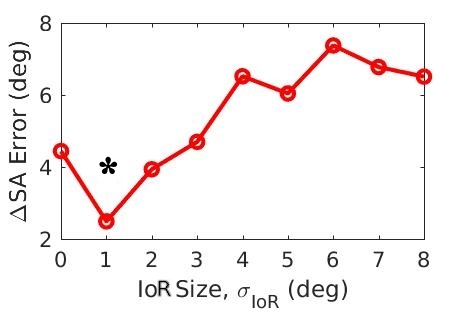}
		\end{subfigure}\unskip\ \vrule\
		\begin{subfigure}{0.20\linewidth}
			\includegraphics[width=\linewidth,height=2.75cm]{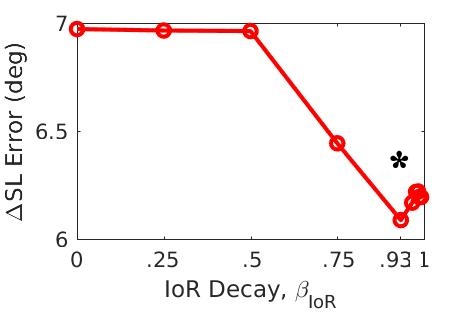}
		\end{subfigure}
		\begin{subfigure}{0.20\linewidth}
			\includegraphics[width=\linewidth,height=2.75cm]{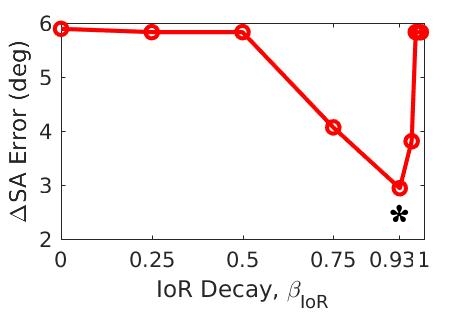}
		\end{subfigure}\\
		%\begin{subfigure}{0.20\linewidth}
		%\includegraphics[width=\linewidth]{figures/ior/statistics/NSWAM+IoR_SAvsBETA.jpg}
		%\end{subfigure}
		\begin{subfigure}{0.20\linewidth}
			\includegraphics[width=\linewidth,height=2.75cm]{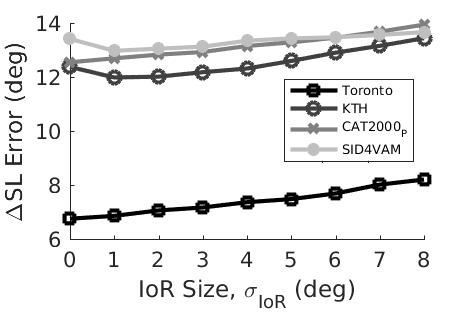}
		\end{subfigure}
		\begin{subfigure}{0.20\linewidth}
			\includegraphics[width=\linewidth,height=2.75cm]{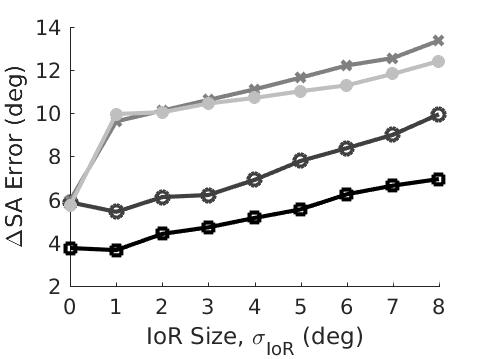}
		\end{subfigure}\unskip\ \vrule\
		\begin{subfigure}{0.20\linewidth}
			\includegraphics[width=\linewidth,height=2.75cm]{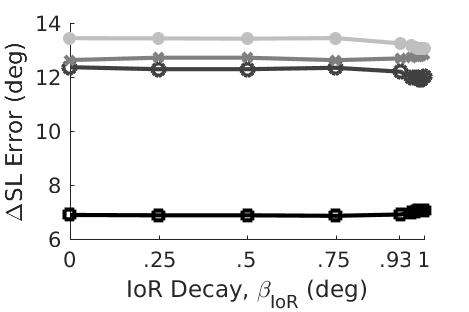}
		\end{subfigure}
		\begin{subfigure}{0.20\linewidth}
			\includegraphics[width=\linewidth,height=2.75cm]{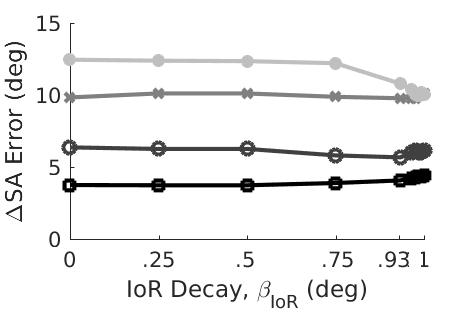}
		\end{subfigure}
		%\begin{subfigure}{0.20\linewidth}
		%\includegraphics[width=\linewidth]{figures/ior/statistics/NSWAM+IoR_SAvsSIGMA.jpg}
		%\end{subfigure}
	\end{adjustwidth}
	\par\vspace{\abovecaptionskip}\parbox{\linewidth}{\raggedright \textbf{*}: Lowest error ($\Delta$SL or $\Delta$SA) at specific parametrization \par}
	\caption{Statistics of scanpath prediction ($\Delta$SL and $\Delta$SA) by the parametrization of IoR decay ($\beta_{\{IoR\}}$) and IoR size ($\sigma_{\{IoR\}}$) in a single sample (\textbf{Top row}, from image scanpaths in Fig. 13) and saliency datasets (\textbf{Bottom row}).}
	%(top row, \hyperref[fig:resultsior1]{Fig. \ref*{fig:resultsior1}})
	%plot_scanpaths_dataset %plot_ior_static_properties
	\label{fig:resultsior2}
\end{figure}

We simulated the inhibition factor for all datasets by substracting the inhibition factor $I_{\{IoR\}}$ to our models' saliency maps (NSWAM+IoR). After computing prediction errors in SA and SL for a single sample (\hyperref[fig:resultsior2]{Fig. \ref*{fig:resultsior2}-Top}), best predictions seem to appear at decay values of $\beta_{\{IoR\}}$ between .93 and .98, which corresponds to 1 to 5 saccades (similarly explained by Samuel \& Kat \cite{Samuel2003} and Berga et al. \cite{Berga2018a}, where takes from 300-1600 ms for the duration of the IoR, corresponding to 1 to 5 times the fixation duration). For the case of the $\sigma_{\{IoR\}}$, lowest prediction error (again, both in SA and SL) is found from 1 to 3 deg (in comparison, LeMeur \& Liu \cite{LeMeur2015} parametrized it by default as 2 deg). Results on $\Delta$SA statistics have similar / slightly increasing performance until ($\beta_{\{IoR\}}<$1) a single fixation time, decreasing at highest decay $\beta_{\{IoR\}}\geq$5th saccade. For $\Delta$SL values, errors in datasets such as KTH and SID4VAM are decreased at higher decay. For the latter, $\Delta$SA errors are shown to decrease progressively at highest decay values ($\beta_{\{IoR\}}\geq$.93). Lastly, when parametrizing the spatial properties of the IoR, saccade prediction performance is highest at lower size (with a near-constant  error in SA and SL increasing about 1 deg for $\sigma_{\{IoR\}}$=1 to 8 deg on all datasets). 

%ior is present on all stimuli, is stimulus point relative to the size or a cognitive / contextual factor

%scanpaths amb/sense ior
%alternativa - table: number of fixations inside the AOI, saccade landing, saccade amplitude
%alternativa - table: probability of fixations (y) i saccade aplitude (x)

\begin{figure}[H] %utilitzar la mateixa imatge, veure com cambia amb ior
	\centering
	\begin{adjustwidth}{-0.4in}{-2in}
		\begin{subfigure}{.3\linewidth}
			\includegraphics[width=1\linewidth]{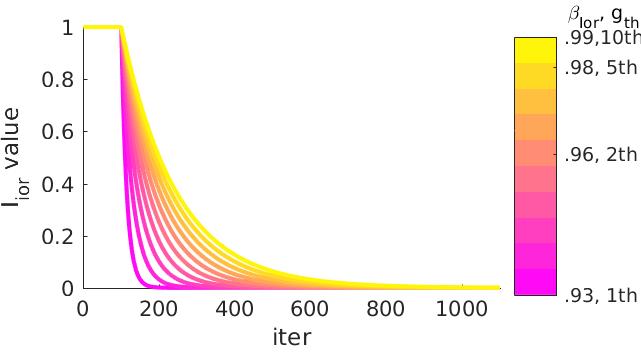}
			\includegraphics[width=1\linewidth]{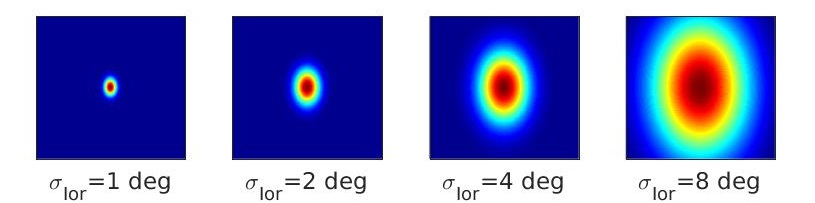}
		\end{subfigure}
		\begin{subfigure}{.6\linewidth}  %diferents decays, scale 2
			\includegraphics[width=0.2\linewidth]{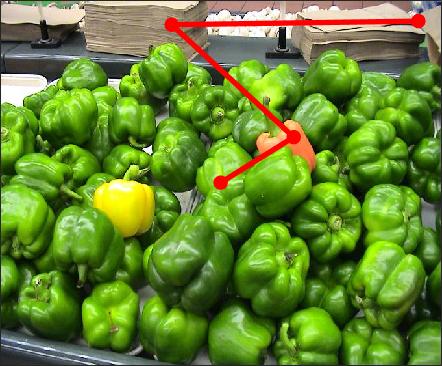}
			\includegraphics[width=0.2\linewidth]{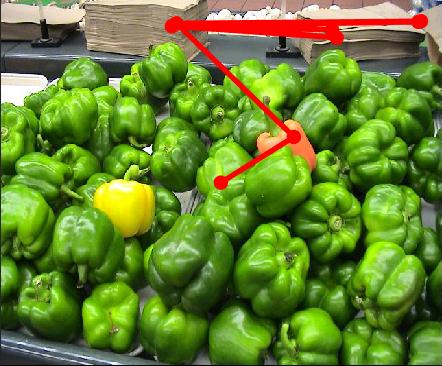}
			\includegraphics[width=0.2\linewidth]{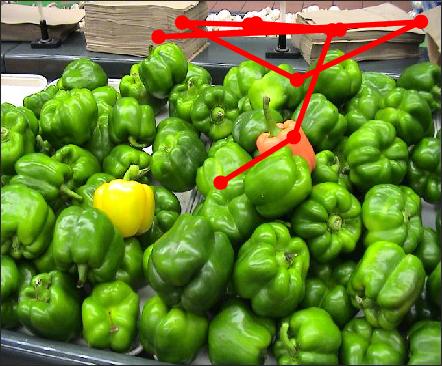}
			\includegraphics[width=0.2\linewidth]{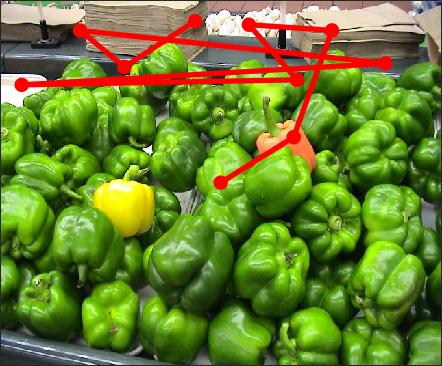}\\
			\makebox[.2\linewidth]{$\beta$=0}
			\makebox[.2\linewidth]{$\beta$=.5}
			\makebox[.2\linewidth]{$\beta$=.93}
			\makebox[.2\linewidth]{$\beta$=1}\\
			%diferents sigmas, decay 1
			\includegraphics[width=0.2\linewidth]{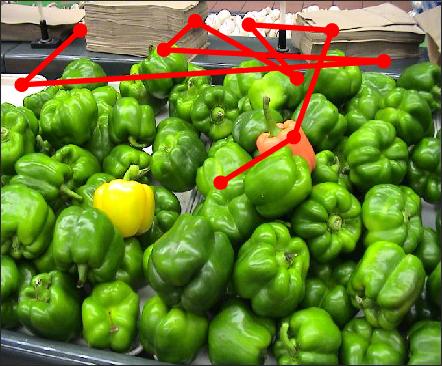}
			\includegraphics[width=0.2\linewidth]{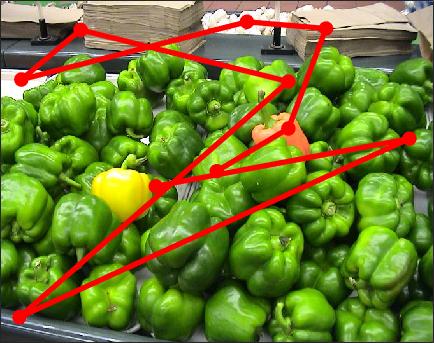}
			\includegraphics[width=0.2\linewidth]{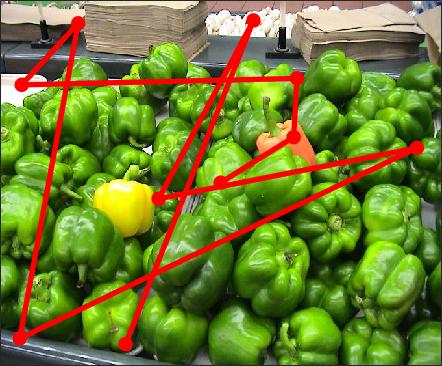}
			\includegraphics[width=0.2\linewidth]{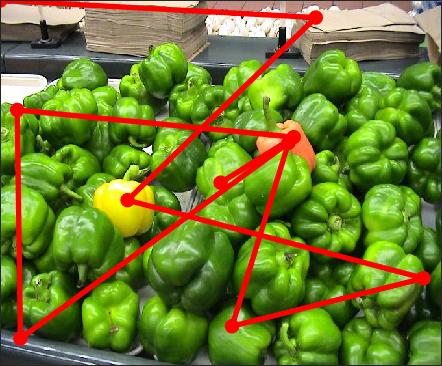}\\
			\makebox[.2\linewidth]{$\sigma$=1 deg}
			\makebox[.2\linewidth]{$\sigma$=2 deg}
			\makebox[.2\linewidth]{$\sigma$=4 deg}
			\makebox[.2\linewidth]{$\sigma$=8 deg}
		\end{subfigure}
	\end{adjustwidth}
	\caption{\textbf{Left}: Evolution of inhibition factor for 100 mem.time (about 1000 iterations), corresponding approximately to performing 10 saccades to the model (top). Spatial representation of the IoR with distinct size (bottom). \textbf{Right}: Examples of scanpaths for different IoR decay factor (top, $\sigma_{\{IoR\}}$=2 deg, $\beta_{\{IoR\}}$=$\{0,.5,.9,1\}$) or distinct IoR size (bottom, $\sigma_{\{IoR\}}$=$\{1,2,4,8\}$ deg,$\beta_{\{IoR\}}$=1).}
	%adapted to maximal scale response ($\sigma_{\{IoR\}}=2^{s+2}$)
	\label{fig:resultsior1}
	
\end{figure} %mostrar ior en static i dinamic

%scanpaths amb/sense ior

\subsubsection{Discussion}

Our model predictions on SA correlate better (i.e. obtain higher $\rho SA$ values) than other scanpath models (in terms of how SA evolves over fixations), however, prediction errors are higher in both SL and SA. We believe that these errors are caused by incorrectly predicting locations of fixations, but not for failing on predictions of the saccade sequence per se. These locations are mainly influenced by systematic tendencies in free-viewing (derived by center biases and/or focal fixations in a particular region of the image). Cortical magnification mechanisms might be responsible for processing higher saliency at regions outside the fovea, generating tendencies of uniquely capturing large saccades. These can be solved by processing high-level feature computations near the fovea, which would increase the probability of fixations at lower SA. Nonetheless, we have to stress that first fixations are long known for being determinants of bottom-up attention \cite{Antes1974,Berga2018a}. Instead, higher inter-participant differences \cite{Tatler2005} and center biases \cite{Rothkegel2017} increase as functions of fixation number, suggested as worse candidates for predicting attention. These parameters appear to specifically affect each stimuli differently (and accounting that each stimulus may convey specific semantic importance between each contextual element), which may relate to top-down attention but not to the image characteristics per se. We also want to stress the importance of foveation in our model. This is a major procedure for determining saccade characteristics (including oculomotor tendencies) and saliency computations, as it determines current human actions during scene visualization. The decrease of spatial resolution at increasing eccentricity provides the aforementioned properties, innate in human vision and invariant to scene semantics.

Adding an IoR mechanism has been seen to affect model activity and therefore scanpath predictions. In \hyperref[fig:resultsior1]{Fig. \ref*{fig:resultsior1}-Left} we show how our inhibition factor ($I_{\{Ior\}}$) decreases over simulation time in relation to the parametrized decay $\beta_{\{IoR\}}$, as well as the projected RF size with respect the gaussian parameter $\sigma_{\{IoR\}}$. These variables (decay and size) affect either location of saccades and its sequence, modulating firing rate activity to already visited locations. It is shown in \hyperref[fig:resultsior1]{Fig. \ref*{fig:resultsior1}-Right} that the initial saccade is focused on the salient region and then it spreads to a specific location in the scene, not repeating with higher value of inhibition decay or field size. In the next section we show how our model can preproduce eye movements beyond free-viewing tasks by modulating of inhibitory top-down signals.

%figura 3. estatic: ior decay (x) vs saccade amplitude diff(y)
%figura 4. estatic: ior decay (x) vs saccade landing (y)
%%figura 5. dinamic: ior decay (x) vs saccade amplitude diff (y)
%%figura 6. dinamic: ior decay (x) vs saccade landing (y)

\subsection{Results on feature and exemplar search}

We have compared our model predictions with bottom-up only (NSWAM $|$ NSWAM-CM) and with top-down inhibitory modulation (NSWAM+VS $|$ NSWAM-CM+VS) for singleton search stimuli (for both real \cite{Bruce2005} and synthetic targets \cite{Berga_2019_ICCV}).
Top-down selection is applied to our low-level feature dimensions (scale, orientation, channel opponency and its polarity). In VS$_M$, inhibition is parametrized considering the feature with the highest activity inside the stimulus ROI (\hyperref[eq:topdown]{Equation \ref*{eq:topdown}-Top}). Besides, inhibitory control in VS$_C$ has been set as the mean wavelet coefficients instead (\hyperref[eq:topdown]{Equation \ref*{eq:topdown}-Bottom}). 

\begin{figure}[H]
	%\makebox[5em]{A}
	\centering
	%\begin{adjustwidth}{0in}{0in}
	\begin{subfigure}{.45\linewidth} 
		\caption*{\centering Synthetic Pattern Search}
		\includegraphics[width=\linewidth]{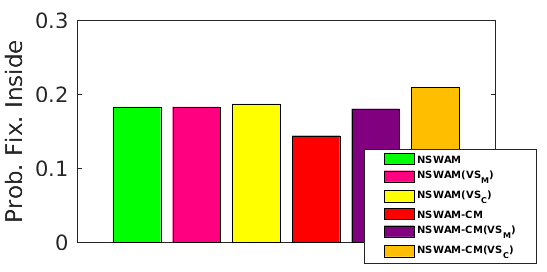}
		\includegraphics[width=\linewidth]{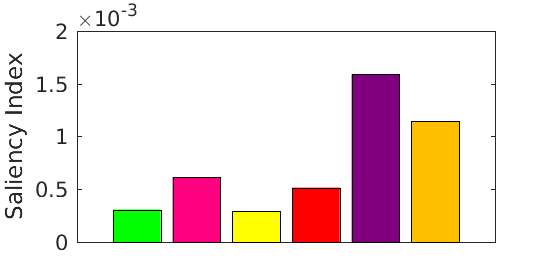}
	\end{subfigure}
	\begin{subfigure}{.45\linewidth} 
		\caption*{\centering Object Search}
		\includegraphics[width=\linewidth]{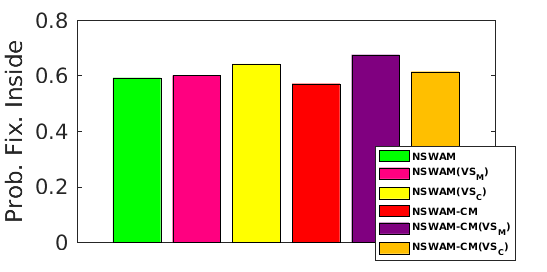}
		\includegraphics[width=\linewidth]{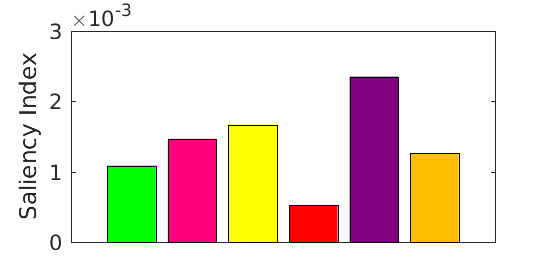}
	\end{subfigure}
	%\end{adjustwidth}
	\caption{Probability of Fixations Inside the ROI (\textbf{Bottom row}) and statistics of Saliency Index (\textbf{Top row}) for synthetic image patterns (\textbf{Left}) and salient object detection regions from real image scenes (\textbf{Right}). %\textcolor{red}{posar ambdos casos (6 barras), i les barres d'error}
	}
	\label{fig:results_search1}
\end{figure}

Results of our model predictions with top-down attention (NSWAM+VS $|$ NSWAM-CM+VS) present higher scores for both SI and PFI (\hyperref[fig:results_search1]{Fig. \ref*{fig:results_search1}}) than the case of bottom-up attention only (NWAM $|$ NSWAM-CM), specially for the case of using cortical magnification NSWAM-CM+VS. Here, there is an increase of fixations inside the ROI: $\Delta(PFI)_{+VS_M}\simeq$1\%, $\Delta(PFI)_{-CM+VS_M}\simeq$10\% and $\Delta(PFI)_{VS_C}\simeq$6\%, $\Delta(PFI)_{-CM+VS_C}\simeq$4\% when searching real objects (\hyperref[fig:results_search1]{Fig.\ref*{fig:results_search1}-Top/Right}) and $\Delta(PFI)_{+VS_M}\simeq$0\%, $\Delta(PFI)_{-CM+VS_M}\simeq$4\% and $\Delta(PFI)_{+VS_C}$ $\simeq$1\%, $\Delta(PFI)_{-CM+VS_C}$ $\simeq$7\% when searching synthetic patterns (\hyperref[fig:results_search1]{Fig.\ref*{fig:results_search1}-Top/Left}). The SI is also seen to increase for both types of images, with differences of $\Delta(SI)_{+VS_M}$=$3.8\times 10^{-4}$, $\Delta(SI)_{-CM+VS_M}$=$1.8\times 10^{-3}$ and $\Delta(SI)_{+VS_C}$=$5.9\times 10^{-4}$, $\Delta(SI)_{-CM+VS_C}$=$7\times 10^{-4}$ for object search (\hyperref[fig:results_search1]{Fig.\ref*{fig:results_search1}-Bottom/Right}) and $\Delta(SI)_{+VS_M}$=$3.1\times 10^{-4}$, $\Delta(SI)_{-CM+VS_M}$=$1.1\times 10^{-3}$ and $\Delta(SI)_{+VS_C}$=$1.3\times 10^{-5}$, $\Delta(SI)_{-CM+VS_C}$=$6\times 10^{-4}$ for psychophysical pattern search (\hyperref[fig:results_search1]{Fig.\ref*{fig:results_search1}-Bottom/Left}). 

\begin{figure}[H]
	%\makebox[5em]{A}
	\begin{adjustwidth}{-0.4in}{-0.8in}
		\footnotesize
		\makebox[5em]{ }
		\makebox[0.10\linewidth]{Image}
		\makebox[0.09\linewidth]{Mask}
		%\makebox[0.15\linewidth]{NSWAM}
		%\makebox[0.15\linewidth]{NSWAM$+VS_M$}
		%\makebox[0.15\linewidth]{NSWAM+$VS_C$}
		%\makebox[0.15\linewidth]{NSWAM-CM}
		%\makebox[0.15\linewidth]{NSWAM-CM+$VS_M$}
		%\makebox[0.15\linewidth]{NSWAM-CM+$VS_C$}
		\makebox[0.10\linewidth]{NSWAM}
		\makebox[0.10\linewidth]{\textbf{NSWAM}}
		\makebox[0.10\linewidth]{\textbf{NSWAM}}
		\makebox[0.10\linewidth]{NSWAM-CM}
		\makebox[0.11\linewidth]{\textbf{NSWAM-CM}}
		\makebox[0.11\linewidth]{\textbf{NSWAM-CM}}\\
		\makebox[5em]{ }
		\makebox[0.10\linewidth]{}
		\makebox[0.09\linewidth]{}
		\makebox[0.10\linewidth]{(saliency)}
		\makebox[0.10\linewidth]{\textbf{+VS$_M$}}
		\makebox[0.10\linewidth]{\textbf{+VS$_C$}}
		\makebox[0.10\linewidth]{(10 sacc.)}
		\makebox[0.11\linewidth]{\textbf{+VS$_M$}}
		\makebox[0.11\linewidth]{\textbf{+VS$_C$}}
		
		%\makebox[3.75em]{}
		\centering
		
		\begin{subfigure}{1\linewidth}
			\makebox[5em]{"Banana"}
			\includegraphics[width=0.10\linewidth,height=1.9cm]{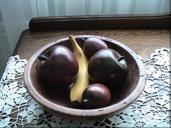} %img
			\includegraphics[width=0.10\linewidth,height=1.9cm]{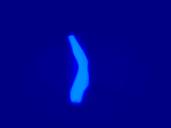} %mask
			\includegraphics[width=0.10\linewidth,height=1.9cm]{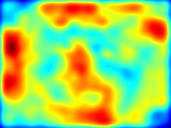} %NSWAM
			\includegraphics[width=0.10\linewidth,height=1.9cm]{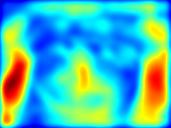} %NSWAM+VSM
			\includegraphics[width=0.10\linewidth,height=1.9cm]{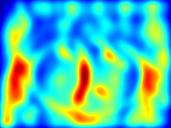} %NSWAM+VSC
			\includegraphics[width=0.10\linewidth,height=1.9cm]{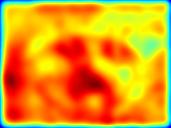} %NSWAM
			\includegraphics[width=0.10\linewidth,height=1.9cm]{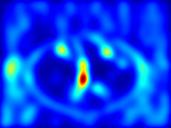} %NSWAM+VSM
			\includegraphics[width=0.10\linewidth,height=1.9cm]{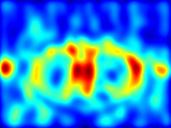} %NSWAM+VSC
		\end{subfigure}
		\begin{subfigure}{1\linewidth} 
			\makebox[5em]{"Bag"}
			\includegraphics[width=0.10\linewidth,height=1.9cm]{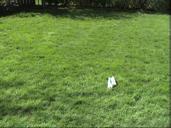} %img
			\includegraphics[width=0.10\linewidth,height=1.9cm]{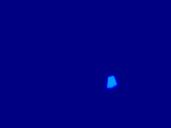} %mask
			\includegraphics[width=0.10\linewidth,height=1.9cm]{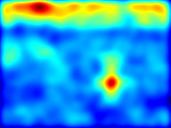} %NSWAM
			\includegraphics[width=0.10\linewidth,height=1.9cm]{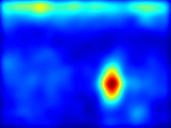} %NSWAM+VSM
			\includegraphics[width=0.10\linewidth,height=1.9cm]{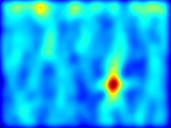} %NSWAM+VSC
			\includegraphics[width=0.10\linewidth,height=1.9cm]{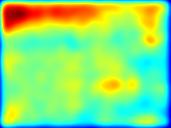} %NSWAM
			\includegraphics[width=0.10\linewidth,height=1.9cm]{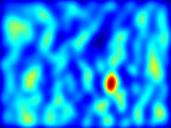} %NSWAM+VSM
			\includegraphics[width=0.10\linewidth,height=1.9cm]{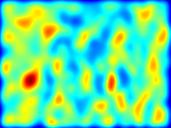} %NSWAM+VSC
		\end{subfigure}
		\begin{subfigure}{1\linewidth} 
			\makebox[5em]{"Bottle"}
			\includegraphics[width=0.10\linewidth,height=1.9cm]{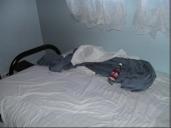} %img
			\includegraphics[width=0.10\linewidth,height=1.9cm]{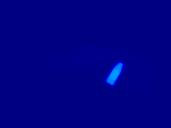} %mask
			\includegraphics[width=0.10\linewidth,height=1.9cm]{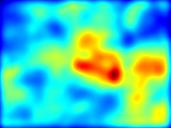} %NSWAM
			\includegraphics[width=0.10\linewidth,height=1.9cm]{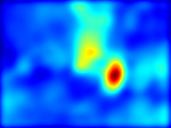} %NSWAM+VSM
			\includegraphics[width=0.10\linewidth,height=1.9cm]{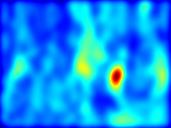}
			%NSWAM+VSC
			\includegraphics[width=0.10\linewidth,height=1.9cm]{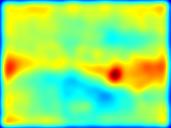} %NSWAM
			\includegraphics[width=0.10\linewidth,height=1.9cm]{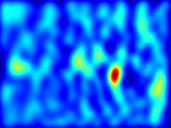} %NSWAM+VSM
			\includegraphics[width=0.10\linewidth,height=1.9cm]{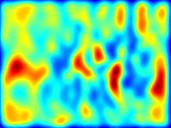}
			%NSWAM+VSC
			%\includegraphics[width=0.35\linewidth,height=2cm]{figures/topdown/plots/SIndex_35.jpg} %plot
		\end{subfigure}
		\begin{subfigure}{1\linewidth} 
			\makebox[5em]{"Traffic"}
			\includegraphics[width=0.10\linewidth,height=1.9cm]{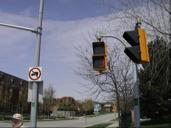} %img
			\includegraphics[width=0.10\linewidth,height=1.9cm]{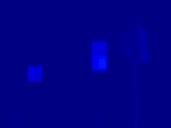} %mask
			\includegraphics[width=0.10\linewidth,height=1.9cm]{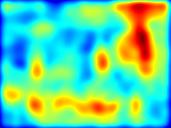} %NSWAM
			\includegraphics[width=0.10\linewidth,height=1.9cm]{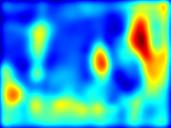} %NSWAM+VSM
			\includegraphics[width=0.10\linewidth,height=1.9cm]{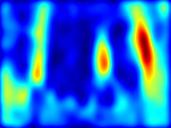} %NSWAM+VSC
			\includegraphics[width=0.10\linewidth,height=1.9cm]{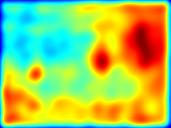} %NSWAM
			\includegraphics[width=0.10\linewidth,height=1.9cm]{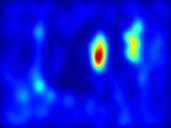} %NSWAM+VSM
			\includegraphics[width=0.10\linewidth,height=1.9cm]{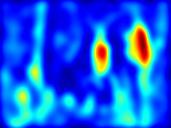} %NSWAM+VSC
		\end{subfigure}
		\begin{subfigure}{1\linewidth} 
			\makebox[5em]{"Lamp"}
			\includegraphics[width=0.10\linewidth,height=1.9cm]{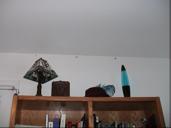} %img
			\includegraphics[width=0.10\linewidth,height=1.9cm]{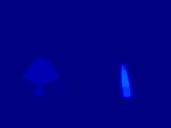} %mask
			\includegraphics[width=0.10\linewidth,height=1.9cm]{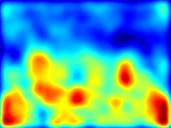} %NSWAM
			\includegraphics[width=0.10\linewidth,height=1.9cm]{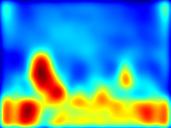} %NSWAM+VSM
			\includegraphics[width=0.10\linewidth,height=1.9cm]{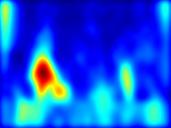}
			%NSWAM+VSC
			\includegraphics[width=0.10\linewidth,height=1.9cm]{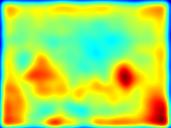} %NSWAM
			\includegraphics[width=0.10\linewidth,height=1.9cm]{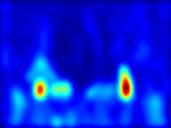} %NSWAM+VSM
			\includegraphics[width=0.10\linewidth,height=1.9cm]{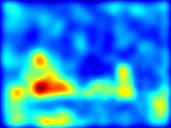}
			%NSWAM+VSC
			%\includegraphics[width=0.35\linewidth,height=2cm]{figures/topdown/plots/SIndex_104.jpg} %plot
		\end{subfigure}
		\begin{subfigure}{1\linewidth} 
			\makebox[5em]{"Green Ball"}
			\includegraphics[width=0.10\linewidth,height=1.9cm]{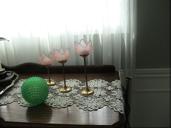} %img
			\includegraphics[width=0.10\linewidth,height=1.9cm]{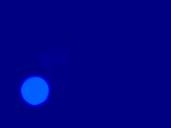} %mask
			\includegraphics[width=0.10\linewidth,height=1.9cm]{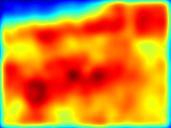} %NSWAM
			\includegraphics[width=0.10\linewidth,height=1.9cm]{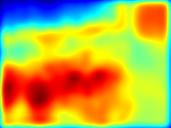} %NSWAM+VSM
			\includegraphics[width=0.10\linewidth,height=1.9cm]{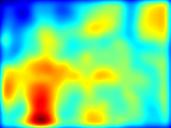}
			%NSWAM+VSC
			\includegraphics[width=0.10\linewidth,height=1.9cm]{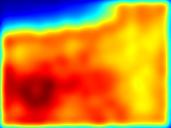} %NSWAM
			\includegraphics[width=0.10\linewidth,height=1.9cm]{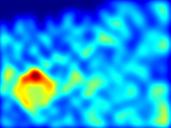} %NSWAM+VSM
			\includegraphics[width=0.10\linewidth,height=1.9cm]{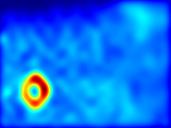}
			%NSWAM+VSC
			%\includegraphics[width=0.35\linewidth,height=2cm]{figures/topdown/plots/SIndex_2.jpg} %plot
		\end{subfigure}
		\begin{subfigure}{1\linewidth} 
			\makebox[5em]{"Person"}
			\includegraphics[width=0.10\linewidth,height=1.9cm]{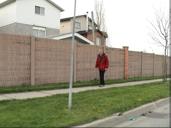} %img
			\includegraphics[width=0.10\linewidth,height=1.9cm]{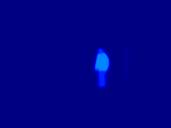} %mask
			\includegraphics[width=0.10\linewidth,height=1.9cm]{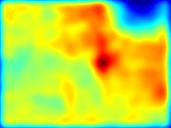} %NSWAM
			\includegraphics[width=0.10\linewidth,height=1.9cm]{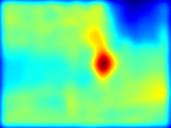} %NSWAM+VSM
			\includegraphics[width=0.10\linewidth,height=1.9cm]{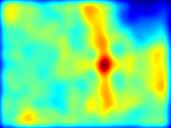}
			%NSWAM+VSC
			\includegraphics[width=0.10\linewidth,height=1.9cm]{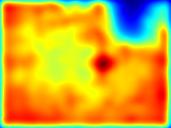} %NSWAM
			\includegraphics[width=0.10\linewidth,height=1.9cm]{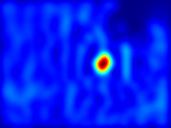} %NSWAM+VSM
			\includegraphics[width=0.10\linewidth,height=1.9cm]{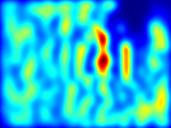}
			%NSWAM+VSC
			%\includegraphics[width=0.35\linewidth,height=2cm]{figures/topdown/plots/SIndex_94.jpg} %plot
		\end{subfigure}
		\begin{subfigure}{1\linewidth} 
			\makebox[5em]{"Magazine"}
			\includegraphics[width=0.10\linewidth,height=1.9cm]{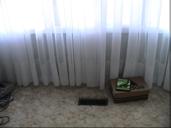} %img
			\includegraphics[width=0.10\linewidth,height=1.9cm]{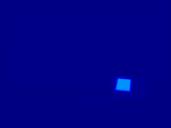} %mask
			\includegraphics[width=0.10\linewidth,height=1.9cm]{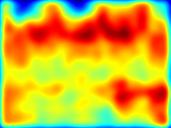} %NSWAM
			\includegraphics[width=0.10\linewidth,height=1.9cm]{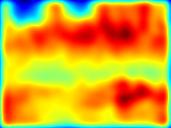} %NSWAM+VSM
			\includegraphics[width=0.10\linewidth,height=1.9cm]{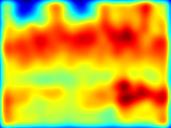}
			%NSWAM+VSC
			\includegraphics[width=0.10\linewidth,height=1.9cm]{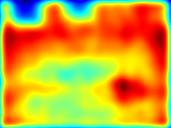} %NSWAM
			\includegraphics[width=0.10\linewidth,height=1.9cm]{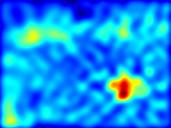} %NSWAM+VSM
			\includegraphics[width=0.10\linewidth,height=1.9cm]{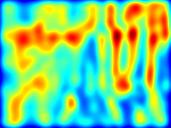}
			%NSWAM+VSC
			%\includegraphics[width=0.35\linewidth,height=2cm]{figures/topdown/plots/SIndex_7.jpg} %plot
		\end{subfigure}
		\begin{subfigure}{1\linewidth} 
			\makebox[5em]{"Tomato"}
			\includegraphics[width=0.10\linewidth,height=1.9cm]{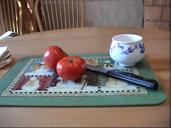} %img
			\includegraphics[width=0.10\linewidth,height=1.9cm]{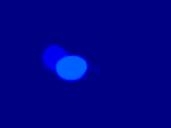} %mask
			\includegraphics[width=0.10\linewidth,height=1.9cm]{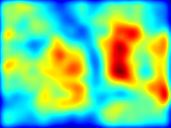} %NSWAM
			\includegraphics[width=0.10\linewidth,height=1.9cm]{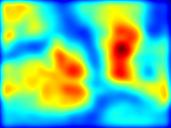} %NSWAM+VSM
			\includegraphics[width=0.10\linewidth,height=1.9cm]{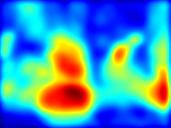}
			%NSWAM+VSC
			\includegraphics[width=0.10\linewidth,height=1.9cm]{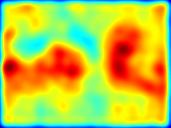} %NSWAM
			\includegraphics[width=0.10\linewidth,height=1.9cm]{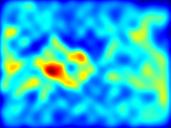} %NSWAM+VSM
			\includegraphics[width=0.10\linewidth,height=1.9cm]{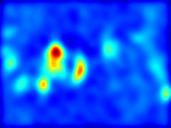}
			%NSWAM+VSC
			%\includegraphics[width=0.35\linewidth,height=2cm]{figures/topdown/plots/SIndex_10.jpg} %plot
		\end{subfigure}
		\begin{subfigure}{1\linewidth} 
			\makebox[5em]{"Car"}
			\includegraphics[width=0.10\linewidth,height=1.9cm]{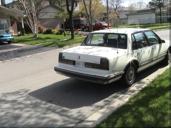} %img
			\includegraphics[width=0.10\linewidth,height=1.9cm]{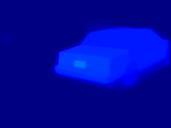} %mask
			\includegraphics[width=0.10\linewidth,height=1.9cm]{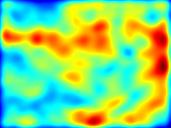} %NSWAM
			\includegraphics[width=0.10\linewidth,height=1.9cm]{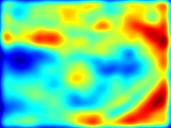} %NSWAM+VSM
			\includegraphics[width=0.10\linewidth,height=1.9cm]{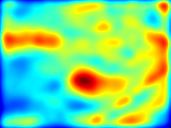}
			%NSWAM+VSC
			\includegraphics[width=0.10\linewidth,height=1.9cm]{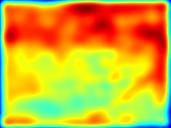} %NSWAM
			\includegraphics[width=0.10\linewidth,height=1.9cm]{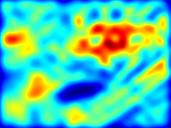} %NSWAM+VSM
			\includegraphics[width=0.10\linewidth,height=1.9cm]{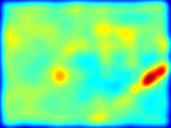}
			%NSWAM+VSC
			%\includegraphics[width=0.35\linewidth,height=2cm]{figures/topdown/plots/SIndex_12.jpg} %plot
		\end{subfigure}
		\begin{subfigure}{1\linewidth} 
			\makebox[5em]{"Telephone"}
			\includegraphics[width=0.10\linewidth,height=1.9cm]{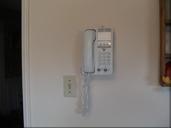} %img
			\includegraphics[width=0.10\linewidth,height=1.9cm]{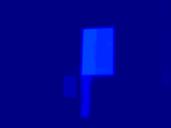} %mask
			\includegraphics[width=0.10\linewidth,height=1.9cm]{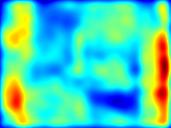} %NSWAM
			\includegraphics[width=0.10\linewidth,height=1.9cm]{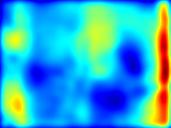} %NSWAM+VSM
			\includegraphics[width=0.10\linewidth,height=1.9cm]{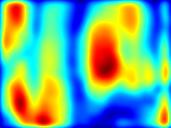}
			%NSWAM+VSC
			\includegraphics[width=0.10\linewidth,height=1.9cm]{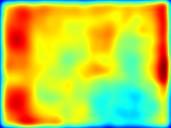} %NSWAM
			\includegraphics[width=0.10\linewidth,height=1.9cm]{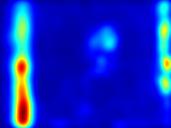} %NSWAM+VSM
			\includegraphics[width=0.10\linewidth,height=1.9cm]{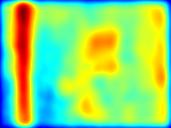}
			%NSWAM+VSC
			%\includegraphics[width=0.35\linewidth,height=2cm]{figures/topdown/plots/SIndex_5.jpg} %plot
		\end{subfigure}
		\vspace{-1em}
		\caption{Search instances with a specific ROI (Mask) based on a category/word exemplar. }
		\label{fig:results_search2}
	\end{adjustwidth}
	
\end{figure}

Some object localization examples are shown in \hyperref[fig:results_search2]{Fig. \ref*{fig:results_search2}}, where the relevance maps (NSWAM+VS $|$ NSWAM-CM+VS) seemingly capture the regions inside the ROI/mask compared to the cases of saliency maps (NSWAM $|$ NSWAM-CM).

%Low-level features to be inhibited in NSWAM+VS$_M$ were selected according to experimental values that gave maximal responses of bottom-up projections for these specific features ($\Omega_{pso\theta}$).

\begin{figure}[H]
	\includegraphics[width=\linewidth]{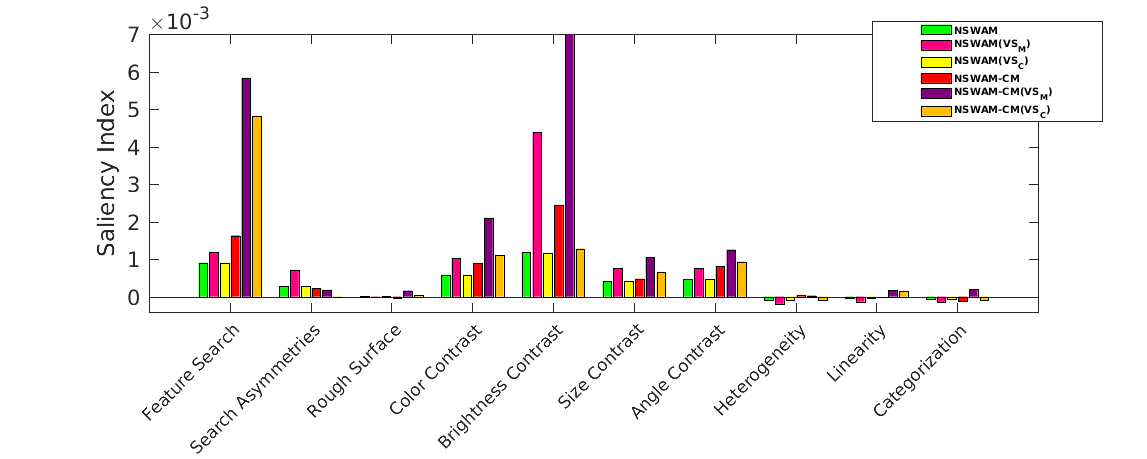}
	\caption{Performance on visual search evaluated on each distinct low-level feature, stimulus instances are from SID4VAM's dataset \cite{Berga2018a,Berga_2019_ICCV}.}
	\label{fig:results_search4}
\end{figure}

\begin{figure}[H]
	\centering
	\begin{adjustwidth}{-0.4in}{-0.75in}
		\begin{subfigure}{.5\linewidth}
			\makebox[0.16\linewidth]{ }
			\includegraphics[width=.10\linewidth,height=.65cm]{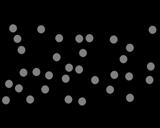}
			\includegraphics[width=.10\linewidth,height=.65cm]{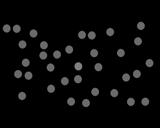}
			\includegraphics[width=.10\linewidth,height=.65cm]{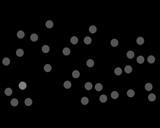}
			\includegraphics[width=.10\linewidth,height=.65cm]{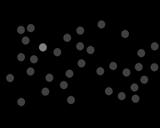}
			\includegraphics[width=.10\linewidth,height=.65cm]{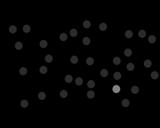}
			\includegraphics[width=.10\linewidth,height=.65cm]{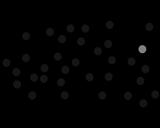}
			\includegraphics[width=.10\linewidth,height=.65cm]{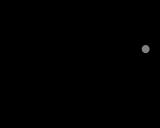}\\
			\makebox[0.16\linewidth]{\scalebox{0.5}{NSWAM}}
			\includegraphics[width=.10\linewidth,height=.65cm]{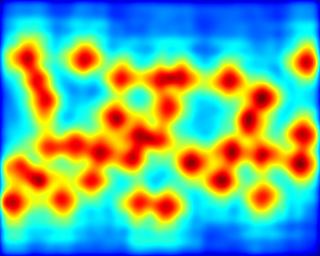}
			\includegraphics[width=.10\linewidth,height=.65cm]{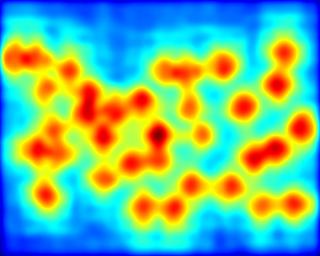}
			\includegraphics[width=.10\linewidth,height=.65cm]{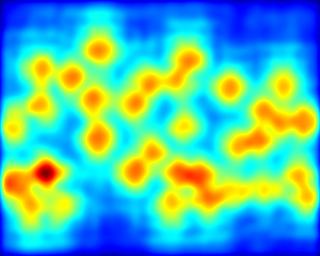}
			\includegraphics[width=.10\linewidth,height=.65cm]{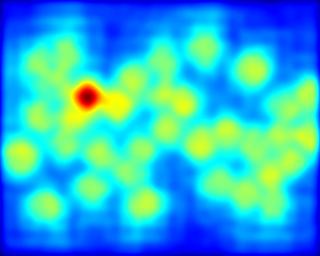}
			\includegraphics[width=.10\linewidth,height=.65cm]{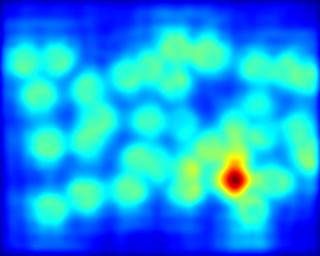}
			\includegraphics[width=.10\linewidth,height=.65cm]{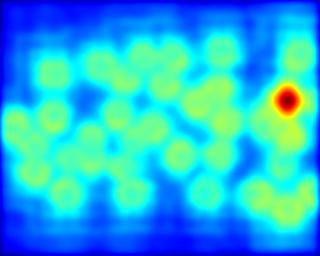}
			\includegraphics[width=.10\linewidth,height=.65cm]{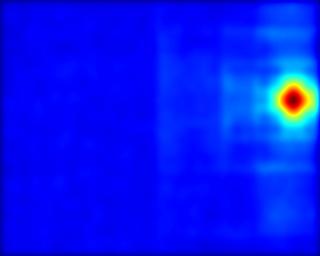}\\
			\makebox[0.16\linewidth]{\scalebox{0.5}{\textbf{NSWAM+VS$_M$}}}
			\includegraphics[width=.10\linewidth,height=.65cm]{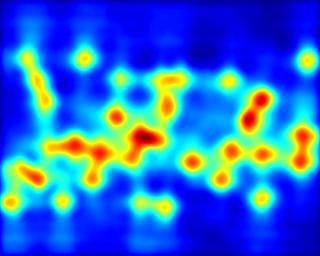}
			\includegraphics[width=.10\linewidth,height=.65cm]{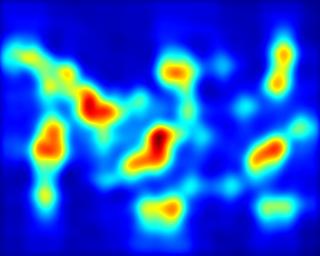}
			\includegraphics[width=.10\linewidth,height=.65cm]{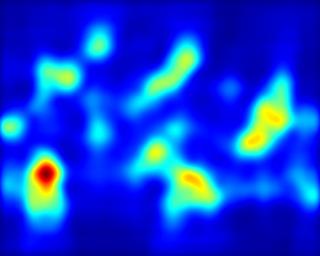}
			\includegraphics[width=.10\linewidth,height=.65cm]{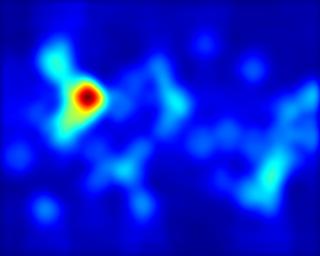}
			\includegraphics[width=.10\linewidth,height=.65cm]{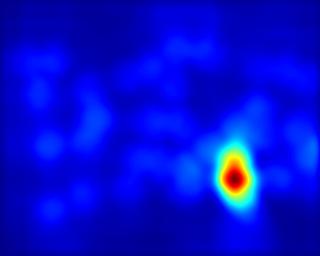}
			\includegraphics[width=.10\linewidth,height=.65cm]{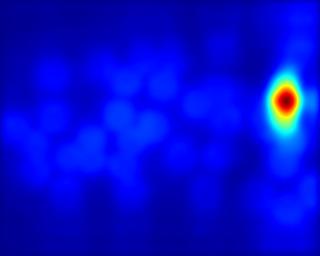}
			\includegraphics[width=.10\linewidth,height=.65cm]{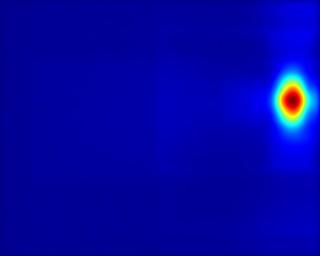}\\
			\makebox[0.16\linewidth]{\scalebox{0.5}{\textbf{NSWAM+VS$_C$}}}
			\includegraphics[width=.10\linewidth,height=.65cm]{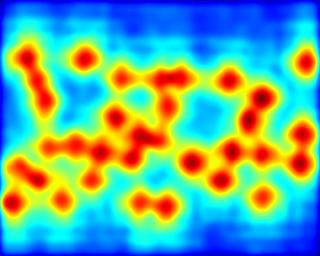}
			\includegraphics[width=.10\linewidth,height=.65cm]{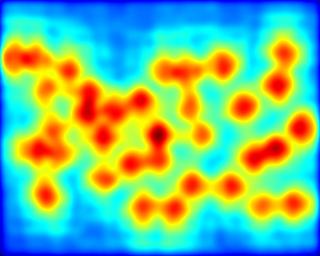}
			\includegraphics[width=.10\linewidth,height=.65cm]{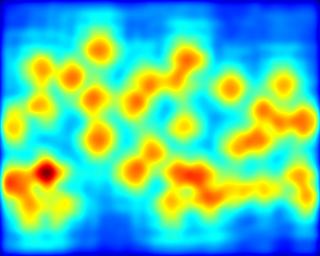}
			\includegraphics[width=.10\linewidth,height=.65cm]{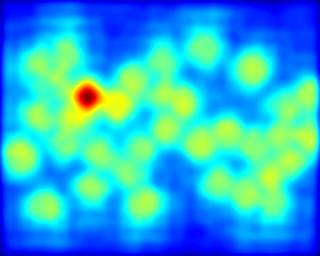}
			\includegraphics[width=.10\linewidth,height=.65cm]{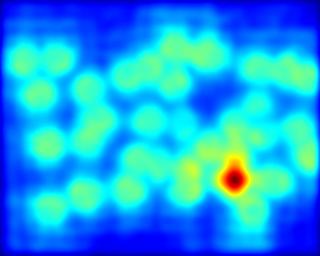}
			\includegraphics[width=.10\linewidth,height=.65cm]{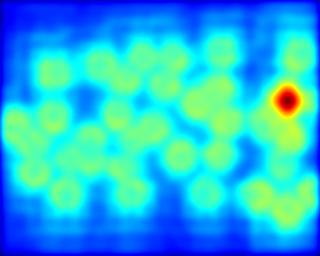}
			\includegraphics[width=.10\linewidth,height=.65cm]{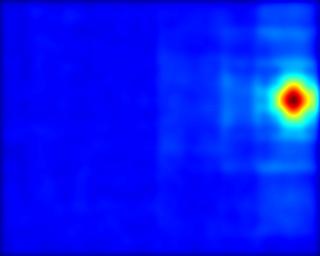}\\
			\makebox[0.16\linewidth]{\scalebox{0.5}{NSWAM-CM}}
			\includegraphics[width=.10\linewidth,height=.65cm]{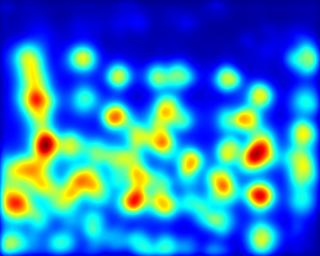}
			\includegraphics[width=.10\linewidth,height=.65cm]{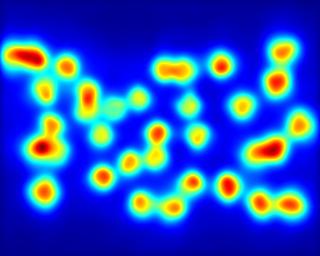}
			\includegraphics[width=.10\linewidth,height=.65cm]{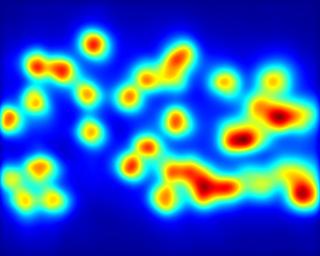}
			\includegraphics[width=.10\linewidth,height=.65cm]{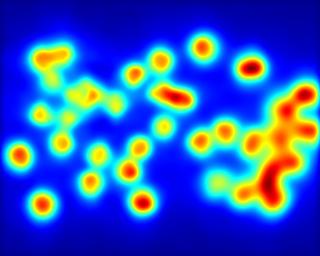}
			\includegraphics[width=.10\linewidth,height=.65cm]{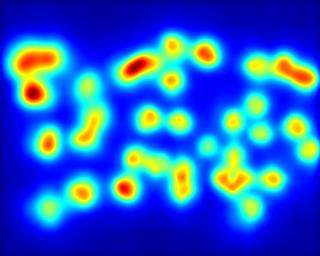}
			\includegraphics[width=.10\linewidth,height=.65cm]{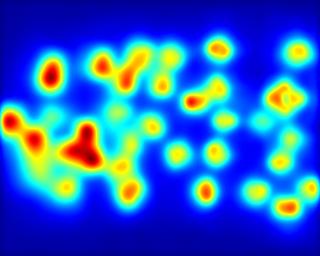}
			\includegraphics[width=.10\linewidth,height=.65cm]{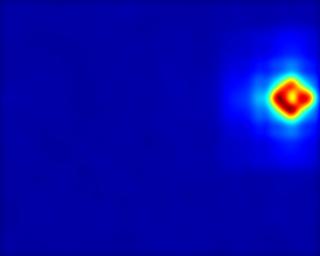}\\
			\makebox[0.16\linewidth]{\scalebox{0.5}{\textbf{NSWAM-CM+VS$_M$}}}
			\includegraphics[width=.10\linewidth,height=.65cm]{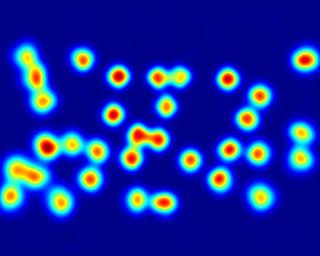}
			\includegraphics[width=.10\linewidth,height=.65cm]{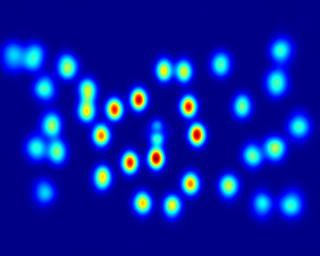}
			\includegraphics[width=.10\linewidth,height=.65cm]{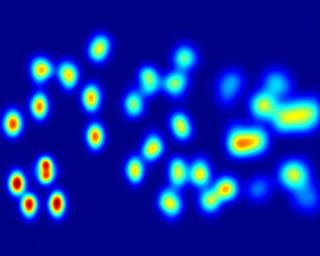}
			\includegraphics[width=.10\linewidth,height=.65cm]{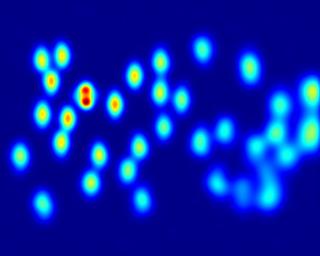}
			\includegraphics[width=.10\linewidth,height=.65cm]{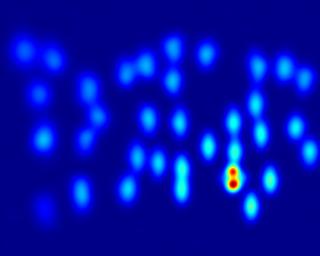}
			\includegraphics[width=.10\linewidth,height=.65cm]{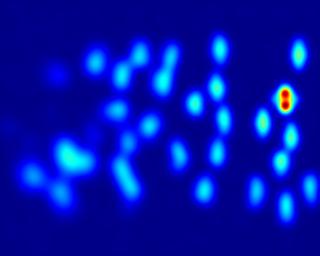}
			\includegraphics[width=.10\linewidth,height=.65cm]{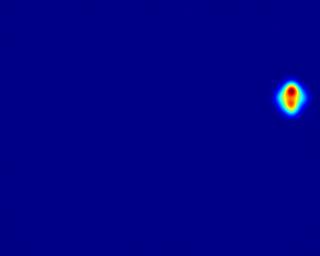}\\
			\makebox[0.16\linewidth]{\scalebox{0.5}{\textbf{NSWAM-CM+VS$_C$}}}
			\includegraphics[width=.10\linewidth,height=.65cm]{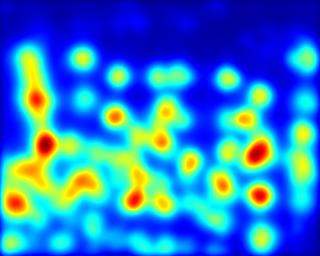}
			\includegraphics[width=.10\linewidth,height=.65cm]{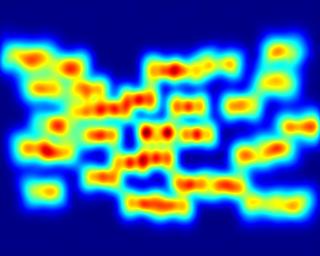}
			\includegraphics[width=.10\linewidth,height=.65cm]{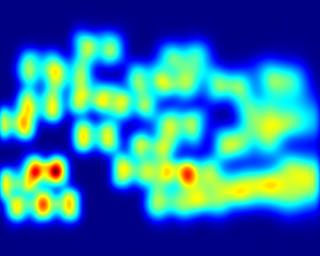}
			\includegraphics[width=.10\linewidth,height=.65cm]{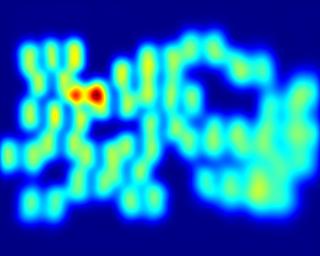}
			\includegraphics[width=.10\linewidth,height=.65cm]{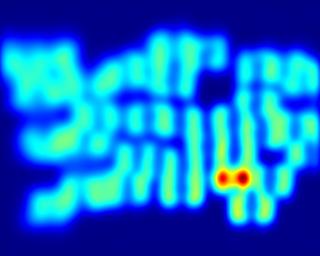}
			\includegraphics[width=.10\linewidth,height=.65cm]{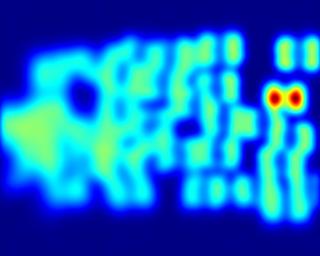}
			\includegraphics[width=.10\linewidth,height=.65cm]{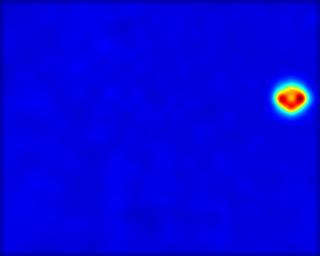}\\
		\end{subfigure} \hspace{-1.5em}
		\begin{subfigure}{.48\linewidth}
			\includegraphics[width=\linewidth,height=4cm]{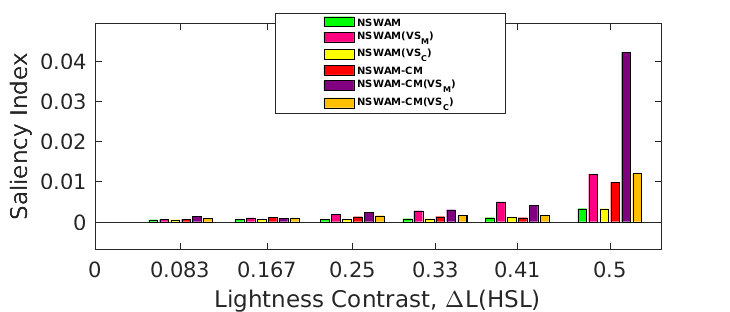}
		\end{subfigure}\\
		\begin{subfigure}{.5\linewidth}
			\makebox[0.16\linewidth]{ }
			\includegraphics[width=.10\linewidth,height=.65cm]{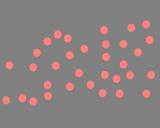}
			\includegraphics[width=.10\linewidth,height=.65cm]{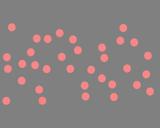}
			\includegraphics[width=.10\linewidth,height=.65cm]{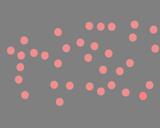}
			\includegraphics[width=.10\linewidth,height=.65cm]{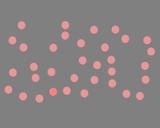}
			\includegraphics[width=.10\linewidth,height=.65cm]{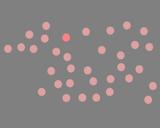}
			\includegraphics[width=.10\linewidth,height=.65cm]{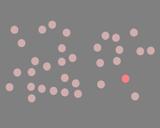}
			\includegraphics[width=.10\linewidth,height=.65cm]{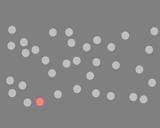}\\
			\makebox[0.16\linewidth]{\scalebox{0.5}{NSWAM}}
			\includegraphics[width=.10\linewidth,height=.65cm]{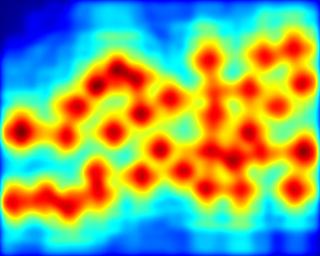}
			\includegraphics[width=.10\linewidth,height=.65cm]{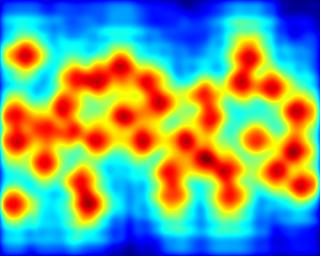}
			\includegraphics[width=.10\linewidth,height=.65cm]{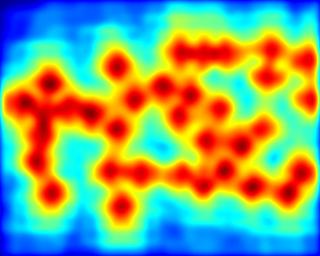}
			\includegraphics[width=.10\linewidth,height=.65cm]{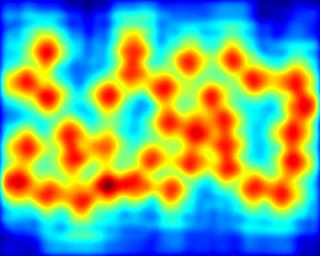}
			\includegraphics[width=.10\linewidth,height=.65cm]{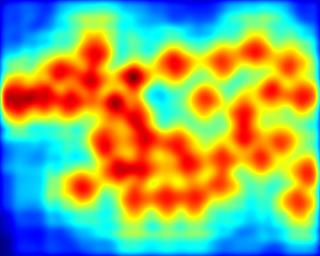}
			\includegraphics[width=.10\linewidth,height=.65cm]{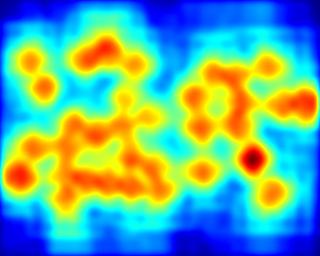}
			\includegraphics[width=.10\linewidth,height=.65cm]{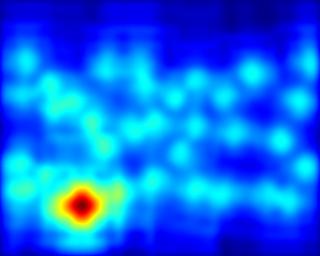}\\
			\makebox[0.16\linewidth]{\scalebox{0.5}{\textbf{NSWAM+VS$_M$}}}
			\includegraphics[width=.10\linewidth,height=.65cm]{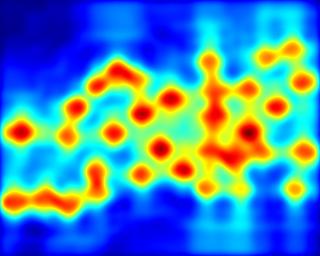}
			\includegraphics[width=.10\linewidth,height=.65cm]{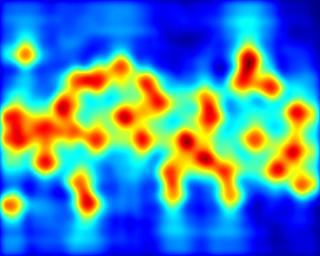}
			\includegraphics[width=.10\linewidth,height=.65cm]{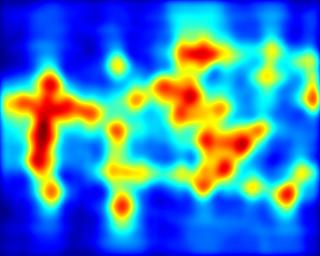}
			\includegraphics[width=.10\linewidth,height=.65cm]{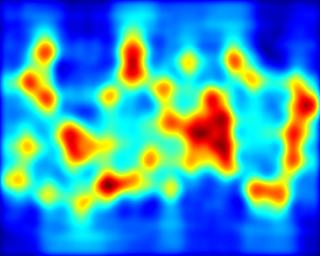}
			\includegraphics[width=.10\linewidth,height=.65cm]{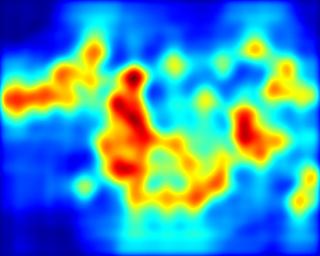}
			\includegraphics[width=.10\linewidth,height=.65cm]{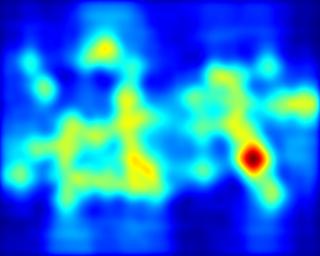}
			\includegraphics[width=.10\linewidth,height=.65cm]{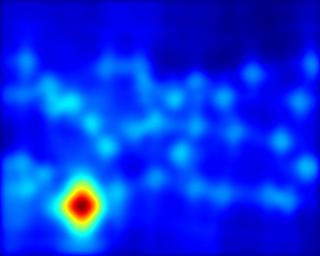}\\
			\makebox[0.16\linewidth]{\scalebox{0.5}{\textbf{NSWAM+VS$_C$}}}
			\includegraphics[width=.10\linewidth,height=.65cm]{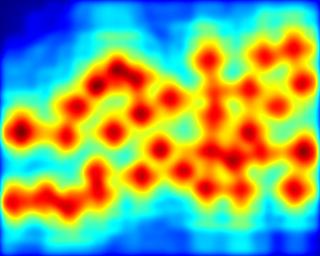}
			\includegraphics[width=.10\linewidth,height=.65cm]{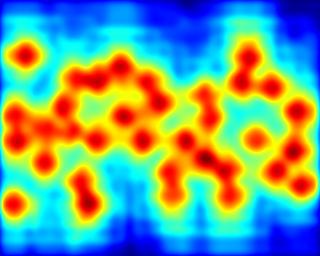}
			\includegraphics[width=.10\linewidth,height=.65cm]{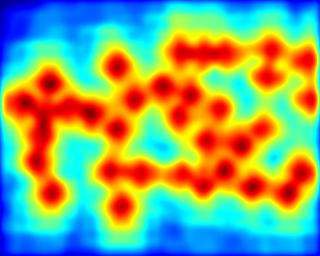}
			\includegraphics[width=.10\linewidth,height=.65cm]{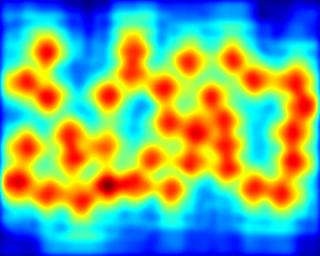}
			\includegraphics[width=.10\linewidth,height=.65cm]{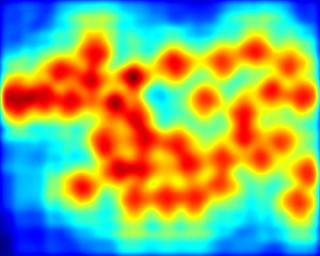}
			\includegraphics[width=.10\linewidth,height=.65cm]{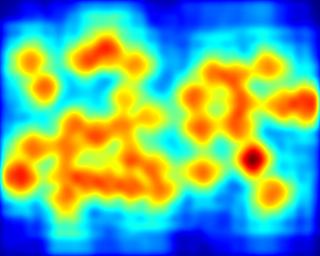}
			\includegraphics[width=.10\linewidth,height=.65cm]{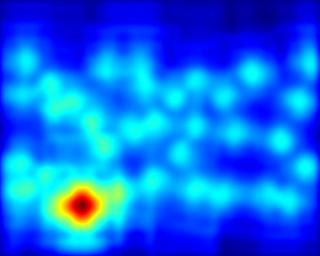}\\
			\makebox[0.16\linewidth]{\scalebox{0.5}{NSWAM-CM}}
			\includegraphics[width=.10\linewidth,height=.65cm]{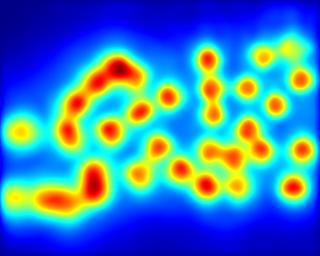}
			\includegraphics[width=.10\linewidth,height=.65cm]{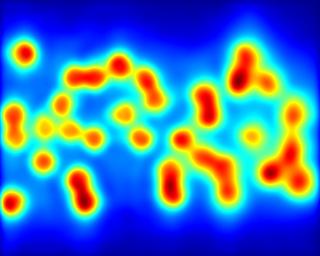}
			\includegraphics[width=.10\linewidth,height=.65cm]{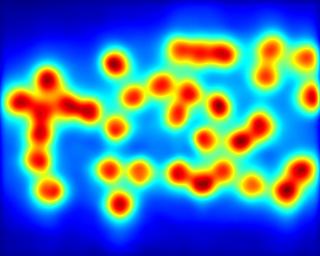}
			\includegraphics[width=.10\linewidth,height=.65cm]{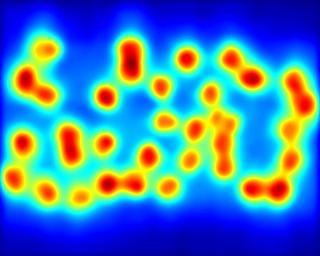}
			\includegraphics[width=.10\linewidth,height=.65cm]{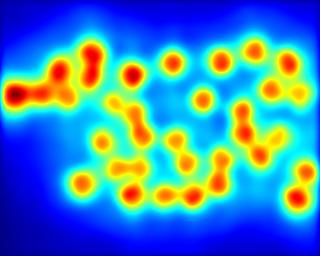}
			\includegraphics[width=.10\linewidth,height=.65cm]{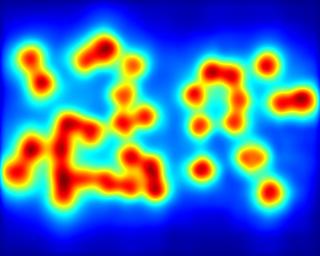}
			\includegraphics[width=.10\linewidth,height=.65cm]{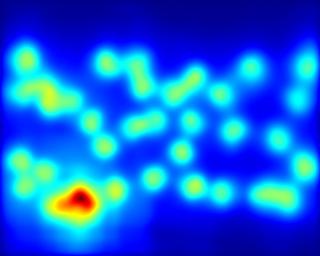}\\
			\makebox[0.16\linewidth]{\scalebox{0.5}{\textbf{NSWAM-CM+VS$_M$}}}
			\includegraphics[width=.10\linewidth,height=.65cm]{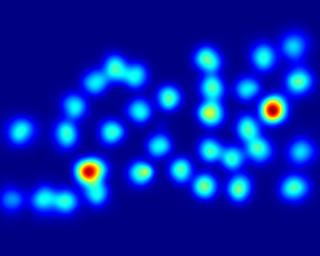}
			\includegraphics[width=.10\linewidth,height=.65cm]{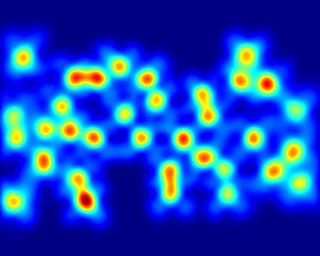}
			\includegraphics[width=.10\linewidth,height=.65cm]{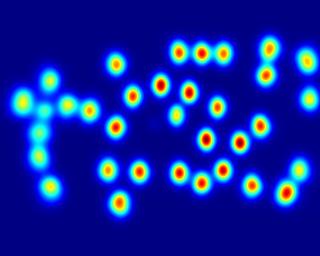}
			\includegraphics[width=.10\linewidth,height=.65cm]{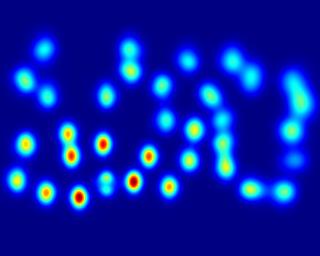}
			\includegraphics[width=.10\linewidth,height=.65cm]{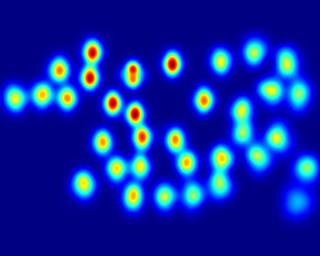}
			\includegraphics[width=.10\linewidth,height=.65cm]{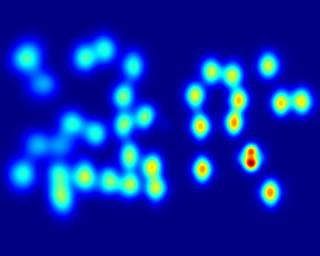}
			\includegraphics[width=.10\linewidth,height=.65cm]{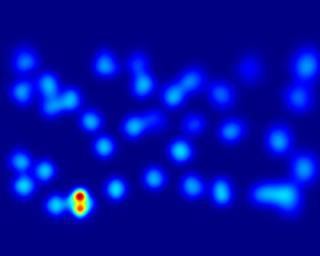}\\
			\makebox[0.16\linewidth]{\scalebox{0.5}{\textbf{NSWAM-CM+VS$_C$}}}
			\includegraphics[width=.10\linewidth,height=.65cm]{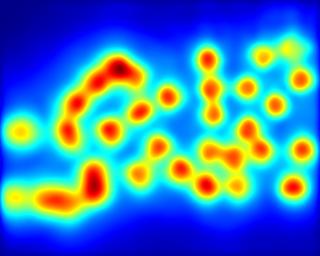}
			\includegraphics[width=.10\linewidth,height=.65cm]{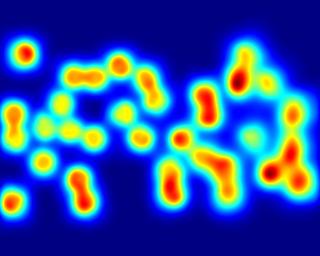}
			\includegraphics[width=.10\linewidth,height=.65cm]{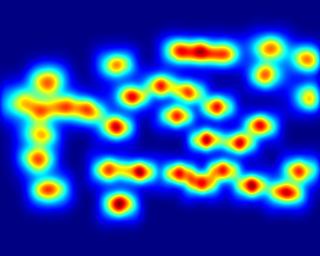}
			\includegraphics[width=.10\linewidth,height=.65cm]{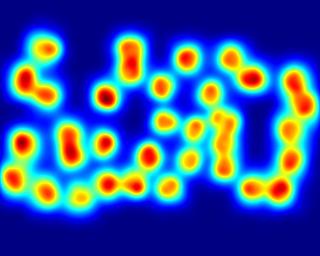}
			\includegraphics[width=.10\linewidth,height=.65cm]{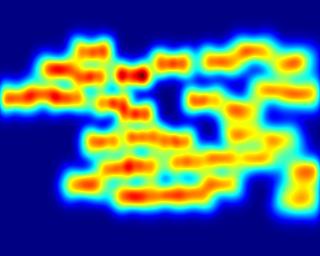}
			\includegraphics[width=.10\linewidth,height=.65cm]{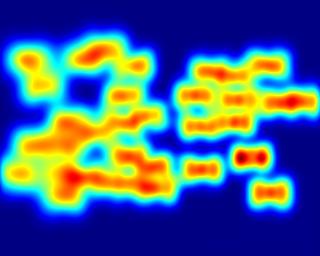}
			\includegraphics[width=.10\linewidth,height=.65cm]{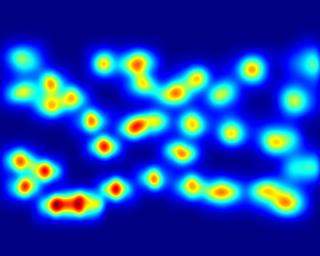}\\
		\end{subfigure}
		\begin{subfigure}{.48\linewidth} \hspace{-2.5em}
			\includegraphics[width=\linewidth,height=4cm]{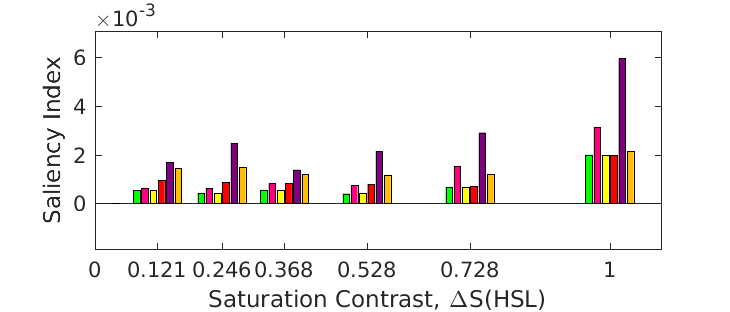}
		\end{subfigure}
		\begin{subfigure}{.5\linewidth}
			\makebox[0.16\linewidth]{ }
			\includegraphics[width=.10\linewidth,height=.65cm]{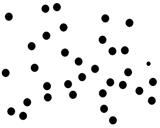}
			\includegraphics[width=.10\linewidth,height=.65cm]{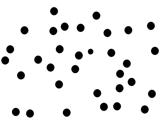}
			\includegraphics[width=.10\linewidth,height=.65cm]{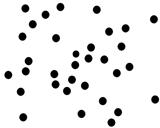}
			\includegraphics[width=.10\linewidth,height=.65cm]{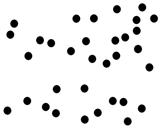}
			\includegraphics[width=.10\linewidth,height=.65cm]{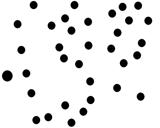}
			\includegraphics[width=.10\linewidth,height=.65cm]{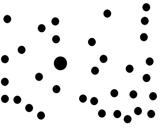}
			\includegraphics[width=.10\linewidth,height=.65cm]{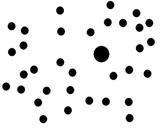}\\
			\makebox[0.16\linewidth]{\scalebox{0.5}{NSWAM}}
			\includegraphics[width=.10\linewidth,height=.65cm]{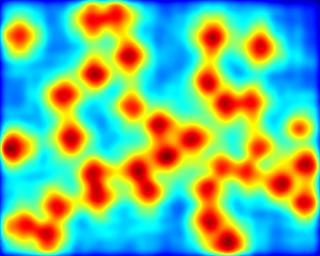}
			\includegraphics[width=.10\linewidth,height=.65cm]{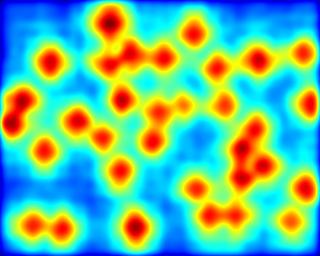}
			\includegraphics[width=.10\linewidth,height=.65cm]{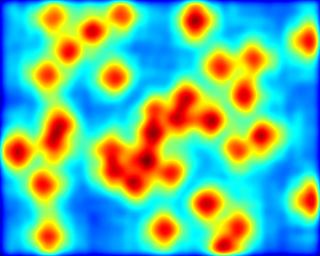}
			\includegraphics[width=.10\linewidth,height=.65cm]{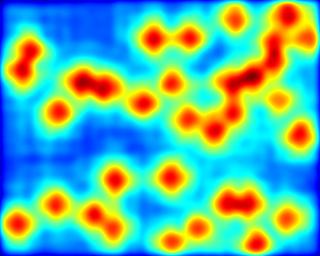}
			\includegraphics[width=.10\linewidth,height=.65cm]{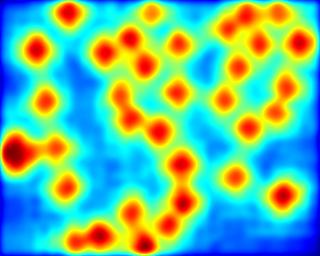}
			\includegraphics[width=.10\linewidth,height=.65cm]{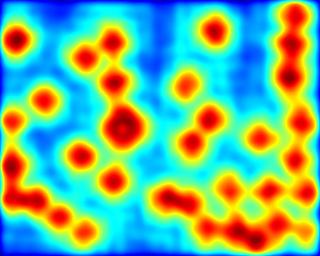}
			\includegraphics[width=.10\linewidth,height=.65cm]{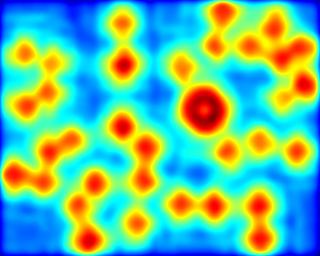}\\
			\makebox[0.16\linewidth]{\scalebox{0.5}{\textbf{NSWAM+VS$_M$}}}
			\includegraphics[width=.10\linewidth,height=.65cm]{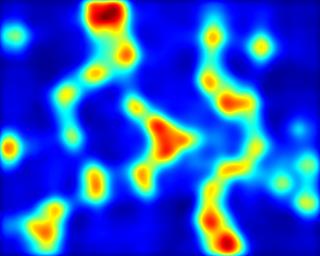}
			\includegraphics[width=.10\linewidth,height=.65cm]{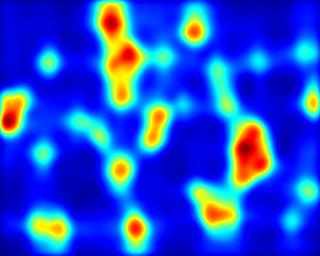}
			\includegraphics[width=.10\linewidth,height=.65cm]{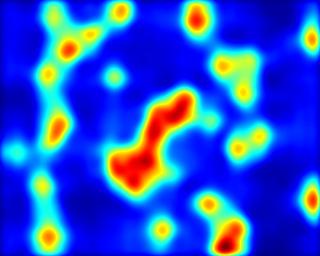}
			\includegraphics[width=.10\linewidth,height=.65cm]{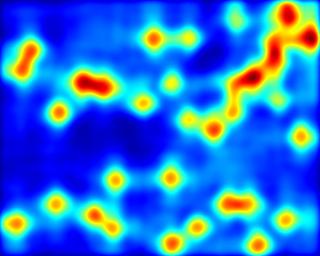}
			\includegraphics[width=.10\linewidth,height=.65cm]{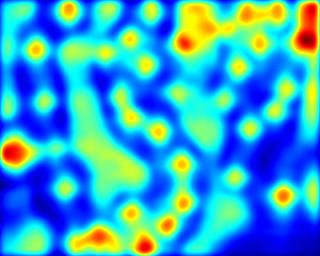}
			\includegraphics[width=.10\linewidth,height=.65cm]{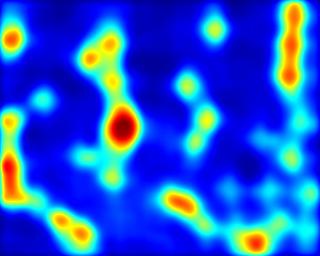}
			\includegraphics[width=.10\linewidth,height=.65cm]{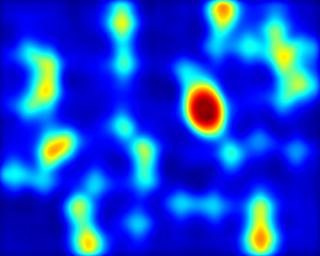}\\
			\makebox[0.16\linewidth]{\scalebox{0.5}{\textbf{NSWAM+VS$_C$}}}
			\includegraphics[width=.10\linewidth,height=.65cm]{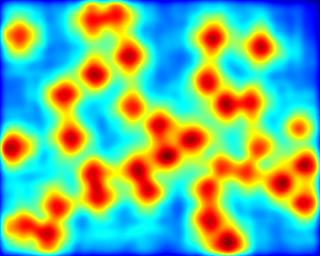}
			\includegraphics[width=.10\linewidth,height=.65cm]{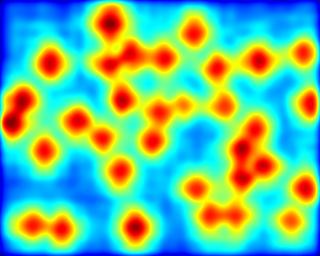}
			\includegraphics[width=.10\linewidth,height=.65cm]{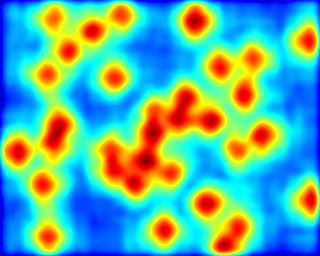}
			\includegraphics[width=.10\linewidth,height=.65cm]{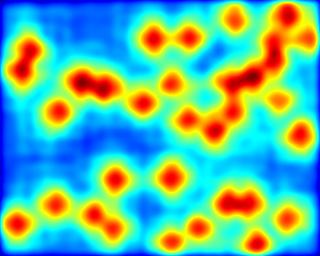}
			\includegraphics[width=.10\linewidth,height=.65cm]{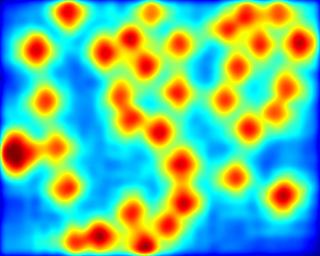}
			\includegraphics[width=.10\linewidth,height=.65cm]{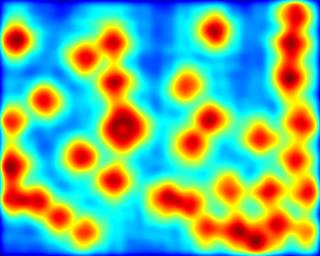}
			\includegraphics[width=.10\linewidth,height=.65cm]{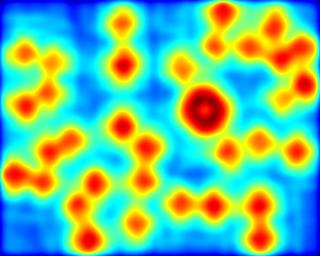}\\
			\makebox[0.16\linewidth]{\scalebox{0.5}{NSWAM-CM}}
			\includegraphics[width=.10\linewidth,height=.65cm]{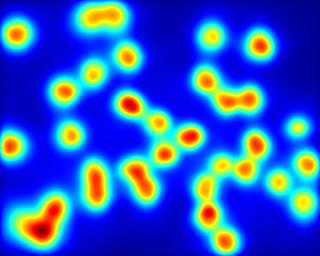}
			\includegraphics[width=.10\linewidth,height=.65cm]{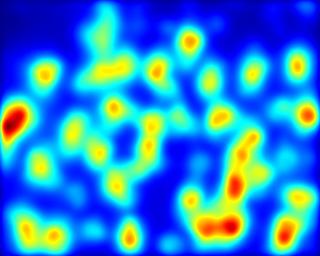}
			\includegraphics[width=.10\linewidth,height=.65cm]{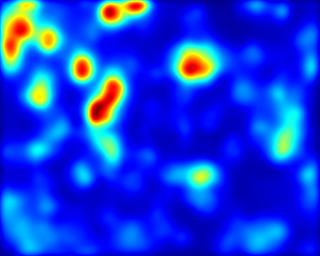}
			\includegraphics[width=.10\linewidth,height=.65cm]{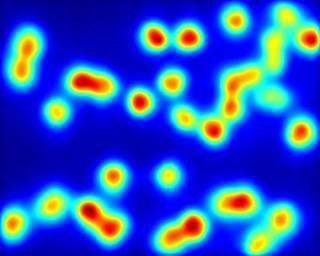}
			\includegraphics[width=.10\linewidth,height=.65cm]{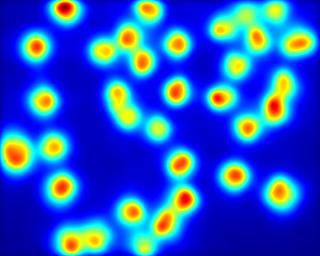}
			\includegraphics[width=.10\linewidth,height=.65cm]{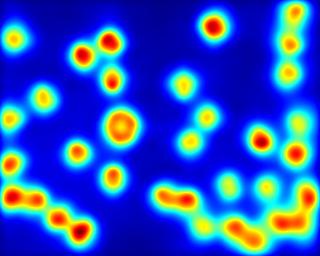}
			\includegraphics[width=.10\linewidth,height=.65cm]{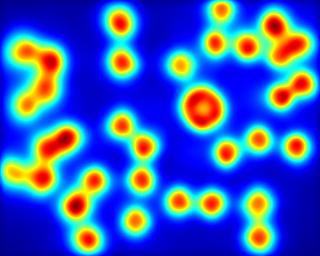}\\
			\makebox[0.16\linewidth]{\scalebox{0.5}{\textbf{NSWAM-CM+VS$_M$}}}
			\includegraphics[width=.10\linewidth,height=.65cm]{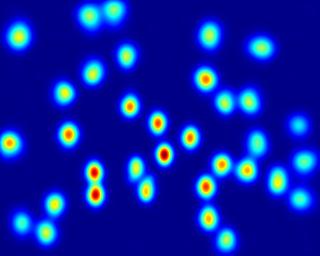}
			\includegraphics[width=.10\linewidth,height=.65cm]{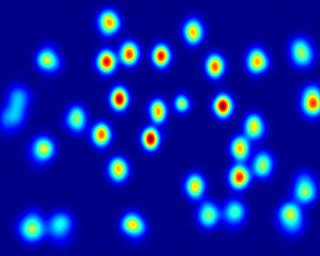}
			\includegraphics[width=.10\linewidth,height=.65cm]{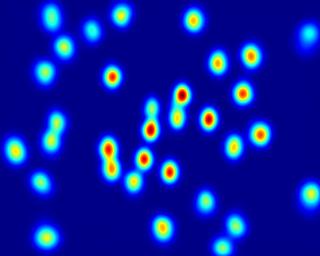}
			\includegraphics[width=.10\linewidth,height=.65cm]{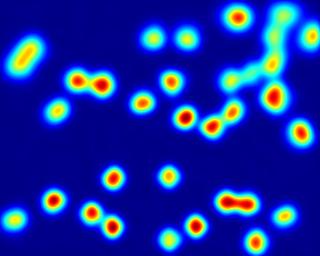}
			\includegraphics[width=.10\linewidth,height=.65cm]{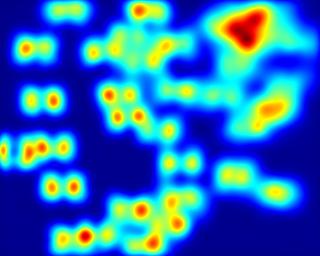}
			\includegraphics[width=.10\linewidth,height=.65cm]{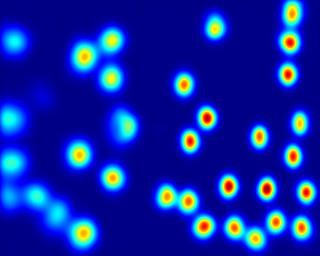}
			\includegraphics[width=.10\linewidth,height=.65cm]{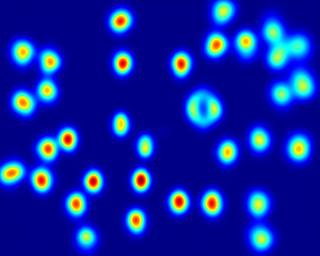}\\
			\makebox[0.16\linewidth]{\scalebox{0.5}{\textbf{NSWAM-CM+VS$_C$}}}
			\includegraphics[width=.10\linewidth,height=.65cm]{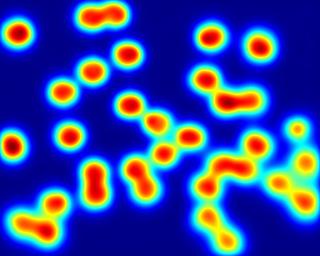}
			\includegraphics[width=.10\linewidth,height=.65cm]{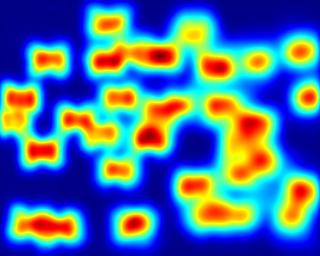}
			\includegraphics[width=.10\linewidth,height=.65cm]{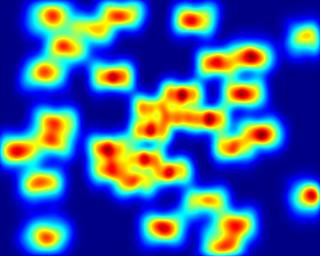}
			\includegraphics[width=.10\linewidth,height=.65cm]{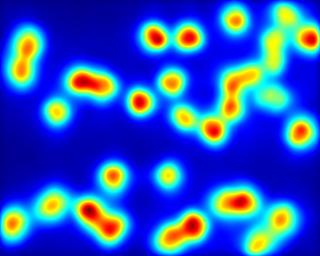}
			\includegraphics[width=.10\linewidth,height=.65cm]{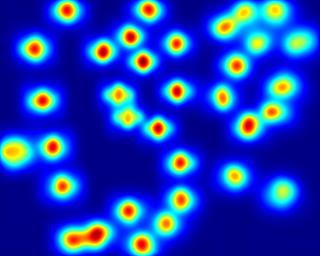}
			\includegraphics[width=.10\linewidth,height=.65cm]{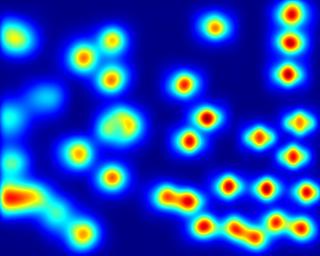}
			\includegraphics[width=.10\linewidth,height=.65cm]{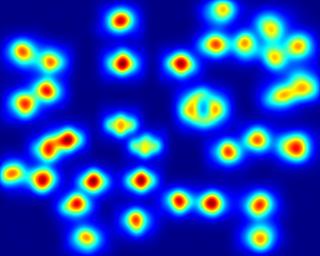}\\
		\end{subfigure}\hspace{-2em}
		\begin{subfigure}{.48\linewidth}
			\includegraphics[width=\linewidth,height=4cm]{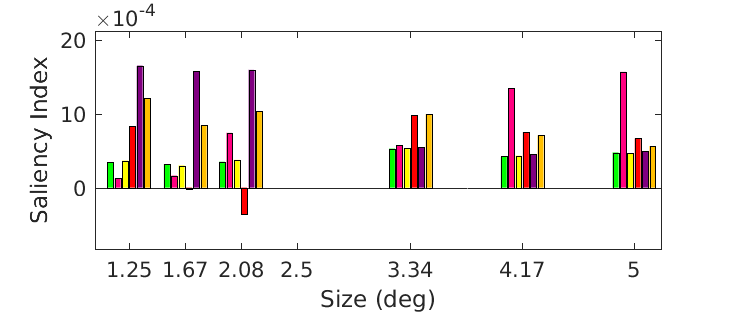}
		\end{subfigure}\\
		\iftrue
		\begin{subfigure}{.5\linewidth}
			\makebox[0.16\linewidth]{ } \includegraphics[width=.10\linewidth,height=.65cm]{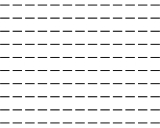}
			\includegraphics[width=.10\linewidth,height=.65cm]{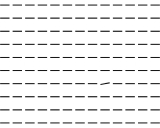}
			\includegraphics[width=.10\linewidth,height=.65cm]{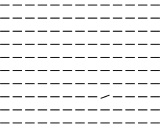}
			\includegraphics[width=.10\linewidth,height=.65cm]{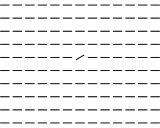}
			\includegraphics[width=.10\linewidth,height=.65cm]{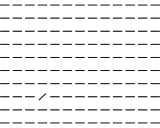}
			\includegraphics[width=.10\linewidth,height=.65cm]{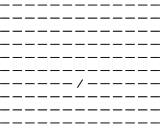}
			\includegraphics[width=.10\linewidth,height=.65cm]{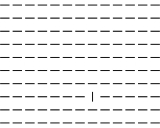}\\
			\makebox[0.16\linewidth]{\scalebox{0.5}{NSWAM}} \includegraphics[width=.10\linewidth,height=.65cm]{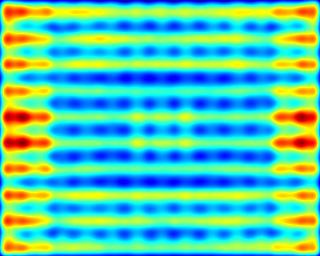}
			\includegraphics[width=.10\linewidth,height=.65cm]{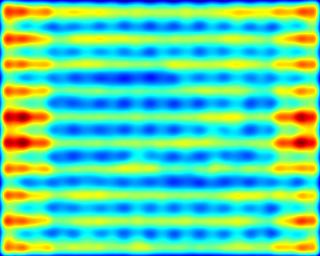}
			\includegraphics[width=.10\linewidth,height=.65cm]{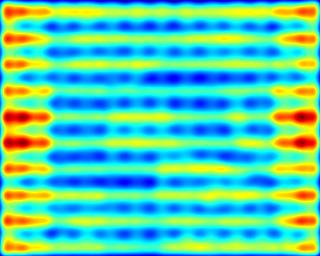}
			\includegraphics[width=.10\linewidth,height=.65cm]{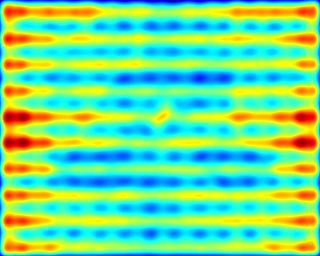}
			\includegraphics[width=.10\linewidth,height=.65cm]{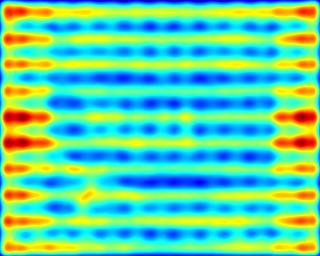}
			\includegraphics[width=.10\linewidth,height=.65cm]{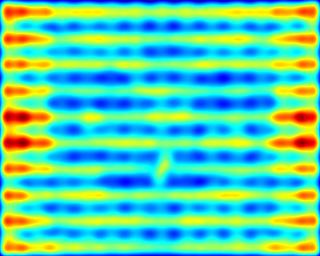}
			\includegraphics[width=.10\linewidth,height=.65cm]{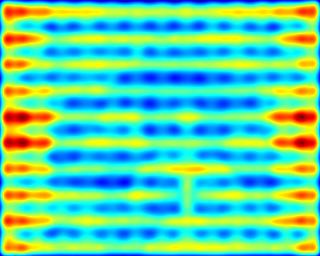}\\
			\makebox[0.16\linewidth]{\scalebox{0.5}{\textbf{NSWAM+VS$_M$}}} \includegraphics[width=.10\linewidth,height=.65cm]{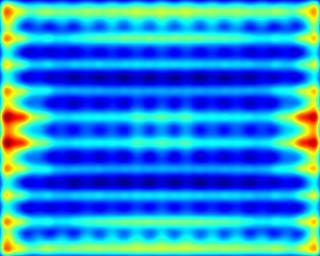}
			\includegraphics[width=.10\linewidth,height=.65cm]{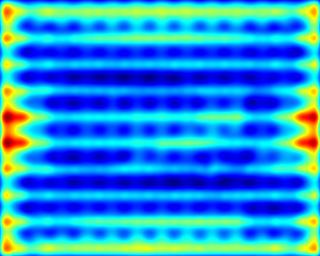}
			\includegraphics[width=.10\linewidth,height=.65cm]{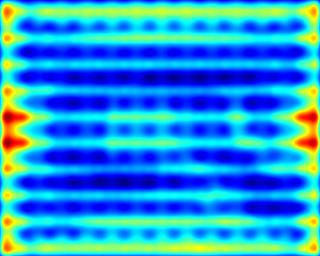}
			\includegraphics[width=.10\linewidth,height=.65cm]{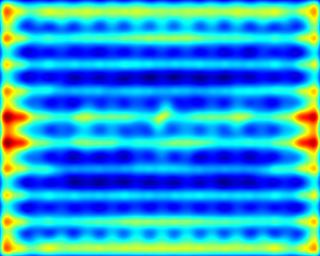}
			\includegraphics[width=.10\linewidth,height=.65cm]{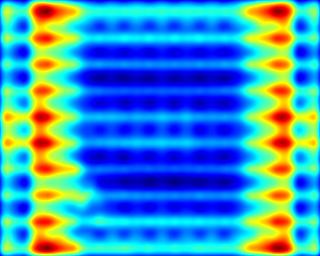}
			\includegraphics[width=.10\linewidth,height=.65cm]{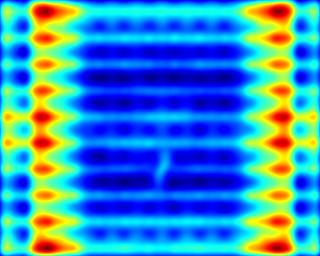}
			\includegraphics[width=.10\linewidth,height=.65cm]{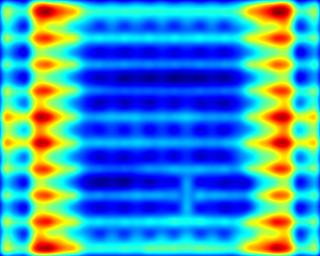}\\
			\makebox[0.16\linewidth]{\scalebox{0.5}{\textbf{NSWAM+VS$_C$}}} \includegraphics[width=.10\linewidth,height=.65cm]{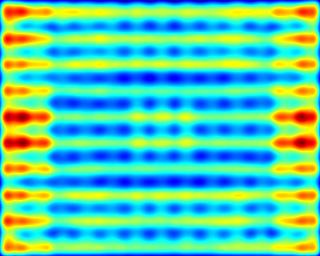}
			\includegraphics[width=.10\linewidth,height=.65cm]{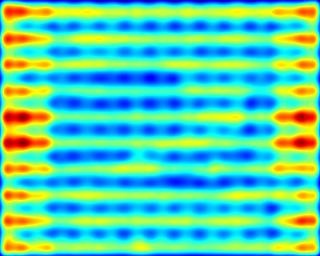}
			\includegraphics[width=.10\linewidth,height=.65cm]{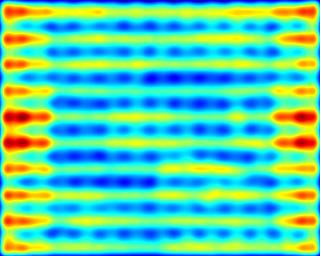}
			\includegraphics[width=.10\linewidth,height=.65cm]{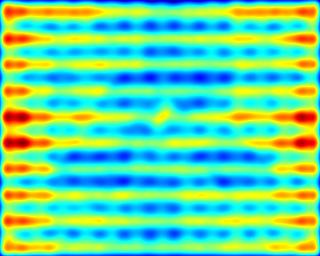}
			\includegraphics[width=.10\linewidth,height=.65cm]{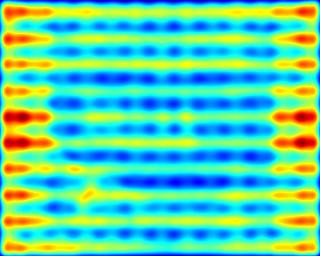}
			\includegraphics[width=.10\linewidth,height=.65cm]{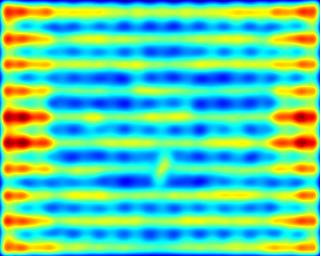}
			\includegraphics[width=.10\linewidth,height=.65cm]{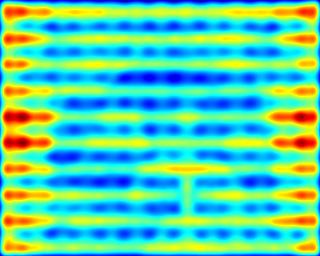}\\
			\makebox[0.16\linewidth]{\scalebox{0.5}{NSWAM-CM}} \includegraphics[width=.10\linewidth,height=.65cm]{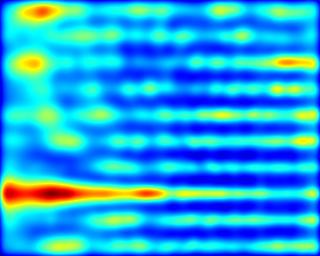}
			\includegraphics[width=.10\linewidth,height=.65cm]{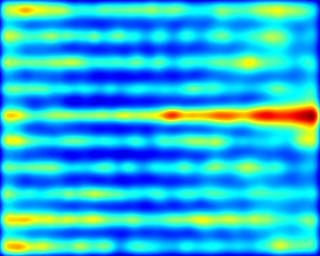}
			\includegraphics[width=.10\linewidth,height=.65cm]{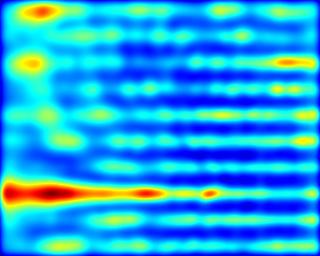}
			\includegraphics[width=.10\linewidth,height=.65cm]{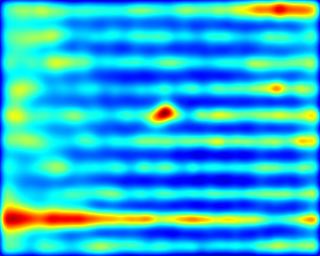}
			\includegraphics[width=.10\linewidth,height=.65cm]{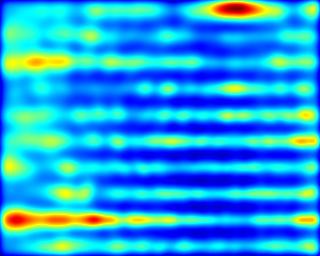}
			\includegraphics[width=.10\linewidth,height=.65cm]{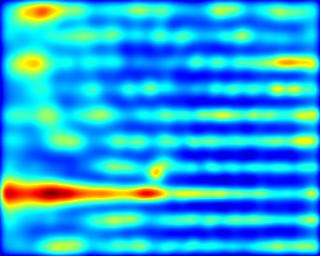}
			\includegraphics[width=.10\linewidth,height=.65cm]{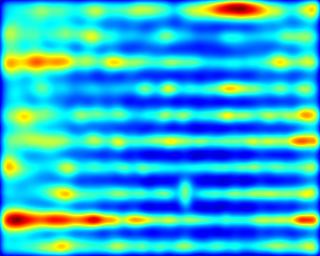}\\
			\makebox[0.16\linewidth]{\scalebox{0.5}{\textbf{NSWAM-CM+VS$_M$}}} \includegraphics[width=.10\linewidth,height=.65cm]{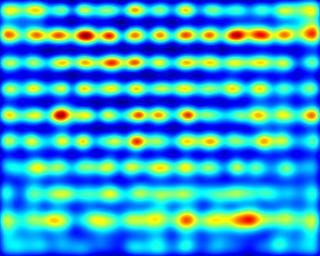}
			\includegraphics[width=.10\linewidth,height=.65cm]{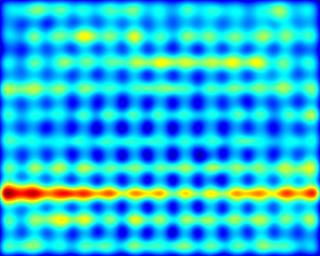}
			\includegraphics[width=.10\linewidth,height=.65cm]{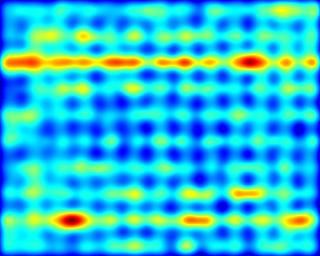}
			\includegraphics[width=.10\linewidth,height=.65cm]{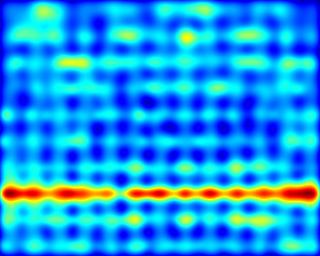}
			\includegraphics[width=.10\linewidth,height=.65cm]{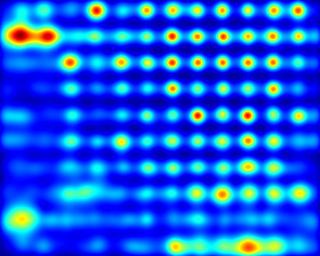}
			\includegraphics[width=.10\linewidth,height=.65cm]{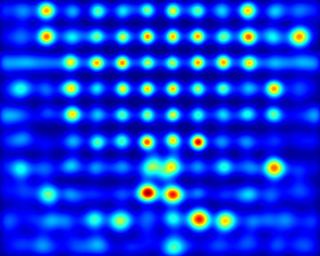}
			\includegraphics[width=.10\linewidth,height=.65cm]{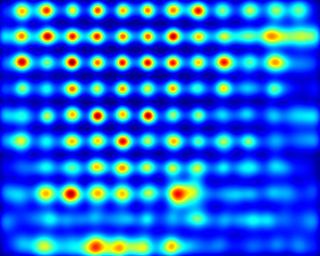}\\
			\makebox[0.16\linewidth]{\scalebox{.5}{\textbf{NSWAM-CM+VS$_C$}}} \includegraphics[width=.10\linewidth,height=.65cm]{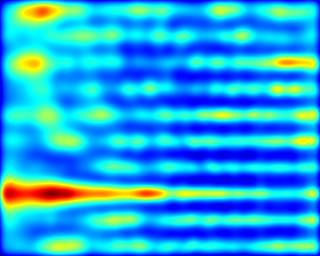}
			\includegraphics[width=.10\linewidth,height=.65cm]{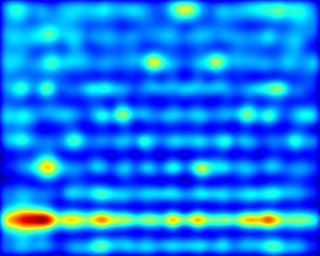}
			\includegraphics[width=.10\linewidth,height=.65cm]{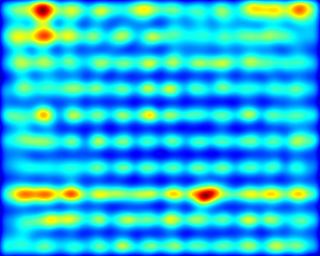}
			\includegraphics[width=.10\linewidth,height=.65cm]{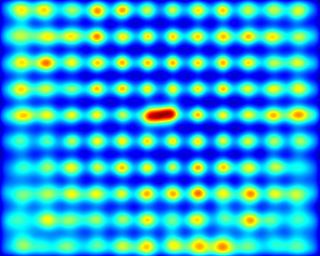}
			\includegraphics[width=.10\linewidth,height=.65cm]{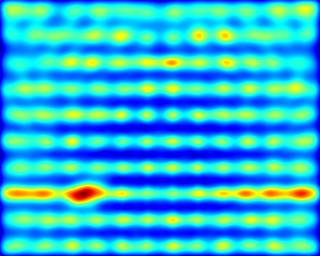}
			\includegraphics[width=.10\linewidth,height=.65cm]{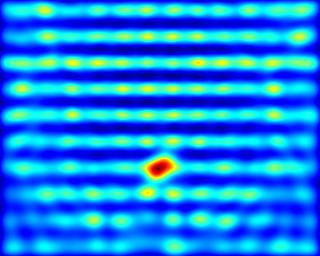}
			\includegraphics[width=.10\linewidth,height=.65cm]{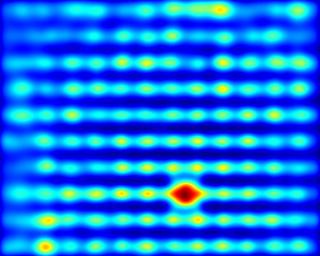}\\
		\end{subfigure}\hspace{-2em}
		\begin{subfigure}{.48\linewidth}
			\includegraphics[width=\linewidth,height=4cm]{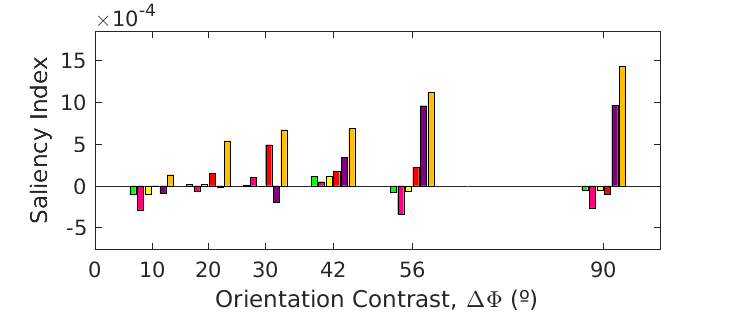}
		\end{subfigure}
		\fi
	\end{adjustwidth}
	\vspace{-1.5em}\caption{Performance on visual search examples with a specific low-level feature contrast (for Brightness, Color, Size and Orientation). We represented 7 instances ordered by search difficulty of each feature sample.}
	\label{fig:results_search3}
\end{figure}

In \hyperref[fig:results_search3]{Fig. \ref*{fig:results_search3}} we illustrated results of PFI and SI in relation to relative target-distractor feature contrast for cases of Brigthness, Color, Size and Orientation differences. After computing SI for each distinct psychophysical stimuli, we can see in \hyperref[fig:results_search4]{Fig. \ref*{fig:results_search4}} that our model performs best for searching objects in stimuli where there are clear differences in brightness, color, size and/or angle, rather than for the case of different combination of features, specially with heterogeneous, nonlinear or categorical angle configurations.

%This phenomena supports the hypothesis of attentional mechanisms select certain maximal RF and inhibiting the rest when deciding saccade shifts \cite{Tsotsos1995,Huang2007}.

%Saliency metrics of sAUC and InfoGain (with Toronto's eye tracking dataset) increase with the search-based strategy \{$\Delta(sAUC)_{VS_M}$=.018, $\Delta(sAUC)_{VS_C}$=.003; $\Delta(InfoGain)_{VS_M}$=.002, $\Delta(InfoGain)_{VS_C}$=.035\}. 

%Free-viewing fixations are seemingly predicted with similar performance in comparison with NSWAM predictions (\hyperref[fig:saliency1]{Fig. \ref*{fig:saliency1}}). For the case of synthetic patterns, saliency metrics are similar or increasing with respect NSWAM for feature singleton search fixations \{$\Delta(sAUC)_{+VS_M}$=3.6$\times10^{-3}$, $\Delta(sAUC)_{+VS_C}$=2.9$\times10^{-3}$; $\Delta(InfoGain)_{+VS_M}$=4.1$\times10^{-2}$, $\Delta(InfoGain)_{+VS_C}$=9.4$\times10^{-4}$\}, but lower or even decreasing for the case of free-viewing \{$\Delta(sAUC)_{+VS_M}$=-12$\times10^{-3}$, $\Delta(sAUC)_{+VS_C}$=-8.7$\times10^{-3}$; $\Delta(InfoGain)_{+VS_M}$=-13.7$\times10^{-2}$, $\Delta(InfoGain)_{+VS_C}$=-3.3$\times10^{-2}$\}. 

%The amount of features to inhibit in categorical search, as 
%(our low-level NSWAM+VS$_M$ computations)

%\hyperref[fig:results_search1]{Figs \ref*{fig:results_search1}-\ref*{fig:results_search2}}

%codi: plot_metrics_synthetic
%FV: results_all_pertask{1}(2:3)-results_all_pertask{1}(1)
%VS: results_all_pertask{2}(2:3)-results_all_pertask{2}(1)

%exemplar search

%feature search

\subsubsection{Discussion}

Overall results show that features computed by the top-down approach seemingly performs better in visual search than saliency, both considering features with maximal cortical activity (VS$_M$) and average statistics of low-level features (VS$_C$). Search in both objects and psychophysical image patterns is significantly more efficient in SI and PFI when selecting maximal feature activations (VS$_M$). Our model is able to localize objects in real scenes, specially when objects are distinct enough from others (in these low-level feature computations). However, the model fails when there are sparse regions of the image that interfere with the selected object (being too salient, such as in \hyperref[fig:results_search2]{Fig. \ref*{fig:results_search2}-"Telephone"}) and when characteristics of some parts of these objects (comprised in the mask) do not significantly pop-out or either coincide with other non-relevant objects (see \hyperref[fig:results_search2]{Fig. \ref*{fig:results_search2}-"Car"}). This could be improved by computing a higher number of features \cite{Mly2016,Berga2019c} (which would represent in more detail each cortical cell sensitivity at higher visual areas of cortex). We can observe that when using both cortical magnification transform and top-down selection (-CM+VS), some non-relevant parts of the image are discriminated easier than using top-down selection alone (see non-relevant artifacts caused by repetitive patterns or wrap-around filtering effects \hyperref[fig:results_search3]{Fig. \ref*{fig:results_search3}-Bottom}). This suggests that using foveation not only can improve performance on localizing objects (\hyperref[fig:results_search1]{Fig. \ref*{fig:results_search1}}) but also that provides biologically-plausible perceptual characteristics not considered in most artificial models. Even if our computations of top-down selection are fed to the model as a constant factor (according to the activity from exemplars), our model's lateral interactions leverage at each saccade the activity from both bottom-up and top-down attention.

\section{General Discussion}

%pathways de saliency cap a SC
Current implementation of our V1 model is based on Li's excitatory-inhibitory firing rate network \cite{Li1998}, following previous hypotheses of pyramidal and interneuron connectivity for orientation selectivity in V1 \cite{Gilbert1992,Weliky1995}. To support and extend this hypothesis, distinct connectivity schemas (following up V1 cell subtype characterization) \cite{Grossberg2016,Lee2017} could be tested (e.g. adding dysynaptic connections between inhibitory interneurons) to better understand V1 intra-cortical computations. Furthermore, modeling intra-layer interactions of V1 cells \cite{Sincich2005} could explain how visual information is parallely processed and integrated by simple and complex cells \cite{Mly2016}, how distinct chromatic opponencies (P-,K- and M-) are computed at each layer \cite{Johnson2008}, and how V1 responses affect SC activity (i.e. from layer 5) \cite{Nhan_Callaway_2011}. Testing contributions of each of these chromatic pathways (at distinct single/double opponencies and polarities), as well as distinct fusion mechanisms regarding feature integration, would define a more detailed description of how visual features affect saliency map predictions. %\cite{MartnezCaada2018}

%saliency, scanpaths and time
Previous and current scanpath model predictions could be considered to be insufficient due to the scene complexity and numerous factors (such as the task specificity, scene semantics, etc.) simultaneously involved in saccade programming. These factors increase overall errors on scanpath predictions, as systematic tendencies increase over time \cite{Egeth1997,Tatler2005,Rothkegel2017,Berga2018a}, making late saccades difficult to predict. In that aspect, in free-viewing tasks (when there is no task definition), top-down attention is likely to be dependent on the internal state of the subject. Further understanding of high level attentional processes have only been approximated through statistical and optimization techniques uniquely with fixation data (yet participant decisions on fixations are not accounted and usually have high variability). It has also been later observed that fixations during free-viewing and visual search have distinct temporal properties. This could explain that saliency and relevance are elicited differently during viewing time. Latest literature on that aspect, discern two distinct patterns of fixations (either ambient or focal) where subjects first observe the scene (possibly towards salient regions), then focus their attention on regions that are relevant to them \cite{Eisenberg2016}, and these influences are mainly temporal. Its modelization for eye movements in combination with memory processing is still under discussion. Current return mechanisms have long been computed by inhibiting the regions of previous fixations (spatially-based), nonetheless, IoR could also have feature-selective properties \cite{Hu2014} to consider. 

%saliency and high-level features
We suggest that not all fixations should have the same importance when evaluating saliency predictions. Nature and synthetic scene images lack of semantic (man-made) information, which might contribute to the aforementioned voluntary (top-down guided) eye movements \cite{Hwang2011}. Acknowledging that objects are usually composed by the combination of several features (either in shape, color, etc.), we should analyze if low-level features are sufficient to perform complex categorical search tasks. Extrastriate computations could allow the usage of object representations at higher-level processing, introducing semantically-relevant information and several image samples per category. Cortical processing of extrastriate areas (from V2 and V3) towards temporal (V4 \& IT) and dorsal (V5 \& MT) pathways \cite[Section~II]{werner2014the}\cite{Sincich2005} could represent cortical activity at these distinct levels of processing, modeling in more detail the computations within the two-stream hypothesis (what \& where pathways). Color, shape and motion processing in each of these areas could generate more accurate representations of SC activity \cite{WhiteMunoz2011}, producing more complex predictions such as microsaccadic and smooth pursuit eye movements with dynamic scenes.

\section{Future Work}

Current and future implementations of the model are able to process dynamic stimuli as to represent attention using videos. By simulating motion energy from V1 cells and MT direction selective cells \cite[Section~2.3.5]{zhaoping2014understanding}, would allow our model to reproduce object motion and flicker mechanisms found in the HVS. Moreover, foveation through more plausible cortical mapping algorithms \cite{Schira2010} could provide better spatial detail of the cortical field organization of foveal and peripheral retinotopic regions and lateralization, currently seen to reproduce V1/V2/V3 physiological responses. Adding to that, hypercolumnar feature computations of geniculocortical pathways could be extended with a higher number of orientation and scale sensitivities with self-invertible 2D Log-Gabor filters \cite{Fischer2007}. In that regard, angle configuration pop-out effects and contour detection computations \cite{Asenov2016,Anzai2007} can be done by changing neuron connectivity and orientation tuning modulations. Spatiotemporal convolutions shown for center-surround RF \cite{Somers1995} could be integrated for mimicking the dynamics and feature tuning at each pre-cortical pathway. %Further implementations could reproduce distinct perceptual effects \cite{Zhaoping2002,Zhaoping2005,zhaoping2014understanding}.

We aim in future implementations to model the impact of feedback in cortico-cortical interactions with respect striate and extrastriate areas in the HVS. Some of these regions project directly to SC, including the intermediate areas (pulvinar and medial dorsal) and basal ganglia \cite{WhiteMunoz2011,PierrotDeseilligny2003,PierrotDeseilligny2004}. Our current implementation can be extended with a large scale network of spiking neurons \cite{Izhikevich2004,tavanaei2018deep}, also being able to learn certain image patterns through spike-timing dependent plasticity (STDP) mechanisms \cite{masquelier2007unsupervised}. With such a network, the same model would be able to perform both psychophysical and electrophysiological evaluations while providing novel biologically-plausible computations with large scale image datasets. %Izhikevich2008,\cite{Kruger2013}, using NEST Simulator software \cite{Gewaltig:NEST},\cite{Shipp2002},Liu2016 %\cite{Beyeler2014,Pearson2016}

%\cite{Pang2018} %mencionar proximity preference (Koch and ullman) / oculomotor biases / LeMeur saccadic biases

\section{Conclusion}

In this study we have presented a biologically-plausible model of visual attention by mimicking visual mechanisms from retina to V1 using real images. From such, computations at early visual areas of the HVS (i.e. RP, RGC, LGN and V1) are performed by following physiological and psychophysical characteristics. Here we state that lateral interactions of V1 cells are able to obtain real scene saliency maps and to predict locations of visual fixations. We have also proposed novel scanpath computations of scene visualization using a cortical magnification function. Our model outperforms other biologically inspired saliency models in saliency predictions (specifically with nature and synthetic images) and has a trend to acquire similar scanpath prediction performance with respect other artificial models, outperforming them in saccade amplitude correlations. The aim of this study, besides from acquiring state-of-the-art results, is to explain how lateral connections can predict visual fixations and how these can explain the role of V1 in this and other visual effects. In addition, we formulated projections of recurrent and selective attention using the same model (simulating frontoparietal top-down inhibition mechanisms). Our implementation of these, included top-down projections from DLPFC, FEF and LIP (regarding visual selection and inhibition of return mechanisms). We have shown how scanpath predictions improve by parametrizing the inhibition of return, with highest performance at a size of 2 deg and a decay time between 1 and 5 fixations. By processing low-level feature representations of real images (considering statistics of wavelet coefficients for each object or feature exemplar) and using them as top-down cues, we have been able to perform feature and object search using the same computational architecture. Two search strategies are presented, and we show that both the probability to gaze inside a ROI and the amount of fixations inside that ROI increase with respect saliency. In previous studies, the same model has been able to reproduce brightness \cite{Penacchio2013} and chromatic \cite{Cerda2016} induction, as well as explaining V1 cortical hyperexcitability as a indicator of visual discomfort \cite{Penacchio2016}. With the same parameters and without any type of training or optimization, NSWAM is also able predict bottom-up and top-down attention for free-viewing and visual search tasks. Model characteristics has been constrained (in both architecture and parametrization) with human physiology and visual psychophysics, and can be considered as a simplified and unified simulation of how low-level visual processes occur in the HVS.

%%%%%%%%%%%%%%%%%%%%%%%%%%%%%%%%%%%%%%%%%%%%%%%%%
%%%%%%%%%%%%%%%%%%%%%%%%%%%%%%%%%%%%%%%%%%%%%%%%%
%%%%%%%%%%%%%%%%%%%%%%%%%%%%%%%%%%%%%%%%%%%%%%%%%

%\section*{Supporting information}
%\paragraph*{S1 Table.}\label{S1_Table}{\bf Hola.} 

\section*{Acknowledgments}
This work was funded by the Spanish Ministry of Economy and Competitivity (DPI2017-89867-C2-1-R), Agencia de Gesti\'o d'Ajuts Universitaris i de Recerca (AGAUR) (2017-SGR-649), and CERCA Programme / Generalitat de Catalunya.

%\nolinenumbers

	\bibliographystyle{unsrt}
	\bibliography{library}
	
\end{document}